\documentclass[12pt, reqno]{amsart}

\usepackage[a4paper, margin=1.in, foot=.3in]{geometry}

\usepackage{graphicx, color}

\definecolor{darkblue}{rgb}{0,0,.5}
\definecolor{darkred}{rgb}{.5,0,0}
\definecolor{darkgreen}{rgb}{0,0.5,0}

\usepackage{hyperref}
\hypersetup {
    pdfstartview=FitH,
    colorlinks,
    citecolor=darkgreen,
    linkcolor=darkred,
    urlcolor=darkblue
}

\usepackage[noBBpl,slantedGreek]{mathpazo}

\let\oldtocsection=\tocsection

\let\oldtocsubsection=\tocsubsection

\let\oldtocsubsubsection=\tocsubsubsection

\renewcommand{\tocsection}[2]{\hspace{0em}\oldtocsection{#1}{#2}}
\renewcommand{\tocsubsection}[2]{\hspace{1em}\oldtocsubsection{#1}{#2}}
\renewcommand{\tocsubsubsection}[2]{\hspace{2em}\oldtocsubsubsection{#1}{#2}}

\allowdisplaybreaks
\numberwithin{equation}{section}

\makeatletter
\def\subsection{\@startsection{subsection}{2}%
  \z@{.5\linespacing\@plus.7\linespacing}{-.5em}%
  {\normalfont\bfseries\mathversion{bold}}}
\makeatother

\usepackage{cite}

\usepackage{amssymb}

\allowdisplaybreaks

\numberwithin{equation}{section}

\def \bm #1{\mbox{\boldmath $#1$}}
\def \eqWithRef #1{\underset {\mbox {\tiny (\ref {#1})}}{=}}

\def \bbC {\mathbb C}

\def \bbO {\mathbb O}
\def \bbQ {\mathbb Q}

\def \bbT {\mathbb T}
\def \bbZ {\mathbb Z}

\def \calC {\mathcal C}
\def \calE {\mathcal E}

\def \calK {\mathcal K}
\def \calL {\mathcal L}
\def \calM {\mathcal M}

\def \calQ {\mathcal Q}
\def \calR {\mathcal R}
\def \calT {\mathcal T}
\def \calV {\mathcal V}

\def \gothb {\mathfrak b}
\def \gothg {\mathfrak g}
\def \gothgl {\mathfrak{gl}}
\def \gothh {\mathfrak h}
\def \gothgl {\mathfrak{gl}}
\def \gothsl {\mathfrak{sl}}

\def \End {\mathrm{End}}
\def \Hom {\mathrm{Hom}}
\def \id {\mathrm{id}}
\def \Mat {\mathrm{Mat}}
\def \op {\mathrm{op}}
\def \Osc {\mathrm{Osc}}
\def \rmd {\mathrm d}
\def \rme {\mathrm e}

\def \rmp {\mathrm p}
\def \tr {\mathrm{tr}}

\newcommand{\uqsl}{U_q(\gothsl_2)}
\newcommand{\uqlsl}{U_q(\calL(\gothsl_2))}
\newcommand{\uqtlsl}{U_q(\widetilde{\calL}(\gothsl_2))}
\newcommand{\hlsl}{\widehat{\calL}(\gothsl_2)}
\newcommand{\uqhlsl}{U_q(\widehat{\calL}(\gothsl_2))}
\newcommand{\uqbp}{U_q(\gothb_+)}
\newcommand{\uqbm}{U_q(\gothb_-)}

\title[Universal $R$-matrix and functional relations]{Universal $\mathbold{R}$-matrix and functional relations}

\author[H. Boos]{Herman Boos}
\address{Fachbereich C -- Physik, Bergische Universit\"at Wuppertal, 42097 Wuppertal, Germany}
\email{boos@physik.uni-wuppertal.de}

\author[F. G\"ohmann]{Frank G\"ohmann}
\address{Fachbereich C -- Physik, Bergische Universit\"at Wuppertal, 42097 Wuppertal, Germany}
\email{goehmann@physik.uni-wuppertal.de}

\author[A. Kl\"umper]{Andreas Kl\"umper}
\address{Fachbereich C -- Physik, Bergische Universit\"at Wuppertal, 42097 Wuppertal, Germany}
\email{kluemper@uni-wuppertal.de}

\author[Kh. S. Nirov]{\vskip .2em Khazret S. Nirov}
\address{Institute for Nuclear Research of the Russian Academy of Sciences, 60th October Ave 7a,
117312 Moscow, Russia} \email{nirov@inr.ac.ru}
 \curraddr{Fachbereich C -- Physik, Bergische
Universit\"at Wuppertal, 42097 Wuppertal, Germany}
\email{nirov@uni-wuppertal.de}

\author[A. V. Razumov]{Alexander V. Razumov}
\address{Institute for High Energy Physics, 142281 Protvino, Moscow region, Russia}
\curraddr{Max-Planck-Institut f\"ur Mathematik, Vivatsgasse, 7, 53111, Bonn, Germany}
\email{Alexander.Razumov@ihep.ru}

\begin{document}

\begin{abstract}
We collect and systematize general definitions and facts on the application of quantum groups to
the construction of functional relations in the theory of integrable systems. As an example, we
reconsider the case of the quantum group $\uqlsl$ related to the six-vertex model. We prove the
full set of the functional relations in the form independent of the representation of the quantum
group in the quantum space and specialize them to the case of the six-vertex model.
\end{abstract}

\maketitle

\tableofcontents

\section{Introduction}

The famous Onsager's solution \cite{Ons44} of the square lattice Ising model was the first
essential result in the field of two-dimensional quantum integrable statistical lattice models. The
next step was made by Lieb \cite{Lie67a, Lie67b, Lie67c, Lie67d} who used the Bethe ansatz
\cite{Bet31} to solve different partial cases of the six-vertex model. His results were generalized
to the general case of the six-vertex model by Sutherland \cite{Sut67}. Later, Baxter proposed the
method of functional relations \cite{Bax71a, Bax72a, Bax73a, Bax73b, Bax73c, Bax82, Bax04} to solve
statistical models which cannot be treated with the help of the Bethe ansatz. The method works for
the cases where the Bethe ansatz can be applied as well. It appears that its main ingredients,
transfer matrices and $Q$-operators, are essential not only for the integration of the
corresponding quantum statistical models in the sense of calculating the partition function in the
thermodynamic limit. One of the remarkable recent applications is their usage in the construction
of the fermionic basis \cite{BooJimMiwSmiTak07, BooJimMiwSmiTak09, JimMiwSmi09, BooJimMiwSmi10} for
the observables of the XXZ spin chain closely related to the six-vertex model.

It seems that the most productive, although not comprehensive, modern approach to the theory of
quantum integrable systems is the approach based on the concept of quantum group invented by
Drinfeld and Jimbo \cite{Dri87, Jim85}. In this approach, all the objects describing the model and
related to its integrability originate from the universal $R$-matrix, and the functional relations
are consequences of the properties of the appropriate representations of the quantum group. It was
clearly realized by Bazha\-nov, Lukyanov and Zamolodchikov \cite{BazLukZam96, BazLukZam97,
BazLukZam99}. The present paper can be considered as an introduction to the application of the
theory of quantum groups to formulation of integrable systems and derivation of the corresponding
functional relations. We were prompted to write it by the absence of a detailed and exhaustive
consideration of the method in the literature. One more reason was a desire to fix the terminology
and notations for our future research.

In section \ref{s:slvm}, we discuss the original approach to formulation and investigation of
quantum square lattice vertex models. We introduce basic objects, and for the case of the
six-vertex model reproduce the Baxter's reasonings for the appearance of the functional relations.
In section \ref{s:odurm}, all the objects describing an integrable lattice vertex model and used to
integrate it are constructed starting from a quantum group. Two representations of the quantum
group should be fixed to describe a model. Here, particularly by historical reasons, the
corresponding representation spaces are called the auxiliary space and the quantum space. In most
cases there is an associated quantum mechanical model defined in the quantum space or its tensor
power. In fact, a lattice model arises when we take finite-dimensional representations, and the
associated quantum mechanical model here is some spin chain. The basic example here is the
six-vertex model and XXZ spin chain, see, for example, the book by Baxter \cite{Bax82}. If the
quantum space is the representation space of certain infinite-dimensional vertex representation of
the quantum group, we have a two-dimensional field theory \cite{BazLukZam96, BazLukZam97,
BazLukZam99, BazHibKho02}. In section \ref{s:e} we consider the case of the quantum group $\uqlsl$.
The full set of functional relations in the universal form independent of the choice of
representation of the quantum group in the quantum space is derived in section \ref{s:fr}.

We assume that the reader is acquainted with the basic facts on quantum groups. Beside the original
papers \cite{Dri87, Jim85}, we recommend for this purpose the book by Chari and Pressley~\cite{ChaPre94}.

Below, a vector space is a vector space over the field $\bbC$ of complex numbers, and an algebra is
a complex associative unital algebra. In fact, all general definitions, given in section \ref{s:odurm},
can easily be extended to the case of algebras and vector spaces over a general field or even a general
commutative unital ring.

The symbol `$\otimes$' is used for the tensor product of vector spaces, for the tensor product of
homomorphisms and for the Kronecker product of matrices. Depending on the context, the symbol `$1$'
means the number one, the unit of an algebra or the unit matrix. We denote by $\calL(\gothg)$ the
loop Lie algebra of a finite-dimensional simple Lie algebra $\gothg$, by $\widetilde \calL(\gothg)$
its standard central extension, and by $\widehat \calL(\gothg)$ the Lie algebra obtained from
$\widetilde \calL(\gothg)$ by adjoining the standard derivation, see, for example, the book by Kac
\cite{Kac90}.

\section{Square lattice vertex models} \label{s:slvm}

\subsection{Vertex models and transfer matrix}

Here we recall the basic facts on integrable two-dimensional square lattice vertex models and show
how functional relations arise in the case of the six-vertex model.

First of all, the models in question are quantum statistical models whose properties in the state
of thermodynamic equilibrium are described by the partition function. Label by $\calC$ the possible
eigenstates of the Hamiltonian\footnote{For lattice models we usually call an eigenstate of the
Hamiltonian a configuration of the system.} of the system under consideration and denote by
$\calE_\calC$ the corresponding energy. The partition function is\footnote{We restrict ourselves to
consideration of the canonical ensemble, see, for example, the book \cite{Hua87}.}
\begin{equation*}
Z = \sum_{\calC} \exp (- \beta \calE_\calC),
\end{equation*}
where $\beta = 1 / k_{\mathrm B} T$ with $k_B$ the Boltzmann constant and $T$ the temperature.
The quantity $\exp (- \beta \calE_\calC)$ is called the Boltzmann weight of the configuration $\calC$.

\begin{figure}[htb]
\centering
\includegraphics{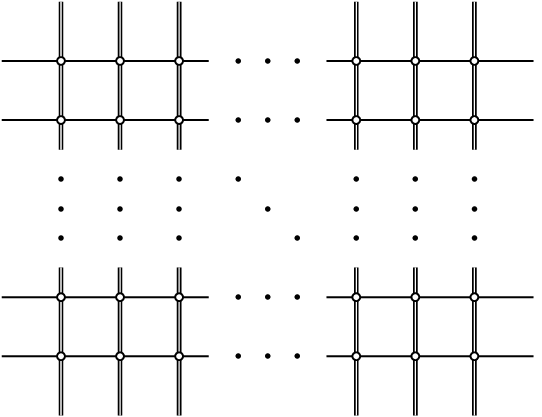}
\caption{Two-dimensional square lattice.}
\label{Lattice}
\end{figure}

Consider now a two-dimensional square lattice, see Figure \ref{Lattice},
and assume that some particles are located at its vertices. Any horizontal bond of the lattice can
be in one of $\ell$ states, and a vertical bond in one of $k$ states. This defines a configuration
of the system. Usually one also imposes boundary conditions. The simplest case here is the periodic
boundary condition, where for each horizontal and vertical row the state of the first bond
coincides with the state of the last bond.

\begin{figure}[htb]
\begin{equation*}
\raise -1.4 cm \hbox{\includegraphics{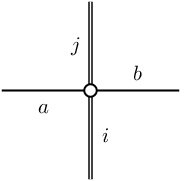}} \quad = \quad M_{a i | b j}
\end{equation*}
\caption{The association of the indices with the bonds.}
\label{Vertex}
\end{figure}

The energy of a configuration is the sum of the energies associated with the vertices. The energy
$\calE_{\calC, \calV}$ associated with a vertex $\calV$ depends on the vertex itself and on the
configuration $\calC$. Therefore, we have
\begin{equation*}
Z = \sum_\calC \exp \Bigl( - \beta \sum_{\calV} \calE_{\calC, \calV} \Bigr) = \sum_\calC
\prod_\calV \exp \Bigl(- \beta \calE_{\calC, \calV} \Bigr).
\end{equation*}
Assume that $\calE_{\calC, \calV}$ depends only on the states of the bonds connecting $\calV$ with
the neighbouring vertices. We label the states of the horizontal and vertical bonds by the integers
from $1$ to $\ell$ and from $1$ to $k$ respectively, and denote the energy associated with a vertex
by $\calE_{a i | b j}$, where the indices correspond to the states of the bonds as is shown in
Figure~\ref{Vertex}.

\begin{figure}[htb]
\centering
\includegraphics{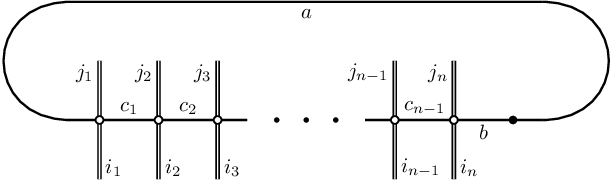}
\caption{Graphical interpretation of the quantities $T_{i_1 i_2 \ldots i_n | j_1 j_2 \ldots j_n}$.}
\label{Transfer}
\end{figure}

\begin{figure}[htb]
\centering
\includegraphics{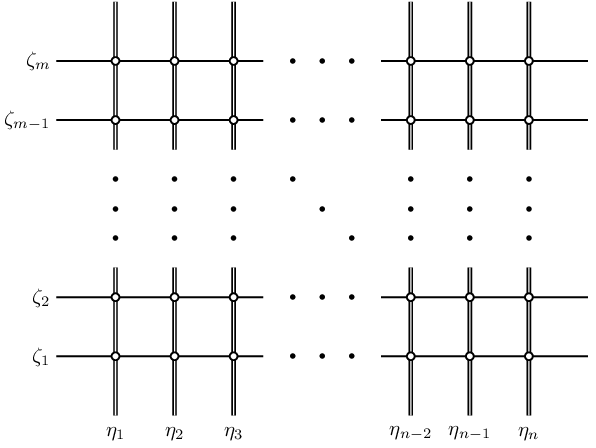}
\caption{Spectral parameters.}
\label{Parameters}
\end{figure}

It is convenient to introduce the Boltzmann weight of a vertex
\begin{equation*}
M_{a i | b j} = \exp (- \beta \, \calE_{a i | b j}).
\end{equation*}
It is clear that the Boltzmann weight of a configuration is the product of the Boltzmann weights of
the vertices, and the summation over the configurations is the summation over the indices
associated with the bonds. One can start with the summation over the indices associated with the
horizontal bonds of a row excluding the first and the last bonds. This gives the quantities
\begin{equation}  \label{Maibj}
M_{a i_1 i_2 \ldots i_n | b j_1 j_2 \ldots j_n} = \sum_{\substack{c_1, c_2, \ldots, c_{n - 1}}}
M_{a i_1 | c_1 j_1} M_{c_1 i_2 | c_2 j_2} \ldots M_{c_{n - 1} i_n | b j_n},
\end{equation}
where $n$ is the number of the vertices in a horizontal row. Now we sum over the states of the
remaining bonds of a horizontal row. If we assume the periodic boundary conditions, we should put
in equation (\ref{Maibj}) $b = a$ and sum over $a$. More generally, one can multiply (\ref{Maibj})
by some quantities $F_{b a}$ and sum over $a$ and $b$ independently. This can be considered as a
generalization of boundary conditions called quasi-periodic or twisted. As the result we obtain the
quantities
\begin{equation*}
T_{i_1 i_2 \ldots i_n | j_1 j_2 \ldots j_n} = \sum_{c_1, c_2, \ldots, c_{n - 1}, a, b} M_{a i_1 |
c_1 j_1} M_{c_1 i_2 | c_2 j_2} \ldots M_{c_{n - 1} i_n | b j_n} F_{b a},
\end{equation*}
which can be graphically interpreted by Figure \ref{Transfer}, where
\begin{equation*}
\raise -.25cm \hbox{\includegraphics{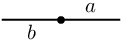}} \quad = \quad F_{b a},
\end{equation*}
and the summation over the indices associated with the internal lines is implied.
Define a $k^n \times k^n$ matrix
\begin{equation*}
\bbT = (T_{i_1 i_2 \ldots i_n | j_1 j_2 \ldots j_n})
\end{equation*}
called the transfer matrix. It is clear that the summation over the states of the horizontal bonds
of two adjacent horizontal rows and over the states of the vertical bonds between them gives the
entries of the matrix $\bbT^2$. Summing over the states of the horizontal bonds of all horizontal
rows and over the states of the vertical bonds between them we obtain the entries of the matrix
$\bbT^m$, where $m$ is the number of the horizontal rows. Assuming the purely periodic boundary
conditions for the vertical rows we see that the summation over the states of the remaining bonds
gives the trace of this matrix. Thus, we come to the equation
\begin{equation*}
Z = \tr (\bbT^m).
\end{equation*}
Recall that the statistical physics describes systems of large numbers of particles. Hence, we are
primarily interested in the case of large $n$ and $m$. If $\lambda_{\mathrm{max}}$ is the maximal
eigenvalue of the transfer matrix $\bbT$ and it is nondegenerate, then for large $m$ we have the
estimation
\begin{equation*}
Z \sim \lambda_{\mathrm{max}}^m.
\end{equation*}
Therefore, the problem of calculating the partition function is reduced to the problem of finding
the maximal eigenvalue of the transfer matrix for large $n$. In some cases it can be done using the
Bethe ansatz or some its modification. In fact, the applicability of the Bethe ansatz is a
manifestation of a rich algebraic structure behind the model under consideration. To reveal this
structure, it is useful to introduce spectral parameters associated with the rows and columns of
the lattice, see Figure \ref{Parameters},
and assume that the Boltzmann weight of a vertex depends on the corresponding row and column
spectral parameters. So we write $\bbT(\zeta | \eta_1, \ldots, \eta_n)$ for the transfer operator.
The case $\eta_1 = \eta_2 = \ldots = \eta_n = 1$ is called homogeneous.

\subsection{Integrable models}

The vertex model under consideration is said to be integrable if we have
\begin{equation}
[\bbT(\zeta_1 | \eta_1, \ldots, \eta_n), \bbT(\zeta_2 | \eta_1, \ldots, \eta_n)] = 0 \label{TTcomm}
\end{equation}
for any $\zeta_1$ and $\zeta_2$. It follows from this equation that the transfer matrix $\bbT(\zeta
| \eta_1, \ldots, \eta_n)$ can be put into Jordan normal form by a similarity transformation which
does not depend on $\zeta$. One can show that equation (\ref{TTcomm}) is valid, in particular, if
there exist $\ell^2 \times \ell^2$ quantities $R_{a_1 a_2 | b_1 b_2}(\zeta_1 | \zeta_2)$ such that
\begin{multline}
\sum_{c_1, c_2, k} R_{a_1 a_2 | c_1 c_2}(\zeta_1 | \zeta_2) M_{c_1 i | b_1 k}(\zeta_1 | \eta)
M_{c_2 k | b_2 j}(\zeta_2 | \eta) \\* = \sum_{c_1, c_2, k} M_{a_2 i | c_2 k}(\zeta_2 | \eta) M_{a_1
k | c_1 j}(\zeta_1 | \eta) R_{c_1 c_2 | b_1 b_2}(\zeta_1 | \zeta_2), \label{RMM}
\end{multline}
and
\begin{equation}
\sum_{c_1, c_2} R_{a_1 a_2 | c_1 c_2}(\zeta_1 | \zeta_2) F_{c_1 b_1}(\zeta_1) F_{c_2 b_2}(\zeta_2)
= \sum_{c_1, c_2} F_{a_1 c_1}(\zeta_1) F_{a_2 c_2}(\zeta_2) R_{c_1 c_2 | b_1 b_2}(\zeta_1 |
\zeta_2). \label{RFF}
\end{equation}
The model possesses the richest algebraic structure if additionally
\begin{multline}
\sum_{c_1, c_2, c_3} R_{a_1 a_2 | c_1 c_2}(\zeta_1 | \zeta_2) R_{c_1 a_3 | b_1 c_3}(\zeta_1 |
\zeta_3) R_{c_2 c_3 | b_2 b_3}(\zeta_2 | \zeta_3) \\ = \sum_{c_1, c_2, c_3} R_{a_2 a_3 | c_2
c_3}(\zeta_2 | \zeta_3) R_{a_1 c_3 | c_1 b_3}(\zeta_1 | \zeta_3) R_{c_1 c_2 | b_1 b_2}(\zeta_1 |
\zeta_2). \label{ybe}
\end{multline}
This is the famous Yang-Baxter equation.

\subsection{Functional relations for the six vertex model}

The standard example of a quantum statistical vertex model is the six-vertex model. Here any
horizontal and vertical bond can be in one of two states labeled by $1$ and $2$, and the Boltzmann
weights $M_{a i | b j}(\zeta | \eta)$ can be arranged into the matrix
\begin{equation}
(M_{a i | b j}(\zeta | \eta)) = \left( \begin{array}{cccc}
a(\zeta \eta^{-1}) & 0 & 0 & 0 \\
0 & b(\zeta \eta^{-1}) & c(\zeta \eta^{-1}) & 0 \\
0 & c(\zeta \eta^{-1}) & b(\zeta \eta^{-1}) & 0 \\
0 & 0 & 0 & a(\zeta \eta^{-1})
\end{array} \right), \label{mm}
\end{equation}
where
\begin{equation}
a(\zeta) = q \, \zeta - q^{-1} \zeta^{-1}, \qquad b(\zeta) = \zeta - \zeta^{-1}, \qquad c(\zeta) =
q - q^{-1}, \label{bw}
\end{equation}
and we order the pairs of indices as $11$, $12$, $21$, $22$. The parameter $q$ is a fixed nonzero
complex number. We see that the Boltzmann weights are different from zero only for six vertex
configurations, hence the name of the model. Equation (\ref{RMM}) is satisfied with
\begin{equation}
R_{a_1 a_2 | b_1 b_2}(\zeta_1 | \zeta_2) = M_{a_1 a_2 | b_1 b_2}(\zeta_1 | \zeta_2). \label{rm}
\end{equation}
The usual choice for $F_{ab}(\zeta)$ is
\begin{equation*}
F_{11}(\zeta) = q^\phi, \qquad F_{12}(\zeta) = 0, \qquad F_{21}(\zeta) = 0,
\qquad F_{22}(\zeta) = q^{-\phi},
\end{equation*}
where $\phi$ is an arbitrary complex number. One can verify that (\ref{RFF}) is satisfied.

To find eigenvectors and eigenvalues of the transfer matrix it is convenient to use the algebraic
Bethe ansatz \cite{TakFad79, SklTakFad79}. This approach shows that there are the eigenvectors of
the transfer matrix with the eigenvalues
\begin{equation}
\lambda(\zeta | \eta_1, \ldots, \eta_n) = q^\phi \prod_{i = 1}^n a(\zeta \eta_i^{-1}) \prod_{\ell =
1}^p \frac{a(\zeta_\ell \zeta^{-1})}{b(\zeta_\ell \zeta^{-1})} + q^{-\phi} \prod_{i = 1}^n b(\zeta
\eta_i^{-1}) \prod_{\ell = 1}^p \frac{a(\zeta \zeta_\ell^{-1})}{b(\zeta \zeta_\ell^{-1})},
\label{lambda}
\end{equation}
where $0 \le p \le n$, and $\zeta_1$, \ldots, $\zeta_p$ satisfy the Bethe equations
\begin{equation*}
q^{2 \phi} \prod_{i = 1}^n \frac{a(\zeta_m^{\mathstrut} \eta_i^{-1})}{b(\zeta_m^{\mathstrut}
\eta_i^{-1})} = (-1)^{p + 1} \prod_{\ell = 1}^p \frac{a(\zeta_m^{\mathstrut}
\zeta_\ell^{-1})}{a(\zeta_\ell^{\mathstrut} \zeta_m^{-1})}, \qquad m = 1, \ldots, p.
\end{equation*}
It can be shown that all eigenvalues can be obtained by the Bethe ansatz, and that the
corresponding eigenvectors form a basis of $\bbC^{2^n}$, see, for example, \cite{Bax02} and
references therein. It is important that these eigenvectors do not depend on $\zeta$, that is in
fact a consequence of equation (\ref{TTcomm}). Now it is clear that there is a matrix $\bbO(\eta_1,
\ldots, \eta_n)$ such that
\begin{equation}
\bbO(\eta_1, \ldots, \eta_n) \bbT_{\mathrm d}(\zeta | \eta_1, \ldots, \eta_n) \bbO^{-1}(\eta_1,
\ldots, \eta_n) =  \bbT(\zeta | \eta_1, \ldots, \eta_n), \label{OTO}
\end{equation}
where $\bbT_{\mathrm d}(\zeta | \eta_1, \ldots, \eta_n)$ is a diagonal matrix.

For a given solution of the Bethe equations we define the function
\begin{equation}
\theta(\zeta | \eta_1, \ldots, \eta_n) = \prod_{\ell = 1}^p b(\zeta_\ell(\eta_1, \ldots, \eta_n)
\zeta^{-1}), \label{theta}
\end{equation}
where the dependence of the Bethe roots $\zeta_1$, \ldots, $\zeta_p$ on the spectral parameters
$\eta_1$, \ldots, $\eta_n$ is shown explicitly. Now we can rewrite relation (\ref{lambda}) as
\begin{multline}
\lambda(\zeta | \eta_1, \ldots, \eta_n) \theta(\zeta | \eta_1, \ldots, \eta_n) \\* = q^\phi
\prod_{i = 1}^n a(\zeta \eta^{-1}_i) \, \theta(q^{-1} \zeta | \eta_1, \ldots, \eta_n) + q^{-\phi}
\prod_{i = 1}^n b(\zeta \eta^{-1}_i) \, \theta(q \zeta | \eta_1, \ldots, \eta_n).
\label{lambdatheta}
\end{multline}

The matrix $\bbT_{\rmd}(\zeta | \eta_1, \ldots, \eta_n)$ in (\ref{OTO}) is a diagonal matrix with
entries being the eigenvalues of $\bbT(\zeta | \eta_1, \ldots, \eta_n)$ of the form (\ref{lambda}).
Denote by $\bbQ_{\rmd}(\zeta | \eta_1, \ldots, \eta_n)$ the diagonal matrix whose entries are the
corresponding functions $\theta(\zeta | \eta_1, \ldots, \eta_n)$ given by~(\ref{theta}). It follows
from (\ref{lambdatheta}) that
\begin{multline}
\bbT_{\rmd}(\zeta | \eta_1, \ldots, \eta_n) \bbQ_{\rmd}(\zeta | \eta_1, \ldots, \eta_n) \\* =
q^\phi \prod_{i = 1}^n a(\zeta \eta^{-1}_i) \, \bbQ_{\rmd}(q^{-1} \zeta | \eta_1, \ldots, \eta_n) +
q^{-\phi} \prod_{i = 1}^n b(\zeta \eta^{-1}_i) \, \bbQ_{\rmd}(q \zeta | \eta_1, \ldots, \eta_n).
\label{TdQd}
\end{multline}
Define the matrix
\begin{equation*}
\bbQ(\zeta | \eta_1, \ldots, \eta_n) = \bbO(\eta_1, \ldots, \eta_n)\bbQ_{\rmd}(\zeta | \eta_1,
\ldots, \eta_n)\bbO^{-1}(\eta_1, \ldots, \eta_n).
\end{equation*}
Since the matrix $\bbQ_{\rmd}(\zeta | \eta_1, \ldots, \eta_n)$ does not depend on $\zeta$, it
follows from (\ref{TdQd}) that
\begin{multline}
\bbT(\zeta | \eta_1, \ldots, \eta_n) \bbQ(\zeta | \eta_1, \ldots, \eta_n) \\* = q^\phi \prod_{i =
1}^n a(\zeta \eta^{-1}_i) \, \bbQ(q^{-1} \zeta | \eta_1, \ldots, \eta_n) + q^{-\phi} \prod_{i =
1}^n b(\zeta \eta^{-1}_i) \, \bbQ(q \zeta | \eta_1, \ldots, \eta_n). \label{TQ}
\end{multline}
This functional equation is called the Baxter's $TQ$-equation. By construction, we also have
\begin{gather}
[\bbQ(\zeta_1 | \eta_1, \ldots, \eta_n), \bbT(\zeta_2 | \eta_1, \ldots, \eta_n)] = 0, \label{TQcomm}
\\[.5em]
[\bbQ(\zeta_1 | \eta_1, \ldots, \eta_n), \bbQ(\zeta_2 | \eta_1, \ldots, \eta_n)] = 0 \label{QQcomm}
\end{gather}
for any $\zeta_1$ and $\zeta_2$. We call (\ref{TTcomm}), (\ref{TQcomm}), (\ref{QQcomm}) and
(\ref{TQ}) functional relations. They are equivalent to the Bethe ansatz in the sense that they can
be used to find the eigenvalues of the transfer matrix, see, for example, the book \cite{Bax82}.

In the next section we explain how the objects necessary for the integration of an integrable model
are related to its background algebraic structure.

\section{\texorpdfstring{Objects defined by the universal $R$-matrix}{Objects defined by the
universal R-matrix}} \label{s:odurm}

In this section $A$ is a $\bbZ$-graded quasitriangular Hopf algebra over the field $\bbC$ with the
comultiplication $\Delta$ and the universal $R$-matrix $\calR$. Some relevant definitions are
reproduced in appendices \ref{a:qha} and \ref{a:zgha}.

\subsection{\texorpdfstring{$R$-operators}{R-operators}}

Let $\varphi$ be a representation of the algebra $A$ in the vector space $V$.\footnote{For the case
of square lattice models the vector space $V$ is the auxiliary space.} Given $\nu \in \bbC^\times$,
we denote
\begin{equation}
\varphi_\nu = \varphi \circ \Phi_\nu, \label{varphizeta}
\end{equation}
where the mapping $\Phi_\nu$ is defined by relation (\ref{aPhinu}). For any $\zeta_1, \zeta_2 \in
\bbC^\times$ we define
\begin{equation}
R_\varphi(\zeta_1 | \zeta_2) = (\varphi_{\zeta_1} \otimes \varphi_{\zeta_2}) (\calR), \label{Rvarphi}
\end{equation}
where $\calR$ is the universal $R$-matrix of $A$. Having in mind the relation to integrable
systems, we call $\zeta_1$ and $\zeta_2$ {\em spectral parameters\/}. It is clear that
$R_\varphi(\zeta_1 | \zeta_2)$ is an element of $\End(V) \otimes \End(V) \cong \End(V \otimes V)$.
We call it an {\em $R$-operator\/}.

It appears often that the universal $R$-matrix $\calR$ of $A$ satisfies the equation
\begin{equation}
(\Phi_\nu \otimes \Phi_\nu)(\calR) = \calR \label{PhiPhiR}
\end{equation}
for any $\nu \in \bbC^\times$. From the point of view of the natural $\bbZ$-gradation of $A \otimes
A$, induced by the $\bbZ$-gradation of $A$, this means that the universal $R$-matrix $\calR$
belongs to the zero grade subalgebra $(A \otimes A)_0$, see appendix \ref{a:zgha}. In this case,
using the equation,
\begin{equation}
\varphi_{\zeta \nu} \eqWithRef{varphizeta} \varphi \circ \Phi_{\zeta \nu} \eqWithRef{aPhinunu}
\varphi_\zeta \circ \Phi_\nu, \label{varphizetanu}
\end{equation}
we obtain
\begin{multline}
R_\varphi(\zeta_1 \nu | \zeta_2 \nu) \eqWithRef{Rvarphi} (\varphi_{\zeta_1 \nu} \otimes
\varphi_{\zeta_2 \nu})(\calR) \\* \eqWithRef{varphizetanu}  ((\varphi_{\zeta_1} \otimes
\varphi_{\zeta_2}) \circ (\Phi_\nu \otimes \Phi_\nu))(\calR) \eqWithRef{PhiPhiR} (\varphi_{\zeta_1}
\otimes \varphi_{\zeta_2}) (\calR) \eqWithRef{Rvarphi} R_\varphi(\zeta_1 | \zeta_2).
\label{zetanuzeta}
\end{multline}
Thus, $R_\varphi(\zeta_1 | \zeta_2)$ depends only on the combination $\zeta_1^{\mathstrut}
\zeta_2^{-1}$, and one can introduce the $R$-operator
\begin{equation*}
R_\varphi(\zeta) = R_\varphi(\zeta | 1)
\end{equation*}
which depends on only one spectral parameter and determines the $R$-operator depending on two
spectral parameters, via the equation
\begin{equation*}
R_\varphi(\zeta_1 | \zeta_2) = R_\varphi(\zeta_1^{\mathstrut} \zeta_2^{-1}).
\end{equation*}

Return to a general situation and apply the mapping $\varphi_{\zeta_1} \otimes \varphi_{\zeta_2}
\otimes \varphi_{\zeta_3}$ to both sides of the Yang-Baxter equation (\ref{aRRR}) for the universal
$R$-matrix. We obtain the Yang--Baxter equation for the $R$-operator,
\begin{equation}
R^{12}_\varphi(\zeta_1 | \zeta_2) \, R^{13}_\varphi(\zeta_1 | \zeta_3) \, R^{23}_\varphi(\zeta_2 |
\zeta_3) = R^{23}_\varphi(\zeta_2 | \zeta_3) \, R^{13}_\varphi(\zeta_1 | \zeta_3) \,
R^{12}_\varphi(\zeta_1 | \zeta_2). \label{RzRzRz}
\end{equation}
In the case where equation (\ref{PhiPhiR}) is valid, for the $R$-operator depending on one spectral
parameter we have
\begin{equation*}
R^{12}_\varphi(\zeta_{12}) \, R^{13}_\varphi(\zeta_{13}) \, R^{23}_\varphi(\zeta_{23}) =
R^{23}_\varphi(\zeta_{23}) \, R^{13}_\varphi(\zeta_{13}) \, R^{12}_\varphi(\zeta_{12}).
\end{equation*}
Here and below we denote $\zeta_{ij} = \zeta_i^{\mathstrut} \zeta_j^{-1}$.

One often uses two operators directly related to the $R$-operator defined by equation
(\ref{Rvarphi}). One of them is defined as
\begin{equation*}
\check R_\varphi(\zeta_1 | \zeta_2) = P R_\varphi(\zeta_1 | \zeta_2),
\end{equation*}
where $P$ is the permutation operator in $V \otimes V$, see appendix \ref{a:sgtp}. Using this
definition, we can rewrite the left hand side of the Yang--Baxter equation in the following way:
\begin{multline*}
R^{12}_\varphi(\zeta_1 | \zeta_2) R^{13}_\varphi(\zeta_1 | \zeta_3) R^{23}_\varphi(\zeta_2 | \zeta_3) \\
= P^{12} \check R^{12}_\varphi(\zeta_1 | \zeta_2) P^{13} \check R^{13}_\varphi(\zeta_1 | \zeta_3)
P^{23} \check R^{23}_\varphi(\zeta_2 | \zeta_3) \\ \eqWithRef{aPM} P^{12} P^{13} P^{23} \check
R^{23}_\varphi(\zeta_1 | \zeta_2) \check R^{12}_\varphi(\zeta_1 | \zeta_3) \check
R^{23}_\varphi(\zeta_2 | \zeta_3).
\end{multline*}
Similarly, we rewrite the right hand side as
\begin{equation*}
R^{23}_\varphi(\zeta_2 | \zeta_3) R^{13}_\varphi(\zeta_1 | \zeta_3) R^{12}_\varphi(\zeta_1 |
\zeta_2) =  P^{23} P^{13} P^{12} \check R^{12}_\varphi(\zeta_2 | \zeta_3) \check
R^{23}_\varphi(\zeta_1 | \zeta_3) \check R^{12}_\varphi(\zeta_1 | \zeta_2).
\end{equation*}
It is not difficult to verify that
\begin{equation*}
P^{12} P^{13} P^{23} = P^{23} P^{13} P^{12},
\end{equation*}
therefore, the Yang--Baxter equation (\ref{RzRzRz}) is equivalent to the equation
\begin{equation*}
\check R^{23}_\varphi(\zeta_1 | \zeta_2) \check R^{12}_\varphi(\zeta_1 | \zeta_3) \check
R^{23}_\varphi(\zeta_2 | \zeta_3) = \check R^{12}_\varphi(\zeta_2 | \zeta_3) \check
R^{23}_\varphi(\zeta_1 | \zeta_3) \check R^{12}_\varphi(\zeta_1 | \zeta_2).
\end{equation*}
This equation can also be written as
\begin{multline*}
(\id \otimes \check R_\varphi(\zeta_1 | \zeta_2)) (\check R_\varphi(\zeta_1 | \zeta_3) \otimes \id)
(\id \otimes \check R_\varphi(\zeta_2 | \zeta_3)) \\= (\check R_\varphi(\zeta_2 | \zeta_3) \otimes
\id) (\id \otimes \check R_\varphi(\zeta_1 | \zeta_3)) (\check R_\varphi(\zeta_1 | \zeta_2) \otimes
\id).
\end{multline*}
Another companion for the $R$-operator is defined as
\begin{equation*}
\hat R_\varphi(\zeta_1 | \zeta_2) = R_\varphi(\zeta_1 | \zeta_2) P,
\end{equation*}
Here the Yang--Baxter equation takes the form
\begin{equation*}
\hat R^{12}_\varphi(\zeta_1 | \zeta_2) \hat R^{23}_\varphi(\zeta_1 | \zeta_3) \hat
R^{12}_\varphi(\zeta_2 | \zeta_3) = \hat R^{23}_\varphi(\zeta_2 | \zeta_3) \hat
R^{12}_\varphi(\zeta_1 | \zeta_3) \hat R^{23}_\varphi(\zeta_1 | \zeta_2),
\end{equation*}
or, equivalently,
\begin{multline*}
(\hat R_\varphi(\zeta_1 | \zeta_2) \otimes \id) (\id \otimes \hat R_\varphi(\zeta_1 | \zeta_3))
(\hat R_\varphi(\zeta_2 | \zeta_3) \otimes \id)) \\= (\id \otimes \hat R_\varphi(\zeta_2 |
\zeta_3)) (\hat R_\varphi(\zeta_1 | \zeta_3) \otimes \id) (\id \otimes \hat R_\varphi(\zeta_1 |
\zeta_2)).
\end{multline*}

Assume now that the vector space $V$ is finite-dimensional of dimension $\ell$. Let $\{e_a\}$ be a
basis of $V$, and $\{E_{ab}\}$ the corresponding basis of $\End(V)$, see appendix \ref{a:ea}. We
have
\begin{equation*}
R_\varphi(\zeta_1 | \zeta_2) = \sum_{a, b, c, d} E_{ab | cd} \, R_{ab | cd}(\zeta_1 | \zeta_2)
\eqWithRef{EeqEE} \sum_{a, b, c, d} E_{ac} \otimes E_{bd} \, R_{ab | cd}(\zeta_1 | \zeta_2),
\end{equation*}
where $R_{ab | cd}(\zeta_1 | \zeta_2)$ are some unique complex numbers. One can verify that the
Yang--Baxter equation (\ref{RzRzRz}) in terms of the quantities $R_{ab | cd}(\zeta_1 | \zeta_2)$
has the form (\ref{ybe}). The $\ell^2 \times \ell^2$ matrix with the entries $R_{ab | cd}(\zeta_1 |
\zeta_2)$ is called an {\em $R$-matrix\/}. We denote it~$\bm R_\varphi(\zeta_1 | \zeta_2)$.

It is not difficult to convince oneself that
\begin{equation*}
P = \sum_{a, b} E_{ab} \otimes E_{ba}.
\end{equation*}
Now, defining the quantities $\check R_{ab | cd}(\zeta_1 | \zeta_2)$ by
\begin{equation*}
\check R_\varphi(\zeta_1 | \zeta_2) = \sum_{a, b, c, d} E_{ac} \otimes E_{bd} \, \check R_{ab |
cd}(\zeta_1 | \zeta_2),
\end{equation*}
we see that
\begin{equation*}
\check R_{ab | cd}(\zeta_1 | \zeta_2) = R_{ba | cd}(\zeta_1 | \zeta_2).
\end{equation*}
Similarly, defining the quantities $\hat R_{ab | cd}(\zeta_1 | \zeta_2)$ by
\begin{equation*}
\hat R_\varphi(\zeta_1 | \zeta_2) = \sum_{a, b, c, d} E_{ac} \otimes E_{bd} \, \hat R_{ab |
cd}(\zeta_1 | \zeta_2),
\end{equation*}
we see that
\begin{equation*}
\hat R_{ab | cd}(\zeta_1 | \zeta_2) = R_{ab | dc}(\zeta_1 | \zeta_2).
\end{equation*}
We denote the matrices with the entries $\check R_{ab | cd}(\zeta_1 | \zeta_2)$ and $\hat R_{ab |
cd}(\zeta_1 | \zeta_2)$ by $\check{\bm R}_\varphi(\zeta_1 | \zeta_2)$ and $\hat{\bm
R}_\varphi(\zeta_1 | \zeta_2)$ respectively.

One can also define an $R$-operator using two different representations, say $\varphi_1$ and
$\varphi_2$. In this case we use the notation
\begin{equation*}
R_{\varphi_1, \varphi_2}(\zeta_1 | \zeta_2) = (\varphi_{1 \zeta_1} \otimes \varphi_{2 \zeta_2})(\calR).
\end{equation*}
In the case when $\varphi_1$ and $\varphi_2$ are representations of $A$ in vector spaces $V_1$ and
$V_2$, respectively, the $R$-operator $\check R_{\varphi_1, \varphi_2}(\zeta_1 | \zeta_2)$ serves
as the intertwiner for the representations $\varphi_{1 \zeta_1} \otimes_\Delta \varphi_{2 \zeta_2}$
and $\varphi_{2 \zeta_2} \otimes_\Delta \varphi_{1 \zeta_1}$ of $A$ in the vector spaces $V_1
\otimes V_2$ and $V_2 \otimes V_1$ respectively.\footnote{We use the notation $\otimes_\Delta$ to
distinguish between the tensor product of representations and the usual tensor product of mappings,
so that $\varphi \otimes_\Delta \psi = (\varphi \otimes \psi) \circ \Delta$.} To prove the
intertwiner property of $\check R_{\varphi_1, \varphi_2}(\zeta)$ we note that
\begin{multline*}
\varphi_{2 \zeta_2} \otimes_\Delta \varphi_{1 \zeta_1}
= (\varphi_{2 \zeta_2} \otimes \varphi_{1 \zeta_1}) \circ \Delta \\
\eqWithRef{Pivarphi} \Pi \circ (\varphi_{1 \zeta_1} \otimes \varphi_{2 \zeta_2})
\circ \Pi \circ \Delta \eqWithRef{deltaop} \Pi \circ (\varphi_{1 \zeta_1} \otimes
\varphi_{2 \zeta_2}) \circ \Delta^\op.
\end{multline*}
Hence, one can write
\begin{multline*}
(\varphi_{2 \zeta_2} \otimes_\Delta \varphi_{1 \zeta_1})(a) \eqWithRef{auniR}
(\Pi \circ (\varphi_{1 \zeta_1} \otimes \varphi_{2 \zeta_2}))(\calR \, \Delta(a) \calR^{-1})
\\= \Pi((\varphi_{1 \zeta_1} \otimes \varphi_{2 \zeta_2})(\calR) (\varphi_{1 \zeta_1}
\otimes_\Delta \varphi_{2 \zeta_2})(a) (\varphi_{1 \zeta_1} \otimes \varphi_{2 \zeta_2})(\calR^{-1})) \\[.5em]
\eqWithRef{aPiM} P R_{\varphi_1, \varphi_2}(\zeta_1 | \zeta_2) ((\varphi_{1 \zeta_1}
\otimes_\Delta \varphi_{2 \zeta_2})(a)) (R_{\varphi_1, \varphi_2}(\zeta_1 | \zeta_2))^{-1} P^{-1}.
\end{multline*}
Finally we come to the declared result,
\begin{equation*}
\check R_{\varphi_1, \varphi_2}(\zeta_1 | \zeta_2) (\varphi_{1 \zeta_1}
\otimes_\Delta \varphi_{2 \zeta_2})(a) = (\varphi_{2 \zeta_2} \otimes_\Delta
\varphi_{1 \zeta_1})(a) \check R_{\varphi_1, \varphi_2}(\zeta_1 | \zeta_2).
\end{equation*}

An explicit form of $R$-matrices was obtained from the corresponding universal
$R$-ma\-trices for certain representations of the quantum groups
$U_q(\calL(\gothsl_2))$ \cite{KhoTol92, LevSoiStu93, ZhaGou94, BraGouZhaDel94, BraGouZha95, BooGoeKluNirRaz10},
$U_q(\calL(\gothsl_3))$ \cite{ZhaGou94, BraGouZhaDel94, BraGouZha95, BooGoeKluNirRaz10} and
$U_q(\calL(\gothsl_3, \mu))$ \cite{KhoTol92, BooGoeKluNirRaz11}, where $\mu$ is the standard
diagram automorphism of $\gothsl_3$ of order 2. In fact, up to a scalar factor they coincide
with the $R$-matrices obtained by other methods. Nevertheless, it is very useful to understand
that they can be obtained from the universal $R$-matrices because this allows one to relate them
to other objects involved into the integration process.

\subsection{Monodromy operators}

\subsubsection{Universal monodromy operator} \label{s:Umm}

Let again $\varphi$ be a representation of $A$ in the vector space $V$. Given $\zeta \in
\bbC^\times$, we define the {\em universal monodromy operator\/} $\calM_\varphi(\zeta)$ by the
equation
\begin{equation*}
\calM_\varphi(\zeta) = (\varphi_\zeta \otimes \id) (\calR),
\end{equation*}
where the mapping $\varphi_\zeta$ is defined by equation (\ref{varphizeta}). It is clear that
$\calM_\varphi(\zeta)$ is an element of the algebra $\End(V) \otimes A$.

Applying the mapping $\varphi_{\zeta_1} \otimes \varphi_{\zeta_2} \otimes \id$ to both sides of the
Yang--Baxter equation (\ref{aRRR}), we obtain the equation
\begin{equation*}
R^{12}_\varphi(\zeta_1 | \zeta_2) \calM^{13}_\varphi(\zeta_1) \calM^{23}_\varphi(\zeta_2) =
\calM^{23}_\varphi(\zeta_2) \calM^{13}_\varphi(\zeta_1) R^{12}_\varphi(\zeta_1 | \zeta_2).
\end{equation*}
Multiply both sides of the above equation by $P^{12} \otimes 1$ and use equation (\ref{aPM}). This
gives
\begin{equation}
\check R^{12}_\varphi(\zeta_1 | \zeta_2) \calM^{13}_\varphi(\zeta_1) \calM^{23}_\varphi(\zeta_2) =
\calM^{13}_\varphi(\zeta_2) \calM^{23}_\varphi(\zeta_1) \check R^{12}_\varphi(\zeta_1 | \zeta_2).
\label{cRMM}
\end{equation}
There is a matrix equivalent of this equation.

Assume that the vector space $V$ is finite-dimensional of dimension $\ell$. Let $\{e_a\}$ be a
basis of $V$ and $\{E_{ab}\}$ the corresponding basis of $\End(V)$, see appendix \ref{a:ea}. One
can write
\begin{equation*}
\calM_\varphi(\zeta) = \sum_{a, b} E_{ab} \otimes \calM_{ab}(\zeta),
\end{equation*}
where $\calM_{ab}(\zeta)$ are some unique elements of the algebra $A$. Denote by
$\bm{\calM}_\varphi(\zeta)$ the matrix with the entries $\calM_{ab}(\zeta)$. The matrix
$\bm{\calM}_\varphi(\zeta)$ is an element of $\Mat_\ell(A)$. We call it a {\em universal monodromy
matrix\/}. Now, it follows from (\ref{cRMM}) that
\begin{equation}
\check {\bm R}_\varphi(\zeta_1 | \zeta_2) (\bm{\calM}_\varphi(\zeta_1) \boxtimes
\bm{\calM}_\varphi(\zeta_2)) = (\bm{\calM}_\varphi(\zeta_2) \boxtimes \bm{\calM}_\varphi(\zeta_1))
\check {\bm R}_\varphi(\zeta_1 | \zeta_2). \label{bmRbmMM}
\end{equation}
Here the operation $\boxtimes$ is defined by equation (\ref{boxtimes}), and, using the canonical
embedding of the field $\bbC$ into $A$, we treat $\check {\bm R}_\varphi(\zeta)$ as an element of
$\Mat_{\ell^2}(A)$. It is worth to remind here that $\boxtimes$ is a natural generalization of the
Kronecker product to the case of matrices with noncommuting entries.

\subsubsection{Monodromy operator} \label{s:Mo}

Let $\varphi$ and $\psi$ be representations of $A$ in the vector spaces $V$ and $U$
respectively.\footnote{For the case of square lattice models the vector space $U$ is the quantum
space.} Given $\zeta, \eta \in \bbC^\times$, we define the {\em monodromy operator\/} $M_{\varphi,
\psi}(\zeta | \eta)$ by the equation
\begin{equation*}
M_{\varphi, \psi}(\zeta | \eta) = (\varphi_\zeta \otimes \psi_\eta) (\calR),
\end{equation*}
where the mapping $\varphi_\zeta$ is defined by equation (\ref{varphizeta}) and the mapping
$\psi_\eta$ is defined in the similar way. It is clear that $M_{\varphi, \psi}(\zeta | \eta)$ is an
element of $\End(V) \otimes \End(U)$.

One should note that the monodromy operator $M_{\varphi, \psi}(\zeta | \eta)$ coincides with the
$R$-operator $R_{\varphi, \psi}(\zeta | \eta)$. Nevertheless, we use different names due to
different roles these objects play in the integration process.

Since $A$ is a bialgebra, one can also define the monodromy operator\footnote{We use the
comultiplication $\Delta^\op$ instead of $\Delta$ to have relations similar to those which usually
arise for integrable systems.}
\begin{equation*}
M_{\varphi, \psi}(\zeta | \eta_1, \ldots, \eta_n) = (\varphi_\zeta \otimes (\psi_{\eta_1}
\otimes_{\Delta^\op} \ldots \otimes_{\Delta^\op} \psi_{\eta_n})) (\calR),
\end{equation*}
where $\eta_1$, \ldots, $\eta_n$ are some nonzero complex numbers. Note that this monodromy
operator is an element of $\End(V) \otimes \End(U)^{\otimes n} \cong \End(V) \otimes
\End(U^{\otimes n})$. In fact, one can use different representations, say $\psi_1$, \ldots,
$\psi_n$, for different factors of the tensor product. This is the case for the construction of the
quantum transfer matrix \cite{Kor07} and for the description of integrable defects \cite{Baj06,
Wes10, CorZam10, Cor11}.

For the opposite comultiplication we have
\begin{equation*}
(\id \otimes \Delta^\op)(\calR) \eqWithRef{deltaop} (\id \otimes \Pi^{23})((\id \otimes
\Delta)(\calR)) \\ \eqWithRef{aidDelta} (\id \otimes \Pi^{23})(\calR^{13} \calR^{12})
\eqWithRef{pim} \calR^{12} \calR^{13}.
\end{equation*}
Therefore, we can see that
\begin{multline*}
M_{\varphi, \psi}(\zeta | \eta_1, \eta_2) = (\varphi_\zeta \otimes (\psi_{\eta_1}
\otimes_{\Delta^\op} \psi_{\eta_2})) (\calR) \\ = ((\varphi_\zeta \otimes \psi_{\eta_1} \otimes
\psi_{\eta_2}) \circ (\id \otimes \Delta^\op))(\calR) = M_{\varphi, \psi}^{12}(\zeta | \eta_1)
M_{\varphi, \psi}^{13}(\zeta | \eta_2).
\end{multline*}
In general, we have
\begin{equation}
M_{\varphi, \psi}(\zeta | \eta_1 \ldots, \eta_n) = M^{1 2}_{\varphi, \psi}(\zeta | \eta_1) \ldots
M^{1, n + 1}_{\varphi, \psi}(\zeta | \eta_n). \label{Mnvarphipsi}
\end{equation}
One often labels the first factor of the tensor product $V \otimes U^{\otimes n}$ by $0$ and the
rest by $1$, \ldots, $n$. In this case the above relation takes the form
\begin{equation*}
M_{\varphi, \psi}(\zeta | \eta_1 \ldots, \eta_n) = M^{0 1}_{\varphi, \psi}(\zeta | \eta_1) \ldots
M^{0 n}_{\varphi, \psi}(\zeta | \eta_n).
\end{equation*}

If equation (\ref{PhiPhiR}) is satisfied, in the same way as for the case of $R$-operators,
see~(\ref{zetanuzeta}), we obtain
\begin{equation*}
M_{\varphi, \psi}(\zeta \nu | \eta \nu) = M_{\varphi, \psi}(\zeta | \eta).
\end{equation*}
Therefore, we can write
\begin{equation*}
M_{\varphi, \psi}(\zeta | \eta) = M_{\varphi, \psi}(\zeta \eta^{-1}),
\end{equation*}
where $M_{\varphi, \psi}(\zeta) = M_{\varphi, \psi}(\zeta | 1)$. Furthermore, in this case equation
(\ref{Mnvarphipsi}) gives
\begin{equation*}
M_{\varphi, \psi}(\zeta \nu | \eta_1 \nu, \ldots, \eta_n \nu) = M_{\varphi, \psi}(\zeta | \eta_1,
\ldots, \eta_n)
\end{equation*}
for any $\nu \in \bbC^\times$.

Assume now that the vector space $V$ is finite-dimensional, $\{e_a\}$ is a basis of $V$, and
$\{E_{ab}\}$ the corresponding basis of $\End(V)$. Represent the monodromy operator as
\begin{equation*}
M_{\varphi, \psi}(\zeta| \eta_1, \ldots, \eta_n) = \sum_{a, b} E_{ab} \otimes M_{ab} (\zeta |
\eta_1, \ldots, \eta_n),
\end{equation*}
where $M_{ab} (\zeta | \eta_1, \ldots, \eta_n)$ are elements of $\End(U)^{\otimes n} \cong
\End(U^{\otimes n})$. It is clear that
\begin{equation}
M_{ab} (\zeta | \eta_1, \ldots, \eta_n) = (\psi_{\eta_1} \otimes_{\Delta^\op} \ldots
\otimes_{\Delta^\op} \psi_{\eta_n})(\calM_{ab}(\zeta)), \label{MabcalMab}
\end{equation}
where $\calM_{ab}(\zeta)$ are the entries of the universal monodromy matrix defined in
section~\ref{s:Umm}. Denote by $\bm M_{\varphi, \psi}(\zeta | \eta_1, \ldots, \eta_n)$ the matrix
with the entries $M_{ab}(\zeta | \eta_1, \ldots, \eta_n)$. It is an element of
$\Mat_\ell(\End(U)^{\otimes n}) \cong \Mat_\ell(\End(U^{\otimes n}))$. Using (\ref{Mnvarphipsi}),
one can show that
\begin{equation}
\bm M_{\varphi, \psi}(\zeta | \eta_1, \ldots, \eta_n) = \bm M_{\varphi, \psi}(\zeta | \eta_1)
\boxdot \ldots \boxdot \bm M_{\varphi, \psi}(\zeta | \eta_n), \label{bmMI}
\end{equation}
where the operation $\boxdot$ is defined by (\ref{boxdot}). Applying the mapping $\psi_{\eta_1}
\otimes_{\Delta^\op} \ldots \otimes_{\Delta^\op} \psi_{\eta_n}$ to the entries of matrices in both
sides of equation (\ref{bmRbmMM}) and taking into account relation (\ref{MabcalMab}), we see that
\begin{multline*}
\check {\bm R}_\varphi(\zeta_1 | \zeta_2) (\bm M_{\varphi, \psi}(\zeta_1 | \eta_1, \ldots, \eta_n)
\boxtimes \bm M_{\varphi, \psi}(\zeta_2 | \eta_1, \ldots, \eta_n))  \\* =  (\bm M_{\varphi,
\psi}(\zeta_2 | \eta_1, \ldots, \eta_n) \boxtimes \bm M_{\varphi, \psi}(\zeta_1 | \eta_1, \ldots,
\eta_n)) \check {\bm R}_\varphi(\zeta_1 | \zeta_2).
\end{multline*}
Relations of this type are the basis of the algebraic Bethe ansatz \cite{TakFad79, SklTakFad79}.

Now assume that the vector space $U$ is finite-dimensional. Let $\{e_i\}$ be a basis of $U$, and
$\{E_{ij}\}$ the corresponding basis of $\End(U)$. Represent the monodromy operator as
\begin{equation*}
M_{\varphi, \psi}(\zeta | \eta_1, \ldots, \eta_n) = \sum_{\substack{i_1, \ldots, i_n \\ j_1,
\ldots, j_n}} M_{i_1 \ldots i_n | j_1 \ldots j_n} (\zeta | \eta_1, \ldots, \eta_n) \otimes E_{i_1
j_1} \otimes \ldots \otimes E_{i_n j_n},
\end{equation*}
where $M_{i_1 \ldots i_n | j_1 \ldots j_n} (\zeta | \eta_1, \ldots, \eta_n)$ are elements of
$\End(V)$. Introducing the matrix
\begin{equation*}
{\mathbb M}_{\varphi, \psi}(\zeta | \eta_1, \ldots, \eta_n) = (M_{i_1 \ldots i_n | j_1 \ldots j_n}
(\zeta | \eta_1, \ldots, \eta_n))
\end{equation*}
and using (\ref{Mnvarphipsi}), we obtain the equation
\begin{equation}
\mathbb M_{\varphi, \psi} (\zeta | \eta_1, \ldots, \eta_n) = \mathbb M_{\varphi, \psi}(\zeta
|\eta_1) \boxtimes \ldots \boxtimes \mathbb M_{\varphi, \psi}(\zeta | \eta_n). \label{bbM}
\end{equation}

If both vector spaces $V$ and $U$ are finite-dimensional, we can write
\begin{equation*}
M_{\varphi, \psi}(\zeta | \eta_1, \ldots, \eta_n) = \sum_{\substack{a, i_1, \ldots, i_n \\ b, j_1,
\ldots, j_n}} M_{a i_1 \ldots i_n | b j_1 \ldots j_n} (\zeta | \eta_1, \ldots, \eta_n) \, E_{a b}
\otimes E_{i_1 j_1} \otimes \ldots \otimes E_{i_n j_n},
\end{equation*}
where $M_{a i_1 \ldots i_n | b j_1 \ldots j_n} (\zeta | \eta_1, \ldots, \eta_n)$ are elements of
the field $\bbC$. Here relation (\ref{Mnvarphipsi}) gives
\begin{multline*}
M_{a i_1 i_2 \ldots i_n | b j_1 j_2 \ldots j_n}(\zeta | \eta_1, \ldots, \eta_n) \\* =
\sum_{\substack{c_1, c_2, \ldots, c_{n - 1}}} M_{a i_1 | c_1 j_1}(\zeta | \eta_1) M_{c_1 i_2 | c_2
j_2}(\zeta | \eta_2) \ldots M_{c_{n - 1} i_n | b j_n}(\zeta | \eta_n).
\end{multline*}
This is equation (\ref{Maibj}) with the dependence on the spectral parameters included.

For usual square lattice models, such as the six-vertex model, the representation $\psi$ coincides
with the representation $\varphi$. Here the monodromy operator $M_{\varphi, \psi}(\zeta | \eta) =
M_{\varphi, \varphi}(\zeta | \eta)$ coincides with the corresponding $R$-operator $R_\varphi(\zeta
| \eta)$.

\subsection{Transfer operators}

The transfer operators are obtained via taking the trace over the representation space $V$ of the
representation $\varphi$ used to define the monodromy operators. Some necessary information on
traces can be found in appendix \ref{a:ta}.

\subsubsection{Universal transfer operator} \label{s:Uto}

Let $\varphi$ be a representation of the algebra $A$ in the vector space $V$, and $t$ a group-like element of $A$,
\begin{equation}
\Delta(t) = t \otimes t. \label{Deltatau}
\end{equation}
We define the {\em universal transfer operator\/} as
\begin{equation*}
\calT_\varphi(\zeta) = (\tr_V \otimes \id)(\calM_\varphi(\zeta)(\varphi_\zeta(t) \otimes 1)) =
((\tr_V \circ \varphi_\zeta) \otimes \id) (\calR (t \otimes 1)),
\end{equation*}
where the mapping $\varphi_\zeta$ is defined by relation (\ref{varphizeta}). We call $t$ a {\em
twist element\/}.

It is easy to see that
\begin{equation*}
\Delta^{\mathrm {op}}(t) \eqWithRef{deltaop} \Pi(\Delta(t)) \eqWithRef{Deltatau} t \otimes t.
\end{equation*}
From the other hand
\begin{equation*}
\Delta^{\mathrm {op}}(t) \eqWithRef{auniR} \calR \, \Delta(t) \calR^{-1} \eqWithRef{Deltatau} \calR
(t \otimes t) \calR^{-1}.
\end{equation*}
Therefore, we have the equation
\begin{equation}
\calR (t \otimes t) = (t \otimes t) \calR. \label{rtautau}
\end{equation}
The above equation can be written as
\begin{equation*}
\calR^{12} t^1 t^2 = t^1 t^2 \calR^{12}.
\end{equation*}
Multiplying the Yang--Baxter equation (\ref{aRRR}) from the right by $t^1 t^2$ and using the above
equation, we obtain
\begin{equation}
(\calR^{13} t^1) (\calR^{23} t^2) = (\calR^{12})^{-1} (\calR^{23} t^2) (\calR^{13} t^1) \calR^{12}.
\label{Rtautau2}
\end{equation}
Applying to both sides of this equation the mapping $(\tr \circ \varphi_{\zeta_1}) \otimes (\tr
\circ \varphi_{\zeta_2}) \otimes \id$, we come to the equation
\begin{equation}
\calT_\varphi(\zeta_1) \calT_\varphi(\zeta_2) = \calT_\varphi(\zeta_2) \calT_\varphi(\zeta_1).
\label{TzetaTzeta}
\end{equation}
Here we use equation (\ref{traid}). More generally, if $\varphi_1$ and $\varphi_2$ are arbitrary
representations of the algebra $A$, then
\begin{equation}
\calT_{\varphi_1}(\zeta_1) \calT_{\varphi_2}(\zeta_2) = \calT_{\varphi_2}(\zeta_2)
\calT_{\varphi_1}(\zeta_1) \label{T1T2}
\end{equation}
for all $\zeta_1, \zeta_2 \in \bbC^\times$.

Let $a$ be an invertible group-like element of $A$ which commutes with $t$. Using the equation
\begin{equation*}
\calR^{12} a^1 a^2 = a^1 a^2 \calR^{12},
\end{equation*}
we obtain
\begin{equation*}
\calR^{12} t^1 a^1 a^2 = a^1 a^2 \calR^{12} t^1.
\end{equation*}
Rewriting this relation as
\begin{equation}
(a^1)^{-1} ((\calR^{12} t^1) \, a^2) \, a^1 = a^2 (\calR^{12} t^1), \label{aRa}
\end{equation}
and applying to both the sides the mapping $(\tr \circ \varphi_\zeta) \otimes \id$, we see that
\begin{equation*}
\calT_\varphi(\zeta) a = a \calT_\varphi(\zeta).
\end{equation*}
for any invertible group-like element $a \in A$ commuting with the twist element $t$.

Assume that the vector space $V$ is finite-dimensional of dimension $\ell$, and $\{e_a\}$ is a
basis of $V$. Denote by $\bm F_\varphi(\zeta) \in \Mat_\ell(\bbC)$ the matrix of $\varphi_\zeta(t)$
with respect to the basis $\{e_a\}$. It is clear that
\begin{equation}
\calT_\varphi(\zeta) = \tr(\bm{\calM}_\varphi(\zeta) \bm F_\varphi(\zeta)), \label{trMF}
\end{equation}
where the matrix $\bm{\calM}_\varphi(\zeta) \in \Mat_\ell(A)$ is defined in section \ref{s:Umm},
and, using the canonical embedding of the field $\bbC$ into $A$, we treat the matrix $\bm
F_\varphi(\zeta)$ as an element of $\Mat_\ell(A)$. Applying to both sides of equation
(\ref{rtautau}) the mapping $\varphi_{\zeta_1} \otimes \varphi_{\zeta_2}$, we obtain in terms of
the corresponding matrices
\begin{equation*}
\bm R_\varphi(\zeta_1 | \zeta_2) (\bm F_\varphi(\zeta_1) \otimes \bm F_\varphi(\zeta_2)) = (\bm
F_\varphi(\zeta_1) \otimes \bm F_\varphi(\zeta_2)) \bm R_\varphi(\zeta_1 | \zeta_2).
\end{equation*}
In terms of matrix entries this equation coincides with equation (\ref{RFF}).

\subsubsection{Transfer operator}

Let $\psi$ be a representation of the algebra $A$ in the vector space~$U$. We define the {\em
transfer operator\/} $T_{\varphi, \psi}(\zeta | \eta_1, \ldots, \eta_n)$ by the relation
\begin{multline*}
T_{\varphi, \psi}(\zeta | \eta_1, \ldots, \eta_n) = (\psi_{\eta_1} \otimes_{\Delta^\op}
\ldots \otimes_{\Delta^\op} \psi_{\eta_n}) (\calT_\varphi(\zeta)) \\
= ((\tr_V \circ \varphi_\zeta) \otimes (\psi_{\eta_1} \otimes_{\Delta^\op}
\ldots \otimes_{\Delta^\op} \psi_{\eta_n})) (\calR (t \otimes 1)),
\end{multline*}
where $\eta_1$, \ldots, $\eta_n$ are nonzero complex numbers, and the mapping $\psi_\eta$ is
defined in the same way as $\varphi_\zeta$. Equation (\ref{TzetaTzeta}) immediately gives
\begin{equation}
T_{\varphi, \psi}(\zeta_1 | \eta_1, \ldots, \eta_n) T_{\varphi, \psi}(\zeta_2 | \eta_1, \ldots,
\eta_n) = T_{\varphi, \psi}(\zeta_2 | \eta_1, \ldots, \eta_n) T_{\varphi, \psi}(\zeta_1 | \eta_1,
\ldots, \eta_n) \label{toto}
\end{equation}
for all $\zeta_1, \zeta_2 \in \bbC^\times$.

In the case when $\varphi$ is a finite-dimensional representation, we see that
\begin{multline*}
T_{\varphi, \psi}(\zeta | \eta_1, \ldots, \eta_n) \eqWithRef{trMF} \tr(\bm M_{\varphi, \psi}(\zeta
| \eta_1, \ldots, \eta_n) \bm F_\varphi(\zeta)) \\* \eqWithRef{bmMI} \tr ((\bm M_{\varphi,
\psi}(\zeta | \eta_1) \boxdot \ldots \boxdot \bm M_{\varphi, \psi}(\zeta | \eta_n)) \bm
F_\varphi(\zeta)).
\end{multline*}
Here the matrix $\bm M_{\varphi, \psi}(\zeta | \eta_1, \ldots, \eta_n)$ is defined in section
\ref{s:Mo} and the matrix $\bm F_\varphi(\zeta)$ in section \ref{s:Uto}. In the case where equation
(\ref{PhiPhiR}) is satisfied, from the above relation it follows, in particular, that
\begin{equation*}
T_{\varphi, \psi}(\zeta \nu | \eta_1 \nu, \ldots, \eta_n \nu)
= T_{\varphi, \psi}(\zeta | \eta_1, \ldots, \eta_n)
\end{equation*}
for any $\nu \in \bbC^\times$.

Assume now that $\psi$ is a finite-dimensional representation, $\{e_i\}$ is a basis of $U$, and
$\{E_{ij}\}$ the corresponding basis of $\End(U)$. We can write
\begin{equation*}
T_{\varphi, \psi}(\zeta | \eta_1, \ldots, \eta_n) = \sum_{\substack{i_1, \ldots, i_n
\\ j_1, \ldots, j_n}} T_{i_1 \ldots i_n | j_1 \ldots j_n} (\zeta | \eta_1, \ldots, \eta_n)
\otimes E_{i_1 j_1} \otimes \ldots \otimes E_{i_n j_n}
\end{equation*}
for some $ T_{i_1 \ldots i_n | j_1 \ldots j_n} (\zeta | \eta_1, \ldots, \eta_n) \in \End(V)$, and
define the matrix
\begin{equation*}
{\mathbb T}_{\varphi, \psi}(\zeta | \eta_1, \ldots, \eta_n) = (T_{i_1 \ldots i_n | j_1 \ldots j_n}
(\zeta | \eta_1, \ldots, \eta_n)).
\end{equation*}
Now we have
\begin{multline}
\mathbb T_{\varphi, \psi} (\zeta | \eta_1, \ldots, \eta_n) = \tr_V (\mathbb M_{\varphi, \psi}
(\zeta | \eta_1, \ldots, \eta_n) \varphi_\zeta(t)) \\ \eqWithRef{bbM} \tr_V (\mathbb M_{\varphi,
\psi}(\zeta |\eta_1) \boxtimes \ldots \boxtimes \mathbb M_{\varphi, \psi}(\zeta | \eta_n)
\varphi_\zeta(t)), \label{bbT}
\end{multline}
where the matrix $\mathbb M_{\varphi, \psi} (\zeta | \eta_1, \ldots, \eta_n)$ is defined in section
\ref{s:Mo}, and the mapping $\tr_V$ is applied to the matrix entries. As follows from (\ref{toto})
the matrices $\mathbb T_{\varphi, \psi} (\zeta | \eta_1, \ldots, \eta_n)$ for different values of
the spectral parameter $\zeta$ commute,
\begin{equation*}
[\mathbb T_{\varphi, \psi} (\zeta | \eta_1, \ldots, \eta_n), \mathbb T_{\varphi, \psi} (\zeta |
\eta_1, \ldots, \eta_n)] = 0,
\end{equation*}
see relation (\ref{TTcomm}).

\subsection{\texorpdfstring{$L$-operators}{L-operators}} \label{ss:lo}

To formulate and prove functional relations we additionally need $Q$-operators. We start with
$L$-operators which play in the construction of $Q$-operators the same role as monodromy operators
in the construction of transfer operators.

\subsubsection{Universal $L$-operator}

First of all, we assume that the universal $R$-matrix $\calR$ of the algebra $A$ is an element of
the tensor product $A_+ \otimes A_-$, where $A_+$ and $A_-$ are proper subalgebras of $A$. In
particular, it is so when $A$ is the quantum group associated with an affine Lie algebra, see, for
example, \cite{TolKho92, KhoTol92, LevSoiStu93, Dam98}. Certainly, any representation of $A$ can be
restricted to representations of $A_+$ and $A_-$. However, this does not give new interesting
objects. To construct $L$-operators one uses representations of $A_+$ which cannot be extended to
representations of $A$. Let $\rho$ be such a representation of $A_+$ in a vector space $W$. We
define the {\em universal $L$-operator\/} by the equation
\begin{equation*}
\calL_\rho(\zeta) = (\rho_\zeta \otimes \id)(\calR),
\end{equation*}
where the mapping $\rho_\zeta$ is defined by the relation similar to (\ref{varphizeta}). It is
clear that $\calL_\rho(\zeta)$ is an element of $\End(W) \otimes A_-$.

In spite of the fact that the definition of the universal $L$-operator is very similar to the
definition of the universal monodromy operator, we could not obtain for the universal $L$-operator all
relations satisfied by the universal monodromy operator. This is due to the fact that, to obtain such
relations, we should have a representation of the whole algebra $A$. Moreover, in all known
interesting cases $\rho$ is an infinite-dimensional representation, so we cannot introduce the
corresponding matrices.

In fact, to come to functional relations, one should choose representations $\rho$, defining
$Q$-operators, to be related to representations $\varphi$, used to define the monodromy operators
and the corresponding transfer operators. Presently, we do not have full understanding of how to do
it. It seems that representations $\rho$ should be obtained from representations $\varphi$ via some
limiting procedure, see \cite{BazHibKho02, HerJim11} and the discussion in section~\ref{s:elo}.

\subsubsection{$L$-operator}

Let $\psi$ be a representation of the subalgebra $A_-$ in the vector space $U$. To come to objects
satisfying functional relations one uses as $\psi$ the restriction to $A_-$ of the representation
used to define the corresponding monodromy and transfer operators. The {\em $L$-operator\/}
$L_{\rho, \psi}(\zeta | \eta_1, \ldots, \eta_n)$ is defined as
\begin{multline*}
L_{\rho, \psi}(\zeta | \eta_1, \ldots, \eta_n) = (\psi_{\eta_1} \otimes_{\Delta^\op} \ldots
\otimes_{\Delta^\op} \psi_{\eta_n})(\calL_\rho(\zeta)) \\= (\rho_\zeta \otimes (\psi_{\eta_1}
\otimes_{\Delta^\op} \ldots \otimes_{\Delta^\op} \psi_{\eta_n}))(\calR),
\end{multline*}
where $\eta_1$, \ldots, $\eta_n$ are some nonzero complex numbers and the mapping $\psi_\eta$ is
defined by the relation similar to (\ref{varphizeta}). It is evident that $L_{\rho, \psi}(\zeta |
\eta_1, \ldots, \eta_n)$ is an element of $\End(W) \otimes \End(U)^{\otimes n} \cong \End(W)
\otimes \End(U^{\otimes n})$. As well as for the monodromy operator, see (\ref{Mnvarphipsi}), one can
see that
\begin{equation}
L_{\rho, \psi}(\zeta | \eta_1 \ldots, \eta_n) = L^{1, 2}_{\rho, \psi}(\zeta | \eta_1) \ldots L^{1,
n + 1}_{\rho, \psi}(\zeta | \eta_n). \label{Lpsirho}
\end{equation}

If equation (\ref{PhiPhiR}) is satisfied, in the same way as for the case of $R$-operators,
see~(\ref{zetanuzeta}), we obtain
\begin{equation*}
L_{\rho, \psi}(\zeta \nu | \eta \nu) = L_{\rho, \psi}(\zeta | \eta),
\end{equation*}
and, therefore,
\begin{equation*}
L_{\rho, \psi}(\zeta | \eta) = L_{\rho, \psi}(\zeta \eta^{-1}),
\end{equation*}
where $L_{\rho, \psi}(\zeta) = L_{\rho, \psi}(\zeta | 1)$. Equation (\ref{Lpsirho}) in this case
gives
\begin{equation*}
L_{\rho, \psi}(\zeta \nu | \eta_1 \nu, \ldots, \eta_n \nu) = L_{\rho, \psi}(\zeta | \eta_1, \ldots, \eta_n).
\end{equation*}

Assume now that the representation $\psi$ is finite-dimensional. Let $\{e_i\}$ be a basis of $U$
and $\{E_{ij}\}$ the corresponding basis of $\End(U)$. We can write
\begin{equation*}
L_{\rho, \psi}(\zeta | \eta_1, \ldots, \eta_n) = \sum_{\substack{i_1, \ldots, i_n
\\ j_1, \ldots, j_n}} L_{i_1 \ldots i_n | j_1 \ldots j_n} (\zeta | \eta_1, \ldots, \eta_n)
\otimes E_{i_1 j_1} \otimes \ldots \otimes E_{i_n j_n},
\end{equation*}
where $L_{i_1 \ldots i_n | j_1 \ldots j_n} (\zeta | \eta_1, \ldots, \eta_n)$ are elements of
$\End(W)$. Now we can introduce the matrix
\begin{equation*}
\mathbb L_{\rho, \psi}(\zeta| \eta_1, \ldots, \eta_n) = (L_{i_1 \ldots i_n | j_1 \ldots j_n} (\zeta
| \eta_1, \ldots, \eta_n)),
\end{equation*}
and be convinced that
\begin{equation}
\mathbb L_{\rho, \psi} (\zeta | \eta_1, \ldots, \eta_n) = \mathbb L_{\rho, \psi}(\zeta | \eta_1)
\boxtimes \ldots \boxtimes \mathbb L_{\rho, \psi}(\zeta | \eta_n). \label{bbL}
\end{equation}

Detailed calculations giving the explicit forms of $L$-operators for the case of the quantum groups
$U_q(\calL(\gothsl_2))$ and $U_q(\calL(\gothsl_3))$ can be found in the paper
\cite{BooGoeKluNirRaz10}, and for the case of the quantum group $U_q(\calL(\gothsl_3, \mu))$, where
again $\mu$ is the standard diagram automorphism of $\gothsl_3$ of order 2, in the paper
\cite{BooGoeKluNirRaz11}.

\pagebreak

\subsection{\texorpdfstring{$Q$-operators}{Q-operators}}

\subsubsection{Universal $Q$-operator} \label{sss:uqo}

Given $\zeta \in \bbC^\times$, we define the {\em universal $Q$-operator\/} by the relation
\begin{equation*}
\calQ_\rho(\zeta) = \tr_W (\calL_\rho(\zeta)(\rho_\zeta(t) \otimes 1)).
\end{equation*}
One can easily see that
\begin{equation*}
\calQ_\rho(\zeta) = ((\tr_W \circ \rho_\zeta) \otimes \id) (\calR (t \otimes 1))
= ((\tr_W \circ \rho_\zeta) \otimes \id) (\calR^{12} t^1).
\end{equation*}
It is clear that $\calQ_\rho(\zeta)$ is an element of the algebra $A_-$.

Applying the mapping $(\tr \circ \rho_{\zeta_1}) \otimes (\tr \circ \varphi_{\zeta2}) \otimes \id$
to both sides of equation (\ref{Rtautau2}), we obtain
\begin{equation}
\calQ_\rho(\zeta_1) \calT_\varphi(\zeta_2) = \calT_\varphi(\zeta_2) \calQ_\rho(\zeta_1). \label{QTTQ}
\end{equation}
Here and below we assume that the same twist element is used to define both the universal
$Q$-operator and the universal transfer operator. Note also that, using equation (\ref{aRa}), one can
show that
\begin{equation*}
\calQ_\rho(\zeta) a = a \calQ_\rho(\zeta)
\end{equation*}
for any invertible group-like element $a$ which commutes with the twist element $t$. However, using
only equation (\ref{Rtautau2}), one cannot prove the commutativity of $\calQ_\rho(\zeta)$ for
different values of the spectral parameter, because $\rho$ cannot be extended to a representation
of the whole algebra $A$. Here the following fact appears to be useful \cite{RosWes02}.

Let $\rho_1$ and $\rho_2$ be two representations of the algebra $A_+$ in vector spaces $W_1$ and
$W_2$, respectively, and $\rho_{1 \zeta_1}$ and $\rho_{2 \zeta_2}$ the mappings constructed by the
relations similar to (\ref{varphizeta}). We have
\begin{multline*}
\calQ_{\rho_1}(\zeta_1) \calQ_{\rho_2}(\zeta_2) = \left[ ((\tr_{W_1} \circ \rho_{1 \zeta_1})
\otimes \id) ( \calR^{13} t^1) ) \right] \left[ ((\tr_{W_2} \circ \rho_{2 \zeta_2}) \otimes \id)
(\calR^{23} t^2) \right] \\* \eqWithRef{traotb} ((\tr_{W_1 \otimes W_2} \circ (\rho_{1 \zeta_1}
\otimes \rho_{2 \zeta_2})) \otimes \id) \left( \calR^{13} t^1 \calR^{23} t^2 \right).
\end{multline*}
Using equations (\ref{aDeltaid}) and (\ref{Deltatau}), we obtain
\begin{equation*}
\calR^{13} t^1 \calR^{23} t^2 = \left[ (\Delta \otimes \id)(\calR) \right] \left[ (\Delta \otimes
\id)(t \otimes 1) \right] = (\Delta \otimes \id)(\calR (t \otimes 1)).
\end{equation*}
Hence, one can write
\begin{equation*}
\calQ_{\rho_1}(\zeta_1) \calQ_{\rho_2}(\zeta_2) = (((\tr_{W_1 \otimes W_2} \circ (\rho_{1 \zeta_1}
\otimes \rho_{2 \zeta_2})) \otimes \id) \circ (\Delta \otimes \id))(\calR (t \otimes 1)),
\end{equation*}
and, finally,
\begin{equation}
\calQ_{\rho_1}(\zeta_1) \calQ_{\rho_2}(\zeta_2) = ((\tr_{W_1 \otimes W_2} \circ (\rho_{1 \zeta_1}
\otimes_\Delta \rho_{2 \zeta_2})) \otimes \id) (\calR (t \otimes 1)). \label{Q1Q2}
\end{equation}
In a similar way one can obtain expressions for other products. For example,
\begin{equation}
\calT_{\varphi}(\zeta_1) \calQ_{\rho}(\zeta_2) = ((\tr_{V \otimes W} \circ (\varphi_{\zeta_1}
\otimes_\Delta \rho_{\zeta_2})) \otimes \id) (\calR (t \otimes 1)). \label{T1Q2}
\end{equation}

\subsubsection{$Q$-operator}

We define the {\em $Q$-operator\/} $Q_{\rho, \psi}(\zeta | \eta_1, \ldots, \eta_n)$ by the relation
\begin{equation*}
Q_{\rho, \psi}(\zeta | \eta_1, \ldots, \eta_n) = \tr_W (L_{\rho, \psi}(\zeta | \eta_1, \ldots, \eta_n)).
\end{equation*}
It is evident that
\begin{multline*}
Q_{\rho, \psi}(\zeta | \eta_1, \ldots, \eta_n) = (\psi_{\eta_1} \otimes_{\Delta^\op}
\ldots \otimes_{\Delta^\op} \psi_{\eta_n}) (\calQ_\rho(\zeta)) \\
= ((\tr_W \circ \rho_\zeta) \otimes (\psi_{\eta_1} \otimes_{\Delta^\op}
\ldots \otimes_{\Delta^\op} \psi_{\eta_n})) (\calR (t \otimes 1)).
\end{multline*}

Assume now that $\psi$ is a finite-dimensional representation, $\{e_i\}$ is a basis of the
representation space $U$, and $\{E_{ij}\}$ the corresponding basis of $\End(U)$. We can write
\begin{equation*}
Q_{\rho, \psi}(\zeta | \eta_1, \ldots, \eta_n) = \sum_{\substack{i_1, \ldots, i_n \\ j_1, \ldots, j_n}}
Q_{i_1 \ldots i_n | j_1 \ldots j_n} (\zeta | \eta_1, \ldots, \eta_n) \otimes E_{i_1 j_1} \otimes \ldots
\otimes E_{i_n j_n}
\end{equation*}
where $Q_{i_1 \ldots i_n | j_1 \ldots j_n} (\zeta | \eta_1, \ldots, \eta_n)$ are the appropriate
elements of $\End(W)$, and define the matrix
\begin{equation*}
{\mathbb Q}_{\rho, \psi}(\zeta | \eta_1, \ldots, \eta_n) = (Q_{i_1 \ldots i_n | j_1 \ldots j_n}
(\zeta | \eta_1, \ldots, \eta_n)).
\end{equation*}
Now we have
\begin{multline*}
\mathbb Q_{\rho, \psi} (\zeta | \eta_1, \ldots, \eta_n) = \tr_W (\mathbb L_{\rho, \psi} (\zeta |
\eta_1, \ldots, \eta_n) \rho_\zeta(t)) \\ \eqWithRef{bbL} \tr_W (\mathbb L_{\rho, \psi}(\zeta
|\eta_1) \boxtimes \ldots \boxtimes \mathbb L_{\rho, \psi}(\zeta | \eta_n) \rho_\zeta(t)),
\end{multline*}
where the matrix $\mathbb L_{\rho, \psi} (\zeta | \eta_1, \ldots, \eta_n)$ is defined in section
\ref{ss:lo}, and $\tr_W$ is applied to the matrix entries.

From equation (\ref{QTTQ}) we obtain
\begin{equation*}
[Q_{\rho, \psi}(\zeta_1 | \eta_1, \ldots, \eta_n), T_{\varphi, \psi}(\zeta_2 | \eta_1, \ldots, \eta_n)] = 0,
\end{equation*}
or, in terms of the corresponding matrices,
\begin{equation*}
[\bbQ_{\rho, \psi}(\zeta_1 | \eta_1, \ldots, \eta_n), \bbT_{\varphi, \psi}(\zeta_2 | \eta_1, \ldots, \eta_n)] = 0,
\end{equation*}
see relation (\ref{TQcomm}).

Further progress in obtaining functional relations can be achieved only by using the properties of
the specific representations of concrete quasitriangular Hopf algebras. The corresponding calculations
were given for the case of the quantum group $U_q(\calL(\gothsl_2))$ in the papers
\cite{BazLukZam97, BazLukZam99, BooJimMiwSmiTak09}, for the case of the quantum group
$U_q(\calL(\gothsl_3))$ \cite{BazHibKho02}, for the case of the quantum group
$U_q(\calL(\gothsl_{2|1}))$ in the paper \cite{BazTsu08}, see also
\cite{AntFei97, Kor03, Kor05b, RosWes02, Koj08}. In the next section we
reconsider the case of $U_q(\calL(\gothsl_2))$, having in mind to fill certain
gaps of \cite{BazLukZam97, AntFei97, BazLukZam99, BooJimMiwSmiTak09} and to derive
the full set of functional relations in the model-independent form.

\section{Example related to the six-vertex model} \label{s:e}

As an example we consider the case of the quantum group $\uqlsl$. To obtain objects
related to integrable systems, we need representations of this quasitriangular Hopf
algebra. The standard method here is to use the Jimbo's homomorphism~\cite{Jim86a}
from $\uqlsl$ to the quantum group $\uqsl$, and then construct representations of
$\uqlsl$ from representations of $\uqsl$.

Depending on the sense of $q$, there are at least three definitions of a quantum group. According
to the first definition, $q = \exp \hbar$, where $\hbar$ is an indeterminate, according to the
second one, $q$ is indeterminate, and according to the third one, $q = \exp \hbar$, where $\hbar$
is a complex number such that $q \ne 0, \pm 1$. In the first case a quantum group is a
$\bbC[[\hbar]]$-algebra, in the second case a $\bbC(q)$-algebra, and in the third case it is just a
complex algebra. It seems that to define traces appropriately, it is convenient to use the third
definition. Therefore, we define the quantum group as a $\bbC$-algebra, see, for example, the books
\cite{JimMiw95, EtiFreKir98}.

\subsection{\texorpdfstring{Quantum group $\uqsl$}{Quantum group Uqsl2}}
\label{s:qgA1}

\subsubsection{Definition}

Let $\hbar$ be a complex number such that $q = \exp \hbar \ne 0, \pm 1$. We assume that $q^\nu$,
$\nu \in \bbC$, means the complex number $\exp (\hbar \nu)$. The quantum group $\uqsl$ is a
unital $\bbC$-algebra generated by the elements $E$, $F$, and $q^{\nu H}$, $\nu \in \bbC$, with the
following defining relations:
\begin{gather*}
q^0 = 1, \qquad q^{\nu_1 H} q^{\nu_2 H} = q^{(\nu_1 + \nu_2)H}, \\
q^{\nu H} E q^{-\nu H} = q^{2 \nu} E, \qquad q^{\nu H} F q^{- \nu H} = q^{- 2 \nu} F, \\
[E, F] = \kappa_q^{-1} (q^H - q^{-H}).
\end{gather*}
Here and below $\kappa_q = q - q^{-1}$. Note that $q^{\nu H}$ is just a notation, there is no an
element $H \in \uqsl$. In fact, it is constructive to identify $H$ with the standard Cartan element
of the Lie algebra $\gothsl_2$, and $\nu H$ with a general element of the Cartan subalgebra $\gothh
= \bbC H$. Using such interpretation, one can say that $q^{\nu H}$ is a set of generators
parametrized by the elements of the Cartan subalgebra $\gothh$.

The quantum group $\uqsl$ is also a Hopf algebra with the comultiplication
\begin{gather*}
\Delta(q^{\nu H}) = q^{\nu H} \otimes q^{\nu H}, \\*
\Delta(E) = E \otimes 1 + q^{-H} \otimes E, \qquad \Delta(F) = F \otimes q^H + 1 \otimes F,
\end{gather*}
and the correspondingly defined counit and antipode.\footnote{There are a few different equivalent
choices for comultiplication, counit and antipode in $\uqsl$. Since we are going to use the
Khoroshkin--Tolstoy expression for the universal $R$-matrix, we follow the convention of the paper
\cite{TolKho92}.}

The monomials $E^r F^s q^{\nu H}$ for $r, s \in \bbZ_{\ge 0}$ and $\nu \in \bbC$ form a basis of
$\uqsl$. There is one more basis defined with the help of the {\em quantum Casimir element\/} $C$
which has the form\footnote{We use a nonstandard, but convenient for our purposes, normalization of
$C$.}
\[
C = q^{H - 1} + q^{- H + 1} + \kappa_q^2 E F = q^{H + 1} + q^{- H - 1} + \kappa_q^2 F E.
\]
Here and below we use the notation $q^{\nu H + \mu} = q^\mu q^{\nu H}$, $\nu, \mu \in \bbC$. One
can verify that $C$ belongs to the centre of $\uqsl$. It is clear that the monomials $E^{r + 1} C^s
q^{\nu H}$, $F^{r + 1} C^s q^{\nu H}$ and $C^s q^{\nu H}$ for $r, s \in \bbZ_{\ge 0}$ and $\nu \in
\bbC$ also form a basis of $\uqsl$. This basis is convenient to define traces on $\uqsl$.

\subsubsection{Verma representation}

Given $\mu \in \bbC$, let $\widetilde V^\mu$ be a free vector space generated by the set $\{v_0,
v_1, \ldots \}$. Introduce the notation
\begin{equation*}
[\nu]_q = \frac{q^\nu - q^{-\nu}}{q - q^{-1}}, \qquad \nu \in \bbC.
\end{equation*}
One can show that the relations
\begin{equation}
q^{\nu H} v_n = q^{\nu(\mu - 2 n)} v_n, \qquad E v_n = [n]_q [\mu - n + 1]_q v_{n - 1}, \qquad F
v_n = v_{n + 1} \label{aHvk}
\end{equation}
endow $\widetilde V^\mu$ with the structure of a left $\uqsl$-module. The module $\widetilde V^\mu$
is isomorphic to the Verma module with the highest weight whose action on $H$ gives $\mu$.

We denote the representation of $\uqsl$ corresponding to the module $\widetilde V^\mu$ by
$\widetilde \pi^\mu$. If $\mu$ equals a non-negative integer $m$, the linear hull of the vectors
$v_n$ with $n > m$ is a submodule of $\widetilde V^m$ isomorphic to the module $\widetilde V^{- m -
2}$. We denote the corresponding finite-dimensional quotient module by $V^m$ and the corresponding
representation by $\pi^{m}$.

It is easy to see that for the quantum Casimir element we have
\begin{equation}
\widetilde \pi^{\mu}(C) = q^{\mu + 1} + q^{- \mu - 1}, \qquad \pi^{m}(C) = q^{m + 1} + q^{- m - 1}
\label{pimuC}
\end{equation}
for any $\mu \in \bbC$ and $m \in \bbZ_{\ge 0}$.

\subsection{\texorpdfstring{Quantum group $\uqlsl$}{Quantum group UqLsl2}} \label{s:qga11}

\subsubsection{Definition}

We start with the quantum group $\uqhlsl$. The reason is that the expression for the universal
$R$-matrix given by Khoroshkin and Tolstoy \cite{TolKho92} is valid for the case of $\uqhlsl$.

First, let us describe the root system of $\hlsl$. The Cartan subalgebra of $\hlsl$ is
\begin{equation*}
\widehat \gothh = \gothh \oplus \bbC c \oplus \bbC d,
\end{equation*}
where $\gothh = \bbC H$ is the standard Cartan subalgebra of $\gothsl_2$, $c$ is the central
element, and $d$ is the derivation \cite{Kac90}. Define the Cartan elements
\begin{equation}
h_0 = c - H, \qquad h_1 = H, \label{h0h1}
\end{equation}
so that one has
\begin{equation*}
\widehat \gothh = \bbC h_0 \oplus \bbC h_1 \oplus \bbC d.
\end{equation*}
The simple positive roots $\alpha_0, \alpha_1 \in \widehat{\gothh}^*$ are given by the equations
\begin{gather*}
\alpha_j(h_i) = a_{ij}, \\
\alpha_0(d) = 1, \qquad \alpha_1(d) = 0,
\end{gather*}
where
\begin{equation*}
(a_{ij}) = \left(\begin{array}{rr}
2 & - 2 \\
-2 & 2
\end{array} \right)
\end{equation*}
is the Cartan matrix of the Lie algebra $\hlsl$. The full root system $\triangle$ of $\hlsl$ is the
disjoint union of the system of positive roots
\begin{multline*}
\triangle_+ = \{ \alpha_0 + k (\alpha_0 + \alpha_1) \mid k \in \bbZ_{\ge 0} \} \cup \{ \alpha_1 + k
(\alpha_0 + \alpha_1) \mid k \in \bbZ_{\ge 0} \} \\*\cup \{ k (\alpha_0 + \alpha_1) \mid k \in
\bbZ_{> 0} \}
\end{multline*}
and the system of negative roots $\triangle_- = - \triangle_+$ \cite{Kac90}.

Let again $\hbar$ be a complex number, such that $q = \exp \hbar \ne 0, \pm 1$. The quantum group
$\uqhlsl$  is a unital $\bbC$-algebra generated by the elements $e_i$, $f_i$, $i = 0, 1$, and $q^x$, $x
\in \widehat{\gothh}$, with the relations
\begin{gather}
q^0 = 1, \qquad q^{x_1} q^{x_2} = q^{x_1 + x_2}, \label{qx} \\
q^x e_i q^{-x} = q^{\alpha_i(x)} e_i, \qquad q^x f_i q^{-x} = q^{-\alpha_i(x)} f_i, \label{qeq} \\
[e_i, f_j] = \kappa_q^{-1} \delta_{ij} \, (q^{h_i} - q^{-h_i}) \label{ef}
\end{gather}
satisfied for all $i$ and $j$, and the Serre relations
\begin{gather}
e_i^3 e_j^{\mathstrut} - [3]_q  e_i^2 e_j^{\mathstrut} e_i^{\mathstrut}
+ [3]_q e_i^{\mathstrut} e_j^{\mathstrut} e_i^2 - e_j^{\mathstrut} e_i^3 = 0, \label{sre} \\
f_i^3 f_j^{\mathstrut} - [3]_q  f_i^2 f_j^{\mathstrut} f_i^{\mathstrut}
+ [3]_q f_i^{\mathstrut} f_j^{\mathstrut} f_i^2 - f_j^{\mathstrut} f_i^3 = 0 \label{srf}
\end{gather}
satisfied for all distinct $i$ and $j$.

The quantum group $\uqhlsl$ is a Hopf algebra with the comultiplication $\Delta$ defined by the
relations
\begin{gather}
\Delta(q^x) = q^x \otimes q^x, \label{dqx} \\
\Delta(e_i) = e_i \otimes 1 + q^{-h_i} \otimes e_i, \qquad \Delta(f_i) = f_i \otimes q^{h_i} + 1
\otimes f_i, \label{defi}
\end{gather}
and with the correspondingly defined counit and antipode.

To give the definition of the quantum group $\uqlsl$, we first introduce the Hopf subalgebra
$\uqtlsl$ of $\uqhlsl$ generated by $e_i$, $f_i$, $i = 0, 1$, and $q^x$, $x \in \widetilde \gothh$,
where
\begin{equation*}
\widetilde \gothh = \bbC h_0 \oplus \bbC h_1 = \bbC H \oplus \bbC c.
\end{equation*}
The quantum croup $\uqlsl$ can be defined as the quotient algebra of $\uqtlsl$ by the two-sided
ideal generated by the elements of the form $q^{\nu c} - 1$, $\nu \in \bbC^\times$. In terms of
generators and relations the quantum group $\uqlsl$ is a $\bbC$-algebra generated by the elements
$e_i$, $f_i$, $i = 0, 1$, and $q^x$, $x \in \widetilde{\gothh}$, with relations
(\ref{qx})--(\ref{srf}) and
\begin{equation}
q^{\nu(h_0 + h_1)} = q^{\nu c} = 1, \label{qh0h1}
\end{equation}
where $\nu \in \bbC^\times$. It is a Hopf algebra with the comultiplication defined by (\ref{dqx}),
(\ref{defi}) and with the correspondingly defined counit and antipode. One of the reasons to use
the quantum group $\uqlsl$ instead of $\uqhlsl$ is that $\uqhlsl$ has no finite-dimensional
representations with a nontrivial action of $q^{\nu (h_0 + h_1)} = q^{\nu c}$.

\subsubsection{Useful basis}

We call a nonzero element $a \in \uqhlsl$ a {\em root element\/} corresponding to the root $\gamma
\in \widehat{\gothh}^*$ if
\begin{equation*}
q^x a \, q^{-x} = q^{\gamma(x)} a
\end{equation*}
for any $x \in \widehat{\gothh}$. It can be shown that for any $\gamma \in \triangle$ there is a
nonzero root element, and this element is unique up to multiplication by a nonzero scalar factor.
It is clear that the generators $e_i$ and $f_i$ correspond to the roots $\alpha_i$ and $-\alpha_i$
respectively. Choose for each root of $\triangle$ a root element, and denote the root element
corresponding to a positive root~$\gamma$ by $e_\gamma$ and the root element corresponding to a
negative root $-\gamma$ by $f_\gamma$. Assume that some total order $\prec$ of positive roots is
fixed. It appears that the monomials of the form
\begin{equation*}
e_{\gamma_1}^{k_1} \ldots e_{\gamma_r}^{k_r} \, f_{\delta_1}^{\ell_1} \ldots f_{\delta_s}^{\ell_s} \, q^x,
\end{equation*}
where $\gamma_1 \prec \ldots \prec \gamma_r$ and $\delta_1 \prec \ldots \prec \delta_s$, form a
basis of $\uqhlsl$.

Let us describe the method to construct the root elements corresponding to the roots of $\triangle$
used by Khoroshkin and Tolstoy \cite{TolKho92}. It can be shown \cite{KhoTol93} that the appearing
root elements are closely related to the quantum group generators introduced by
Drinfeld~\cite{Dri88}.

It is customary to denote $\alpha = \alpha_1$ and $\delta = \alpha_0 + \alpha_1$, so that the
simple positive roots are now~$\delta - \alpha$ and~$\alpha$. Then the system of positive roots is
\begin{equation*}
\triangle_+ = \{ \alpha + k \delta \mid k \in \bbZ_{\ge 0} \} \cup \{ k \delta \mid k \in \bbZ_{>
0} \} \cup \{ \delta - \alpha + k \delta \mid k \in \bbZ_{\ge 0} \}.
\end{equation*}
For the simple roots we choose
\begin{equation}
e_\alpha = e_1, \qquad e_{\delta - \alpha} = e_0, \qquad f_\alpha = f_1,
\qquad f_{\delta - \alpha} = f_0. \label{eafa}
\end{equation}
Now we define the root element corresponding to the root $\delta$ putting
\begin{equation}
e'_\delta = e_\alpha \, e_{\delta - \alpha} - q^{-2} e_{\delta - \alpha} \, e_\alpha. \label{epd}
\end{equation}
Here we use the prime because to construct the universal $R$-matrix we redefine the root elements
corresponding to the roots $k \delta$ and $- k \delta$ and denote by $e_{k \delta}$ and $f_{k
\delta}$ the result of the redefinition. The remaining root elements corresponding to the positive
roots are defined recursively by the relations
\begin{gather}
e_{\alpha + k \delta} = [2]_q^{-1} \bigl( e_{\alpha + (k - 1) \delta} \, e'_\delta - e'_\delta \,
e_{\alpha + (k - 1) \delta} \bigr), \label{eamd} \\[.5em]
e_{(\delta - \alpha) + k \delta} = [2]_q^{-1} \bigl(  e'_\delta \,
e_{(\delta - \alpha) + (k - 1) \delta} - e_{(\delta - \alpha) + (k - 1) \delta} \, e'_\delta \bigr),
\label{edmamd} \\[.5em]
e'_{k \delta} = e_{\alpha + (k - 1) \delta} \, e_{\delta - \alpha} - q^{-2}  e_{\delta - \alpha} \,
e_{\alpha + (k - 1) \delta}. \label{epkd}
\end{gather}
The root elements corresponding to the negative roots are defined with the help of the relations
\begin{gather}
f'_\delta = f_{\delta - \alpha} \, f_\alpha \, - q^2 f_\alpha \, f_{\delta - \alpha}, \label{fpd} \\[.5em]
f_{\alpha + k \delta} = [2]_q^{-1} \bigl( f'_\delta \, f_{\alpha + (k - 1) \delta}
- f_{\alpha + (k - 1) \delta} \, f'_\delta\bigr), \label{famd} \\[.5em]
f_{(\delta - \alpha) + k \delta} = [2]_q^{-1}
\bigl(  f_{(\delta - \alpha) + (k - 1) \delta} \, f'_\delta
- f'_\delta \, f_{(\delta - \alpha) + (k - 1) \delta} \bigr) , \label{fdmamd} \\[.5em]
f'_{k \delta} = f_{\delta - \alpha} \, f_{\alpha + (k - 1) \delta} \, - q^2 f_{\alpha + (k - 1)
\delta} \, f_{\delta - \alpha}. \label{fpkd}
\end{gather}
The coefficients in (\ref{eamd}), (\ref{edmamd}), (\ref{famd}) and (\ref{fdmamd}) are chosen in
such a way that for $\gamma = \alpha$ and $\gamma = \delta - \alpha$ we have
\begin{equation*}
[e_{\gamma + k \delta}, f_{\gamma + k \delta}] = \kappa_q^{-1} (q^{h_\gamma} - q^{-h_\gamma}),
\end{equation*}
where $h_\alpha = h_1$ and $h_{\delta - \alpha} = h_0$.

The root elements $e_{k \delta}$ needed for the construction of the universal $R$-matrix are
related to the root elements $e'_{k \delta}$ by the equation
\begin{equation}
\kappa_q \, e_ \delta(x) = \log (1 + \kappa_q \, e'_\delta(x)), \label{edx}
\end{equation}
where
\begin{equation*}
e'_\delta(x) = \sum_{k = 1}^\infty e'_{k \delta} x^{-k}, \qquad
e_\delta(x) = \sum_{k = 1}^\infty e_{k \delta} x^{-k}.
\end{equation*}
The root elements $f_{k \delta}$ are defined with the help of the equation
\begin{equation}
- \kappa_q \, f_ \delta(x) = \log (1 - \kappa_q \, f'_\delta(x)), \label{fdx}
\end{equation}
where
\begin{equation*}
f'_\delta(x) = \sum_{k = 1}^\infty f'_{k \delta} x^{-k}, \qquad
f_\delta(x) = \sum_{k = 1}^\infty f_{k \delta} x^{-k}.
\end{equation*}

\subsubsection{Universal $R$-matrix}

We follow here the approach developed by Khoroshkin and Tolstoy \cite{TolKho92}. Although this is
not clearly stated in the paper \cite{TolKho92}, Khoroshkin and Tolstoy define a quantum group as a
$\bbC[[\hbar]]$-algebra. In fact, one can use the expression for the universal $R$-matrix from the
paper \cite{TolKho92} also for the case of a quantum group defined as a $\bbC$-algebra having in
mind that in this case a quantum group is quasitriangular only in some restricted sense. Namely,
all the relations involving the universal $R$-matrix should be considered as valid only for the
weight $\uqhlsl$-modules, see in this respect the paper \cite{Tan92} and the discussion below.
Remind that a $\uqhlsl$-module $V$ is a {\em weight module\/} if
\begin{equation*}
V = \bigoplus_{\lambda \in \widehat{\gothh}^*} V_\lambda,
\end{equation*}
where
\begin{equation*}
V_\lambda = \{v \in V \mid q^x v = q^{\lambda(x)} v \mbox{ for any } x \in \widehat{\gothh} \}.
\end{equation*}
The same terminology is used for the corresponding representations.

According to the paper \cite{TolKho92}, one starts with choosing some normal order $\prec$ of the
positive roots of $\hlsl$. In general, one says that a system of positive roots is supplied with a
normal order if its roots are totally ordered in such a way that

\begin{itemize}
\item[(i)] all multiple roots follow each other in an arbitrary order;
\item[(ii)] each non-simple root $\alpha + \beta$, where $\alpha$ is not proportional to $\beta$,
is placed between $\alpha$ and $\beta$.
\end{itemize}

In our case a normal order is fixed uniquely if we require
\begin{equation*}
\alpha + k \delta \prec \ell \delta \prec (\delta - \alpha) + m \delta,
\end{equation*}
and in accordance with it the roots go as
\begin{equation*}
\alpha, \, \alpha + \delta, \, \ldots, \, \alpha + k \delta, \, \ldots, \, \delta, \, 2 \delta, \,
\ldots, \, \ell \delta, \, \ldots, \, \ldots, \, (\delta - \alpha) + m \delta, \, \ldots, \,
(\delta - \alpha) + \delta, \, \delta - \alpha.
\end{equation*}

The expression for the universal $R$-matrix, obtained by Khoroshkin and Tolstoy, has the form
\begin{equation}
\calR = \calR_{\prec \delta} \, \calR_{\sim \delta} \, \calR_{\succ \delta} \, \calK. \label{RRRK}
\end{equation}
The first factor is the product over $k \in \bbZ_{\ge 0}$ of the $q$-exponentials
\begin{equation*}
\calR_{\alpha, \, k} = \exp_{q^{-2}} \left( \kappa_q \, e_{\alpha + k
\delta} \otimes f_{\alpha + k \delta} \right)
\end{equation*}
in the order coinciding with the chosen normal order of the roots $\alpha + k \delta$. Here the
$q$-exponential is defined as
\begin{equation*}
\exp_q (x) = \sum_{n = 1}^\infty \frac{x^n}{(n)_q},
\end{equation*}
where
\begin{equation*}
(n)_q! = (1)_q (2)_q \ldots (n)_q, \qquad (n)_q = \frac{q^n - 1}{q - 1}.
\end{equation*}
The factor $\calR_{\sim \delta}$ is given by the expression
\begin{equation}
\calR_{\sim \delta} = \exp \left( \kappa_q \sum_{k = 1}^\infty \frac{k}{[2 k]_q} \, e_{k \delta}
\otimes f_{k \delta} \right). \label{Rsd}
\end{equation}
The factor $\calR_{\succ \delta}$ is the product over $k \in \bbZ_{\ge 0}$ of the $q$-exponentials
\begin{equation*}
\calR_{\delta - \alpha, \, k} = \exp_{q^{-2}} \left( \kappa_q \, e_{(\delta - \alpha) + k \delta}
\otimes f_{(\delta - \alpha) + k \delta} \right) \label{rmdma1}
\end{equation*}
in the order coinciding with the chosen normal order of the roots $(\delta - \alpha) + k \delta$.

The last factor $\calK$ is not defined as an element of $\uqhlsl \otimes \uqhlsl$. However, one can
define its action on the tensor product of  any two weight $\uqhlsl$-modules. Let $V$ and $U$ be
weight $\uqhlsl$-modules with the weight decompositions
\begin{equation*}
V = \bigoplus_{\lambda \in \widehat{\gothh}^*} V_\lambda, \qquad U = \bigoplus_{\mu \in \widehat{\gothh}^*} U_\mu.
\end{equation*}
The action of $\calK$ on $V \otimes U$ is defined by the relation
\begin{equation}
\calK \, v \otimes u = q^{\lambda(h_\alpha) \mu(h_\alpha)/ 2 + \lambda(c) \mu(d) + \lambda(d)
\mu(c)} \, v \otimes u, \label{kuvh}
\end{equation}
where $v \in V_\lambda$ and $u \in U_\mu$. Slightly abusing the notation, we denote the
corresponding operator by $(\varphi \otimes \psi)(\calK)$, where $\varphi$ and $\psi$ are the
representations corresponding to the modules $V$ and $U$ respectively. If the module $U$ is
finite-dimensional, $\{e_r\}$ is a basis of $U$ formed by weight vectors, and $\{E_{rs}\}$ is the
corresponding basis of $\End(U)$, we have
\begin{equation}
(\varphi \otimes \psi)(\calK) = \sum_r \varphi(q^{\mu_r(h_\alpha) h_\alpha / 2 + \mu_r(c) d +
\mu_r(d) c}) \otimes E_{rr}, \label{vppkh}
\end{equation}
where $\mu_r$ is the weight of $e_r$.

Note that in the case when we define a quantum group as a $\bbC[[\hbar]]$-algebra, $\calK$ is an
element of its tensor square of the form
\begin{equation*}
\calK = q^{h_\alpha \otimes h_\alpha / 2 + c \otimes d + d \otimes c},
\end{equation*}
and the notation $(\varphi \otimes \psi)(\calK)$ has a straightforward sense.

In the case of the quantum group $\uqlsl$ we again define the elements $e_\gamma$ and $f_\gamma$,
$\gamma \in \triangle_+$, by relations (\ref{eafa})--(\ref{fdx}). The universal $R$-matrix is again
defined by equation (\ref{RRRK}), where the factors $\calR_{\prec \delta}$, $\calR_{\sim \delta}$
and $\calR_{\succ \delta}$ are defined in the same way as in the case of the quantum group
$\uqhlsl$, while for the factor $\calK$ we have
\begin{equation}
\calK \, v \otimes u = q^{\lambda(h_\alpha) \mu(h_\alpha)/ 2} \, v \otimes u \label{kuv}
\end{equation}
instead of equation (\ref{kuvh}), and
\begin{equation}
(\varphi \otimes \psi)(\calK) = \sum_r \varphi(q^{\mu_r(h_\alpha) h_\alpha / 2}) \otimes E_{rr} \label{vppk}
\end{equation}
instead of equation (\ref{vppkh}).

\subsection{\texorpdfstring{$R$-operators}{R-operators}} \label{s:ero}

First of all, we assume that actual spectral parameters are complex numbers $u$ and $v_i$ such that
\begin{equation}
\zeta = q^u = \rme^{\hbar u}, \qquad \eta_i = q^{v_i} = \rme^{\hbar v_i}. \label{zetazetaw}
\end{equation}
This convention allows us to uniquely define arbitrary complex powers of $\zeta$ and $\eta_i$.

To construct $R$-operators we need representations of the quantum group $\uqlsl$. We start with the
Jimbo's homomorphism~\cite{Jim86a}
\begin{equation*}
\varphi: \uqlsl \to \uqsl
\end{equation*}
defined by the equations
\begin{gather}
\varphi(q^{\nu h_\alpha}) = q^{\nu H}, \qquad \varphi(e_\alpha) = E, \qquad \varphi(f_\alpha)
= F, \label{Jimboa} \\
\varphi(q^{\nu h_{\delta - \alpha}}) = q^{- \nu H}, \qquad \varphi(e_{\delta - \alpha})
= F, \qquad \varphi(f_{\delta - \alpha}) = E. \label{Jimbob}
\end{gather}
Let $\widetilde \pi^{\mu}$ be the highest weight infinite-dimensional representation of $\uqsl$
with the highest weight $\mu$ described above. We define a representation $\widetilde
\varphi^{\mu}$ of $\uqlsl$ as
\begin{equation*}
\widetilde \varphi^{\mu} = \widetilde \pi^{\mu} \circ \varphi.
\end{equation*}
One more necessary ingredient is a $\bbZ$-gradation of $\uqlsl$. We define it  assuming that the
generators $q^x$ belong to the zero grade subspace, the generators $e_i$ belong to the graded
subspaces with the grading indices $s_i$, and the generators $f_i$ belong to the graded subspaces
with the grading indices $-s_i$. For the mapping $\Phi_\nu$, defined by (\ref{aPhinu}), we have
\begin{equation}
\Phi_\nu(q^x) = q^x, \qquad \Phi_\nu(e_i) = \nu^{s_i} e_i, \qquad \Phi_\nu(f_i) = \nu^{-s_i} f_i. \label{Phinuh}
\end{equation}
Note that with this definition of a $\bbZ$-gradation of $\uqlsl$ the universal $R$-matrix
(\ref{RRRK}) satisfies equation (\ref{PhiPhiR}). Below we denote $s = s_0 + s_1$. Now, given $\zeta
\in \bbC^\times$, we define the representation $\widetilde \varphi^{\mu}_\zeta$ as
\begin{equation*}
\widetilde \varphi^{\mu}_\zeta = \widetilde \pi^{\mu} \circ \varphi \circ \Phi_\zeta.
\end{equation*}
Slightly abusing the notation, we denote the corresponding $\uqlsl$-modules by~$\widetilde V^\mu$
and~$\widetilde V^\mu_\zeta$. Taking into account (\ref{aHvk}), (\ref{Jimboa}), (\ref{Jimbob}), and
(\ref{Phinuh}), we see that for the module $\widetilde V^\mu_\zeta$ one has
\begin{align}
&q^{\nu h_\alpha} \, v_n = q^{\nu (\mu - 2 n)} \, v_n, && q^{\nu h_{\delta - \alpha}} \, v_n = q^{-
\nu (\mu - 2 n)} \, v_n, \label{Vh} \\* &e_\alpha \, v_n = \zeta^{s_1} [n]_q [\mu - n + 1]_q \,
v_{n - 1}, && e_{\delta - \alpha} \, v_n = \zeta^{s_0} \, v_{n + 1}, \label{Ve} \\* &f_\alpha \,
v_n = \zeta^{-s_1} \, v_{n + 1}, && f_{\delta - \alpha} \, v_n = \zeta^{-s_0} [n]_q [\mu - n + 1]_q
\, v_{n - 1}. \label{Vf}
\end{align}
In the case when $\mu$ equals a non-negative integer $m$ we denote the corresponding
finite-dimensional representations by $\varphi^m$ and $\varphi^m_\zeta$, and the modules by $V^m$
and $V^m_\zeta$.

Now we denote
\begin{equation*}
R_m(\zeta_{12}) = (\varphi^{m}_{\zeta_1} \otimes \varphi^{m}_{\zeta_2}) (\calR)
\end{equation*}
and shortly describe how to find an explicit expression for $R(\zeta) = R_1(\zeta)$. We refer the
reader to the paper \cite{BooGoeKluNirRaz10} for more details.

The representation $\pi^1$ of $\uqsl$ is two-dimensional and we have
\begin{equation}
\pi^1(q^{\nu H}) = q^\nu E_{11} + q^{-\nu} E_{22}, \qquad \pi^1(E) = E_{12}, \qquad \pi^1(F) =
E_{21}. \label{pi1}
\end{equation}
Here and below $E_{ab}$ are elements of the basis of $\End(\bbC^2)$ corresponding to the standard
basis $\{e_a\}$ of $\bbC^2$. Using (\ref{Jimboa}), (\ref{Jimbob}) and (\ref{Phinuh}), we come to
the relations
\begin{align}
&\varphi^1_\zeta(q^{\nu h_\alpha}) = q^\nu E_{11} + q^{- \nu} E_{22}, && \varphi^1_\zeta(q^{\nu
h_{\delta - \alpha}}) = q^{- \nu} E_{11} + q^\nu E_{22},\label{vha} \\* &\varphi^1_\zeta(e_\alpha)
= \zeta^{s_1} E_{12}, && \varphi^1_\zeta(e_{\delta - \alpha}) = \zeta^{s - s_1} E_{21},\label{pi1e}
\\* &\varphi^1_\zeta(f_\alpha) = \zeta^{-s_1} E_{21}, && \varphi^1_\zeta(f_{\delta - \alpha}) =
\zeta^{- s + s_1} E_{12}. \label{pi1f}
\end{align}

It follows from (\ref{epd}) and (\ref{fpd}) that
\begin{equation*}
\varphi^1_\zeta(e'_\delta) = \zeta^s (E_{11} - q^{-2} E_{22}), \qquad \varphi^1_\zeta(f'_\delta) =
\zeta^{-s} (E_{11} - q^2 E_{22}),
\end{equation*}
and the recursive definitions (\ref{eamd}), (\ref{edmamd}), (\ref{famd}) and (\ref{fdmamd}) give
\begin{gather}
\varphi^1_\zeta(e_{\alpha + k \delta}) = (-1)^k q^{-k} \zeta^{s_1 + k s} E_{12},
\qquad \varphi^1_\zeta(e_{(\delta - \alpha) + k \delta})
= (-1)^k q^{-k} \zeta^{(s - s_1) + k s} E_{21}, \label{eapmd} \\
\varphi^1_\zeta(f_{\alpha + k\delta}) = (-1)^k q^k \zeta^{- s_1 - k s} E_{21}, \qquad
\varphi^1_\zeta(f_{(\delta - \alpha) + k \delta})
= (-1)^k q^k \zeta^{- (s - s_1) - k s} E_{12}. \label{fapmd}
\end{gather}
Starting from (\ref{epkd}) and (\ref{eapmd}), we come to the equation
\[
\varphi^1_\zeta(e'_{k \delta}) = (-1)^{k - 1} q^{-k + 1} \zeta^{k s} (E_{11} - q^{-2} E_{22}).
\]
Taking into account (\ref{edx}), we obtain
\begin{equation}
\varphi^1_\zeta(e_{k\delta}) = (-1)^{k - 1} [k]_q \frac{\zeta^{k s}}{k} (E_{11} - q^{- 2 k} E_{22}).
\label{emd}
\end{equation}
In a similar way, starting from (\ref{fpkd}) and (\ref{fapmd}) and taking into account (\ref{fdx}),
we determine that
\begin{equation}
\varphi^1_\zeta(f_{k\delta}) = (-1)^{k - 1} [k]_q \, \frac{\zeta^{- k s}}{k} (E_{11} - q^{2 k}  E_{22}).
\label{fmd}
\end{equation}

Now we can obtain expressions for the images of the factors entering the Khorosh\-kin--Tolstoy
formula (\ref{RRRK}) for the universal $R$-matrix. To find expressions for the images of the
factors $\calR_{\prec \delta}$ and $\calR_{\succ \delta}$, we use the identities
\begin{equation}
(E_{12})^n = 0, \qquad (E_{21})^n = 0
\label{eek}
\end{equation}
valid for any integer $n > 1$.  Using (\ref{eapmd}) and (\ref{fapmd}) and summing up the arising
geometric series, we find
\begin{align}
(\varphi^1_{\zeta_1} \otimes \varphi^1_{\zeta_2}) ( \calR_{\prec \delta}) &= 1 \otimes 1 + \kappa_q
\frac{\zeta^{s_1}_{12}}{1 - \zeta^s_{12}} \, E_{12} \otimes E_{21},
\label{frlda1} \\
(\varphi^1_{\zeta_1} \otimes \varphi^1_{\zeta_2}) ( \calR_{\succ \delta})
&= 1 \otimes 1 + \kappa_q \frac{\zeta^{s - s_1}_{12}}{1 - \zeta^s_{12}} \, E_{21} \otimes E_{12}.
\label{frgda1}
\end{align}
Using (\ref{emd}) and (\ref{fmd}), we obtain for the image of the factor $\calR_{\sim \delta}$ the
expression
\begin{multline}
(\varphi^1_{\zeta_1} \otimes \varphi^1_{\zeta_2}) (\calR_{\sim \delta}) =
\rme^{\lambda_2(q \zeta^s_{12}) - \lambda_2(q^{-1} \zeta^s_{12})} \biggl[ E_{11} \otimes  E_{11} \\*
+ \frac{1 - q^2 \zeta^s_{12}}{1 - \zeta^s_{12}} \, E_{11} \otimes E_{22}
+ \frac{1 - \zeta^s_{12}}{1 - q^{-2} \zeta^s_{12}} \, E_{22} \otimes E_{11} + E_{22} \otimes E_{22} \biggr],
\label{frpda1}
\end{multline}
where
\begin{equation}
\lambda_2(\zeta) = \sum_{k \in \bbZ_{>0}} \, \frac{1}{q^k + q^{-k}} \, \frac{\zeta^k}{k}
= \sum_{k = 1}^\infty \, \frac{1}{[2]_{q^k}} \, \frac{\zeta^k}{k}.
\label{fa1}
\end{equation}
The simplest part of the calculations is to obtain an expression for the action of the factor
$\calK$ on the space $\bbC^2 \otimes \bbC^2$. As before, slightly abusing the notation, we denote
the corresponding operator by $\varphi^1_{\zeta_1} \otimes \varphi^1_{\zeta_2} (\calK)$. Let
$\mu_1$ and $\mu_2$ be the weights of the vectors $e_1$ and $e_2$ forming the standard basis of
$\bbC^2$. It follows from (\ref{vha}) that
\begin{equation}
\mu_1(h_\alpha) = 1, \qquad \mu_2(h_\alpha) = -1, \label{muh}
\end{equation}
and equation (\ref{vppk}) gives
\begin{multline}
(\varphi^1_{\zeta_1} \otimes \varphi^1_{\zeta_2}) (\calK) = q^{1/2} E_{11} \otimes E_{11} +
q^{-1/2} E_{11} \otimes E_{22} \\* + q^{-1/2} E_{22} \otimes E_{11} + q^{1/2} E_{22} \otimes
E_{22}. \label{fka1}
\end{multline}

Now we have the expressions for all factors necessary to obtain the expression for $R(\zeta) =
R_1(\zeta)$. After simple calculations we determine that
\begin{multline}
R(\zeta) = q^{1/2} \rme^{\lambda_2(q \zeta^s) - \lambda_2(q^{-1} \zeta^s)} \biggl[ E_{11} \otimes
E_{11} + E_{22} \otimes E_{22} \\* + \frac{q^{-1} (1 - \zeta^s)}{1 - q^{-2} \zeta^s} (E_{11}
\otimes E_{22} + E_{22} \otimes E_{11}) \\* + \frac{1 - q^{-2}}{1 - q^{-2} \zeta^s} (\zeta^{s_1}
E_{12} \otimes E_{21} + \zeta^{s_0} E_{21} \otimes E_{12}) \biggr]. \label{Rzeta1}
\end{multline}
It is instructive, using the identity
\begin{equation}
\lambda_2(q \zeta) + \lambda_2(q^{-1} \zeta) = - \log(1 - \zeta), \label{l2l2}
\end{equation}
to rewrite the expression for $R(\zeta)$ as
\begin{multline}
R(\zeta) = q^{-1/2} \zeta^{s/2} \rme^{\lambda_2(q \zeta^s) + \lambda_2(q^{-3} \zeta^s)} \biggl[ (q
\,  \zeta^{-s/2} - q^{-1} \zeta^{s/2})(E_{11} \otimes E_{11} + E_{22} \otimes E_{22}) \\* +
(\zeta^{-s/2} - \zeta^{s/2}) (E_{11} \otimes E_{22} + E_{22} \otimes E_{11}) \\*+ \kappa_q
(\zeta^{-(s_0 - s_1)/2} E_{12} \otimes E_{21} + \zeta^{(s_0 - s_1)/2} E_{21} \otimes E_{12})
\biggr]. \label{Rzeta2}
\end{multline}
Note that we come to the most frequently used symmetric $R$-operator putting $s_0 = -1$ and $s_1 =
-1$ and omitting the factor before the square bracket, compare with relations (\ref{rm}),
(\ref{mm}) and (\ref{bw}).

\subsection{Monodromy operators}

Now we find an expression for the monodromy operator
\begin{equation*}
M_{\varphi, \psi}(\zeta) = M_{\varphi, \psi}(\zeta | 1) = (\varphi_\zeta \otimes \psi) (\calR)
\end{equation*}
choosing as $\varphi$ the Jimbo's homomorphism and as $\psi$ the representation $\varphi^1$. In
fact, we extend the notion of the monodromy operator allowing for using general homomorphisms
instead of representations. To simplify notation, we write instead of $M_{\varphi,
\varphi^1}(\zeta)$ just $M(\zeta)$.

Using relations (\ref{epd}), (\ref{Jimboa}) and (\ref{Jimbob}), we obtain
\[
\varphi_\zeta(e'_\delta) = \kappa_q^{-1} \, q^{-1} \left[ C - (q + q^{-1}) \, q^{- H} \right] \zeta^s.
\]
Now, using definition (\ref{eamd}), we come to the expression
\begin{equation}
\varphi_\zeta(e_{\alpha + k \delta}) = (-1)^k q^{-k H} E \, \zeta^{s_1 + k s}, \label{qgeakd}
\end{equation}
while definition (\ref{edmamd}) gives
\begin{equation}
\varphi_\zeta(e_{(\delta - \alpha) + k \delta}) = (-1)^k F \, q^{- k H} \zeta^{(s - s_1) + k s}. \label{qgedmakd}
\end{equation}
Relation (\ref{epkd}) together with (\ref{Jimbob}) and (\ref{qgeakd}) lead to the equation
\begin{multline*}
\varphi_\zeta({e'_{k \delta}}) = \kappa_q^{-2} \, (-1)^{k - 1} q^{- k} \biggl[ (q^k - q^{-k}) \, C
\, q^{-(k - 1)H} \\* - (q^{k - 1} - q^{- k + 1}) \, q^{- (k - 2)H} - (q^{k + 1} - q^{- k - 1}) \,
q^{- k H} \biggr] \zeta^{k s},
\end{multline*}
and we have
\begin{multline*}
\varphi_\zeta(1 + \kappa_q e'(x)) \\
= (1 + q^{-1} C \, \zeta^s x^{-1}  + q^{-2} \zeta^{2 s} x^{-2}) (1 + q^{-H} \zeta^s x^{-1})^{-1} (1
+ q^{-H-2} \zeta^s x^{-1})^{-1}.
\end{multline*}
Using relation (\ref{edx}) and the equation
\[
\log (1 + x) = \sum_{k = 1}^\infty (-1)^{k - 1} \frac{x^k}{k},
\]
we obtain
\begin{equation}
\varphi_\zeta(e_{k \delta}) =  \kappa_q^{-1} \, (-1)^{k - 1} q^{-k} \left[ C_k - (q^k + q^{-k}) \,
q^{- k H} \right] \frac{\zeta^{k s}}{k}. \label{qgekd}
\end{equation}
Here the elements $C_k \in U_q(\gothsl_2)$ are defined by the generating function
\begin{equation}
C(x) = \sum_{k = 1}^\infty (-1)^{k - 1} C_k \frac{x^{- k}}{k} = \log(1 + C x^{-1} + x^{-2}). \label{Cx}
\end{equation}
In particular, we have
\begin{equation*}
C_1 = C, \qquad C_2 = C^2 - 2, \qquad C_3 = C^3 - 3 C, \qquad C_4 = C^4 - 4 C^2 + 2.
\end{equation*}
Note that all $C_k$ belong to the center of $U_q(\gothsl_2)$.

Below we need expressions for $\widetilde \pi^\mu(C_k)$ and $\pi^m(C_k)$. To obtain them we apply
$\widetilde \pi^\mu$ to both sides of (\ref{Cx}). Taking into account the first relation of
(\ref{pimuC}), we see that
\begin{multline*}
\pi^{\mu}(C(x)) = \log (1 + (q^{\mu + 1} + q^{-(\mu + 1)}) x^{-1} + x^{-2}) \\*
= \log (1 + q^{\mu + 1} x^{-1}) + \log (1 + q^{- (\mu + 1)} x^{-1}) \\*
= \sum_{k = 1}^\infty (-1)^{k - 1} \bigl[ q^{k(\mu + 1)} + q^{-k(\mu + 1)} \bigr] \frac{x^{-k}}{k}.
\end{multline*}
Hence, we have
\begin{equation}
\widetilde \pi^{\mu}(C_k) = q^{k(\mu + 1)} + q^{- k(\mu + 1)}. \label{Clambdam}
\end{equation}
In the same way we come to the equation
\begin{equation}
\pi^m(C_k) = q^{k(m + 1)} + q^{- k(m + 1)}. \label{pimCk}
\end{equation}

To find the expressions for the images of $\calR_{\prec \delta}$ and $\calR_{\succ \delta}$, we
again use identities (\ref{eek}). Taking into account relation (\ref{qgeakd}) and the first
equation of (\ref{fapmd}), we obtain
\begin{equation*}
(\varphi_\zeta \otimes \varphi^1) (\calR_{\prec \delta}) = 1 \otimes 1 + \kappa_q E (1 - q^{- H -
1} \zeta^s)^{-1} \zeta^{s_1} \otimes E_{21}.
\end{equation*}
In a similar way, relation (\ref{qgedmakd}) and the second equation of (\ref{fapmd}) give
\begin{equation*}
(\varphi_\zeta \otimes \varphi^1) (\calR_{\succ \delta}) = 1 \otimes 1 + \kappa_q (1 - q^{- H - 1}
\zeta^s)^{-1} F \, \zeta^{s - s_1} \otimes E_{12}.
\end{equation*}
Using relations (\ref{fmd}) and (\ref{qgekd}), we come to the equation
\begin{multline*}
(\varphi_\zeta \otimes \varphi^1) \biggl( \kappa_q \sum_{k = 1}^\infty \frac{k}{[2k]_q} e_{k
\delta} \otimes f_{k \delta} \biggr) \\* = \left( \Lambda(q^{-1} \zeta^s) + \log(1 - q^{- H - 1}
\zeta^s) \right) \otimes E_{11} \\- \left( \Lambda(q \, \zeta^s) + \log(1 - q^{- H + 1} \zeta^s)
\right) \otimes E_{22},
\end{multline*}
where
\[
\Lambda(\zeta) = \sum_{k = 1}^\infty \frac{1}{q^k + q^{-k}} \, C_k \frac{\zeta^k}{k}.
\]
It is easy to see that
\begin{equation}
\Lambda(q \, \zeta) + \Lambda(q^{-1} \zeta) = - \log(1 - C \zeta + \zeta^2). \label{Lambdaqzeta}
\end{equation}
Hence, we have
\begin{multline*}
(\varphi_\zeta \otimes \varphi^1) (\calR_{\sim \delta})
= (\rme^{\Lambda(q^{-1} \zeta^s)} \otimes 1) \bigl( (1 - q^{- H - 1} \zeta^s) \otimes E_{11} \\[.5em]
+ (1 - C \zeta^s + \zeta^{2s})(1 - q^{- H + 1} \zeta^s)^{-1} \otimes E_{22} \bigr).
\end{multline*}
Finally, for the action of $\calK$ we have
\begin{multline*}
(\varphi_\zeta \otimes \varphi^1) (\calK) \eqWithRef{vppk}
\sum_i \varphi_\zeta(q^{\mu_i(h_\alpha) / 2}) \otimes E_{ii}
\\ \eqWithRef{muh} \varphi_\zeta(q^{h_\alpha / 2}) \otimes E_{11} + \varphi_\zeta(q^{- h_\alpha / 2})
\otimes E_{22} \eqWithRef{Jimboa} q^{H/2} \otimes E_{11} + q^{-H/2} \otimes E_{22}.
\end{multline*}
Collecting all necessary factors, we come to the expression
\begin{multline}
M(\zeta) = (\rme^{\Lambda(q^{-1} \zeta^s)} \otimes 1)
\bigl[ (q^{H/2} - q^{-1} q^{-H/2} \, \zeta^s) \otimes E_{11} \\*[.5em]
+ \kappa_q F \, q^{-H/2} \, \zeta^{s_0} \otimes E_{12} + \kappa_q E \, q^{H/2} \, \zeta^{s_1}
\otimes E_{21} \\*[.5em]
+ (q^{-H/2} - q^{-1} q^{H/2} \, \zeta^s) \otimes E_{22} \bigr]. \label{Mzeta}
\end{multline}
The corresponding matrix $\mathbb M(\zeta)$ has the form
\begin{equation}
\mathbb M(\zeta) = \rme^{\Lambda(q^{-1} \zeta^s)} \left( \begin{array}{cc}
q^{H/2} - q^{-1} q^{-H/2} \, \zeta^s & \kappa_q F \, q^{-H/2} \, \zeta^{s_0} \\[.5em]
\kappa_q E \, q^{H/2} \, \zeta^{s_1} & q^{-H/2} - q^{-1} q^{H/2} \, \zeta^s
\end{array} \right). \label{bbMzeta}
\end{equation}

For any non-negative $m$ we can define the monodromy operator
\begin{equation*}
M_m(\zeta) = (\varphi^m_\zeta \otimes \varphi^1) (\calR) = (\pi^m \otimes \id) (M(\zeta)),
\end{equation*}
where the $(m + 1)$-dimensional representation $\pi^m$ of the quantum group $\uqsl$ is defined in
section \ref{s:qgA1}. Note that here we should have
\begin{equation}
M_1(\zeta) = (\pi^1 \otimes \id) (M(\zeta)) = R(\zeta), \label{RzetaMzeta}
\end{equation}
where the $R$-operator $R(\zeta)$ is given by (\ref{Rzeta1}) or (\ref{Rzeta2}). To show this we
first of all need the expression for $\pi^1(\Lambda( q^{-1}\zeta^s))$. In fact, for an arbitrary
non-negative integer $m$ equation (\ref{pimCk}) gives
\begin{equation}
\Lambda^m(\zeta) = \pi^m(\Lambda(\zeta)) = \sum_{k = 1}^\infty \frac{q^{k(m + 1)} + q^{-k(m +
1)}}{q^k + q^{-k}} \, \frac{\zeta^k}{k}. \label{Lambdalambda}
\end{equation}
Hence, we have
\begin{equation*}
\Lambda^m(\zeta) = \lambda_2(q^{m + 1} \zeta) + \lambda_2(q^{- m - 1} \zeta),
\end{equation*}
and, in particular,
\begin{equation*}
\Lambda^1(q^{-1} \zeta) = \lambda_2(q \zeta) + \lambda_2(q^{-3} \zeta).
\end{equation*}
Using (\ref{pi1}), we obtain
\begin{gather*}
\pi^1(q^{\pm H/2}) = q^{\pm 1/2} E_{11} + q^{\mp 1/2} E_{22}, \\
\pi^1(F \, q^{-H/2}) = q^{-1/2} E_{21}, \qquad \pi^1(E \, q^{H/2}) = q^{-1/2} E_{12}.
\end{gather*}
Now, applying $\pi^1$ to both sides of (\ref{Mzeta}) and comparing the obtained result with
(\ref{Rzeta2}), we see that equation (\ref{RzetaMzeta}) is valid.

More generally, for any $\mu \in \bbC$ we denote
\begin{equation*}
\widetilde M_\mu(\zeta) = (\widetilde \varphi^\mu_\zeta \otimes \varphi^1) (\calR) = (\widetilde
\pi^\mu \otimes \id) (M(\zeta)),
\end{equation*}
where the infinite-dimensional representation $\widetilde \pi^\mu$ of the quantum group $\uqsl$ is
defined in section \ref{s:qgA1}.

\subsection{Transfer operators}

To construct transfer operators we should first choose a twist element $t$. We assume that
\begin{equation*}
t = q^{\phi h_\alpha},
\end{equation*}
where $\phi$ is a complex number. It is clear that $t$ is a group-like element as is required.
Denoting $T_m(\zeta) = T_{\varphi^m, \, \varphi^1}(\zeta)$, and having in mind that
\begin{equation*}
\varphi_\zeta(t) = q^{\phi H},
\end{equation*}
we obtain
\begin{equation*}
T_m(\zeta) = (\tr_{V^m} \otimes \id) (M_m(\zeta)(\varphi^m_\zeta(t) \otimes 1)) = (\tr_m \otimes
\id) (M(\zeta) (q^{\phi H} \otimes 1)),
\end{equation*}
where
\begin{equation*}
\tr_m = \tr_{V^m} \circ \pi^m.
\end{equation*}
The mapping $\tr_m$ is a trace on the algebra $\uqsl$. It is clear that in terms of the
corresponding matrices $\mathbb T_m(\zeta) = \mathbb T_{\varphi^m, \, \varphi^1}(\zeta)$ and
$\mathbb M(\zeta)$ we have
\begin{equation*}
\mathbb T_m(\zeta) = \tr_m (\mathbb M(\zeta) \, q^{\phi H}),
\end{equation*}
where $\tr_m$ is applied to the matrix entries. For the higher transfer matrices one obtains
\begin{equation*}
\mathbb T_m(\zeta | \eta_1, \ldots, \eta_n) = \tr_m ((\mathbb M(\zeta \eta_1^{-1}) \boxtimes \ldots
\boxtimes \mathbb M(\zeta \eta_n^{-1})) \, q^{\phi H}),
\end{equation*}
see relation (\ref{bbT}). To find an explicit form of the transfer matrices $\mathbb T_m(\zeta |
\eta_1, \ldots, \eta_n)$ we have to know the traces for the elements of some basis of $\uqsl$. An
easy calculation gives
\begin{gather}
\tr_m (E^{r + 1} C^s q^{\nu H}) = 0, \qquad \tr_m (F^{r + 1} C^s q^{\nu H}) = 0, \label{trm1} \\
\tr_m (C^s q^{\nu H}) = (q^{m + 1} + q^{- m - 1})^s q^{m \nu} \frac{1 - q^{- 2 \nu (m + 1)}}{1 -
q^{- 2 \nu}} = (q^{m + 1} + q^{- m - 1})^s [m + 1]_{q^\nu}, \label{trm2}
\end{gather}
where $\nu \in \bbC$ and $r, s \in \bbZ_{\ge 0}$. For the simplest case $n = 1$ we determine that
\begin{equation*}
\mathbb T_m(\zeta) = \rme^{\Lambda^m (q^{-1} \zeta^s)} \left( \begin{array}{cc}
[m + 1]_{q^{1/2 + \phi}} - [m + 1]_{q^{- 1/2 + \phi}} \, q^{-1} \zeta^s\hspace{-6em} & 0 \\[.5em]
0 \hspace{-6em} & [m + 1]_{q^{- 1/2 + \phi}} - [m + 1]_{q^{1/2 + \phi}} \, q^{-1} \zeta^s
\end{array} \right).
\end{equation*}
This transfer matrix is finite in the limit where the twist parameter $\phi$ tends to zero. This is
evidently true also for the higher transfer matrices $\mathbb T_m(\zeta | \eta_1, \ldots, \eta_n)$.

The case of the infinite-dimensional representations $\widetilde \pi^\mu$ is more subtle. Here we denote
\begin{equation*}
\widetilde \tr_\mu = \tr_{\widetilde V^\mu} \circ \widetilde \pi^\mu,
\end{equation*}
and for $|q^{-2 \nu}| < 1$ obtain
\begin{gather*}
\widetilde \tr_\mu (E^{r + 1} C^s  q^{\nu H}) = 0, \qquad \widetilde \tr_\mu (F^{r + 1} C^s q^{\nu H}) = 0, \\
\widetilde \tr_\mu (C^s q^{\nu H}) = (q^{\mu + 1} + q^{- \mu - 1})^s \frac{q^{\mu \nu}}{1 - q^{- 2 \nu}}.
\end{gather*}
For $|q^{-2 \nu}| > 1$ the trace $\widetilde \tr_\mu$ is defined with the help of analytic continuation.

Again, for the simplest case we have
\begin{multline*}
\widetilde{\mathbb T}_\mu(\zeta) = \widetilde \tr_\mu (\mathbb M(\zeta) \, q^{\phi H} ) \\*
=\rme^{\Lambda^{\mu}(q^{-1} \zeta^s)} \left( \begin{array}{cc}
\displaystyle \frac{q^{(1/2 + \phi) \mu}}{1 - q^{-1 - 2 \phi}}
- \frac{q^{-(1/2 - \phi) \mu}}{1 - q^{1 - 2 \phi}} \, q^{-1} \zeta^s \hspace{-1em} & 0 \\[.5em]
0 & \hspace{-1em} \displaystyle \frac{q^{-(1/2 - \phi) \mu}}{1 - q^{1 - 2 \phi}}
- \frac{q^{(1/2 + \phi) \mu}}{1 - q^{-1 - 2 \phi}} \, q^{-1} \zeta^s
\end{array} \right),
\end{multline*}
where
\begin{equation*}
\Lambda^\mu(\zeta) = \widetilde \pi^\mu(\Lambda(\zeta)) = \sum_{k = 1}^\infty \frac{q^{k(\mu + 1)}
+ q^{-k(\mu + 1)}}{q^k + q^{-k}} \, \frac{\zeta^k}{k}.
\end{equation*}
The above transfer matrix is finite in the limit where the twist parameter tends to zero. This is
not the case for all the higher transfer matrices
\begin{equation*}
\widetilde{\mathbb T}_\mu(\zeta | \eta_1, \ldots, \eta_n) = \widetilde \tr_\mu ((\mathbb M(\zeta
\eta_1^{-1}) \boxtimes \ldots \boxtimes \mathbb M(\zeta \eta_n^{-1})) \, q^{\phi H}).
\end{equation*}
It is clear that the nonzero contributions to $\widetilde{\mathbb T}_\mu(\zeta | \eta_1, \ldots,
\eta_n)$ are given by the trace of the elements
\begin{equation*}
C^r q^{(n - 2 r) H / 2}, \quad  C^r q^{(n - 2 r - 2) H / 2}, \quad \ldots, \quad C^r q^{- (n - 2 r) H / 2},
\end{equation*}
where $r = 0, 1, \ldots, [n / 2]$. For an even $n$ we have to take the trace of $C^r$, $r = 0,
\ldots, n / 2$, and the result is evidently singular in the zero-twist limit. For an odd $n$ there
are no singularities.

It is worth to note that
\begin{equation*}
\mathbb T_m (\zeta | \eta_1, \ldots, \eta_n) = \widetilde{\mathbb T}_m(\zeta | \eta_1, \ldots,
\eta_n) - \widetilde{\mathbb T}_{- m - 2}(\zeta | \eta_1, \ldots, \eta_n)
\end{equation*}
and the zero-twist limit is nonsingular for the right hand side for all $n$. This equation suggests
that we should define
\begin{equation*}
\mathbb T_\mu (\zeta | \eta_1, \ldots, \eta_n) = \widetilde{\mathbb T}_\mu(\zeta | \eta_1, \ldots,
\eta_n) - \widetilde{\mathbb T}_{- \mu - 2}(\zeta | \eta_1, \ldots, \eta_n)
\end{equation*}
for an arbitrary $\mu \in \bbC$. It is not difficult to prove that $\mathbb T_\mu (\zeta | \eta_1,
\ldots, \eta_n)$ is finite in the zero-twist limit for any $n$. One can say that $\mathbb T_\mu
(\zeta | \eta_1, \ldots, \eta_n)$ is the transfer matrix defined by the trace
\begin{equation}
\tr_\mu = \widetilde \tr_\mu - \widetilde \tr_{-\mu - 2} \label{trmu}
\end{equation}
on the algebra $\uqsl$. It is instructive to compare the definition of $\tr_\mu$ with the
definition of trace given in the papers \cite{BooJimMiwSmiTak06b, BooJimMiwSmiTak06c}, see also
\cite{Pro00, BooJimMiwSmiTak06a, BazLukMenSta10} for the limiting case $q = 1$.

\subsection{\texorpdfstring{$L$-operators}{L-operators}} \label{s:elo}

Remind that with the definition of a quantum group used by us, the quasitriangularity is understood
in some restricted sense, see section \ref{s:qga11}. We can only define the action of the universal
$R$-matrix in the tensor product of weight representations. It is easy to see that to determine the
action of the universal $R$-matrix (\ref{RRRK}) on the tensor product of two representation spaces,
it suffices to use for the first factor representations of the subalgebra $\uqbp$ and for the
second one representations of the subalgebra $\uqbm$. Here the Borel subalgebra $\uqbp$ is
generated by $e_0$, $e_1$ and $q^x$, $x \in \widetilde{\gothh}$, and the Borel subalgebra $\uqbm$
is generated by $f_0$, $f_1$ and $q^x$, $x \in \widetilde{\gothh}$.

It is clear that any representation of the algebra $\uqlsl$ generates representations of the
subalgebras $\uqbp$ and $\uqbm$. However, this does not give new objects. There are other methods
to construct representations of $\uqbp$ and $\uqbm$ from representations of $\uqlsl$. We restrict
ourselves by the case of the Borel subalgebra $\uqbp$.

First note that if $\varphi$ is a representation of $\uqlsl$ and $\xi \in \widetilde \gothh^*$,
then the mapping $\varphi[\xi]$ defined by the equations
\begin{equation*}
\varphi[\xi](e_i) = \varphi(e_i), \qquad \varphi[\xi](q^x) = q^{\xi(x)} \varphi(q^x)
\end{equation*}
is a representation of $\uqbp$ called a {\em shifted representation\/}. It follows from
(\ref{qh0h1}) that we have to assume that
\begin{equation*}
\xi(h_{\delta - \alpha}) = - \xi(h_\alpha).
\end{equation*}
Taking into account relation (\ref{kuv}), we see that for any weight representations $\varphi$ and
$\psi$ one has
\begin{equation*}
(\varphi[\xi] \otimes \psi)(\calK) = (\varphi\otimes \psi)(\calK)(1 \otimes \psi(q^{\xi(h_\alpha)
h_\alpha / 2})).
\end{equation*}
Therefore, for the universal $R$-matrix we can write
\begin{equation*}
(\varphi[\xi] \otimes \psi)(\calR) = (\varphi\otimes \psi)(\calR)(1 \otimes \psi(q^{\xi(h_\alpha)
h_\alpha / 2})).
\end{equation*}
In fact, we can even write
\begin{equation*}
(\varphi[\xi] \otimes \id)(\calR) = (\varphi\otimes \id)(\calR)(1 \otimes q^{\xi(h_\alpha) h_\alpha
/ 2}),
\end{equation*}
having in mind that this equation is true only for weight representations. Under the same
assumption, we have for the universal monodromy operator the equation
\begin{equation*}
\calM_{\varphi[\xi]}(\zeta) = \calM_\varphi(\zeta) (1 \otimes q^{\xi(h_\alpha) h_\alpha / 2}),
\end{equation*}
and for the universal transfer operator the equation
\begin{equation}
\calT_{\varphi[\xi]}(\zeta) = \calT_\varphi(\zeta) \, q^{\xi(h_\alpha) (h_\alpha + 2 \phi) / 2}.
\label{tsh}
\end{equation}
Thus, the use of shifted representations does not give anything really new. Nevertheless, we meet
universal transfer matrices corresponding to shifted representations when proving the functional
relations.

Now let us describe how to obtain the representation necessary for the construction of
$L$-operators. Starting with the representation $\widetilde \varphi^\mu_\zeta$ of $\uqlsl$ defined
by equations (\ref{Vh})--(\ref{Vf}), we obtain the shifted representation $\widetilde
\varphi^\mu_\zeta[\xi]$ of $\uqbp$ defined by the relations
\begin{align}
&q^{\nu h_\alpha} v_n = q^{\nu(\mu - 2n + \xi(h_\alpha))}
v_n, && q^{\nu h_{\delta - \alpha}} v_n = q^{- \nu(\mu - 2n - \xi(h_{\delta - \alpha}))} v_n, \label{shqh} \\
&e_\alpha v_n = \zeta^{s_1} [n]_q [\mu - n + 1]_q v_{n - 1}, && e_{\delta - \alpha} v_n
= \zeta^{s_0} v_{n + 1}, \label{she}
\end{align}
We denote the corresponding $\uqbp$-module by $\widetilde V^\mu_\zeta[\xi]$. Assume that
\begin{equation*}
\xi(h_\alpha) = - \xi(h_{\delta - \alpha}) = - \mu.
\end{equation*}
Relations (\ref{shqh}) take the form
\begin{equation}
q^{\nu h_\alpha} v_n = q^{- 2 \nu n}  v_n, \qquad
q^{\nu h_{\delta - \alpha}} v_n = q^{2 \nu n} v_n. \label{ash}
\end{equation}
Note that we can multiply the operators corresponding to the generators
$e_0$ and $e_1$ by arbitrary nonzero complex numbers. This again gives a representation of $\uqbp$.
Represent the first relation of (\ref{she}) as
\begin{equation*}
q^{-\mu - 1} e_\alpha \, v_n = \zeta^{s_1} \kappa_q^{-1} (q^{- n} - q^{- 2 \mu + n - 2}) [n]_q v_{n - 1}.
\end{equation*}
Now we rescale the operator corresponding to $e_\alpha$ as $e_\alpha \to q^{\mu + 1} e_\alpha$
and consider the limit $\mu \to \infty$ along the real axis. This gives instead of (\ref{she}) the relations
\begin{equation}
e_\alpha \, v_n = \zeta^{s_1} \kappa_q^{-1} q^{- n} [n]_q v_{n - 1}, \qquad
e_{\delta - \alpha} \, v_n = \zeta^{s_0} v_{n + 1}. \label{ase}
\end{equation}
Relations (\ref{ash}) and (\ref{ase}) define a representation of $\uqbp$. Note that this
representation cannot be extended to a representation of the full quantum group $\uqlsl$.
It is useful to give an interpretation of (\ref{ash}) and (\ref{ase}) in terms of $q$-oscillators.
Let us remind the necessary definitions, see, for example the book \cite{KliSch97}.

Let $\hbar$ be a complex number such that $q = \exp \hbar \ne 0, \pm 1$. The $q$-oscillator algebra
$\Osc_q$ is a unital associative $\bbC$-algebra with generators $b^\dagger$, $b$, $q^{\nu N}$, $\nu
\in \bbC$, and relations
\begin{gather*}
q^0 = 1, \qquad q^{\nu_1 N} q^{\nu_2 N} = q^{(\nu_1 + \nu_2)N}, \\
q^{\nu N} b^\dagger q^{-\nu N} = q^\nu b^\dagger, \qquad q^{\nu N} b q^{-\nu N} = q^{-\nu} b, \\
b^\dagger b = \kappa_q^{-1} (q^N - q^{-N}), \qquad b b^\dagger = \kappa_q^{-1} (q^{N + 1} - q^{- N - 1}).
\end{gather*}
There are two interesting for us representations of $\Osc_q$. First, let $W^+$ be a free vector
space generated by the set $\{ v_0, v_1, \ldots \}$. One can show that the relations
\begin{gather}
q^{\nu N} v_n = q^{\nu n} v_n, \label{Nvk} \\*
b^\dagger v_n = v_{n + 1}, \qquad b \, v_n = [n]_q v_{n - 1}, \label{bdvk}
\end{gather}
where we assume that $v_{-1} = 0$, endow $W^+$ with the structure of an $\Osc_q$-module. We denote
the corresponding representation of the algebra $\Osc_q$ by $\chi^+$. Further, let $W^-$ be a free
vector space generated by the set $\{ u_0, u_1, \ldots \}$. The relations
\begin{gather}
q^{\nu N} u_n = q^{- \nu (n + 1)} u_n, \label{Nwk} \\
b \, u_n = u_{n + 1}, \qquad b^\dagger u_n = - [n]_q u_{n - 1}, \label{bwk}
\end{gather}
where we assume that $u_{-1} = 0$, endow $W^-$ with the structure of an $\Osc_q$-module. We denote
the corresponding representation of $\Osc_q$ by $\chi^-$.

Return again to relations (\ref{ash}) and (\ref{ase}). Assume that the operators $N$, $b^\dagger$
and $b$ act in the representation space in accordance with (\ref{Nvk}) and (\ref{bdvk}). This
allows one to write (\ref{ash}) and (\ref{ase}) as
\begin{align*}
&q^{\nu h_\alpha} v_n = q^{-2 \nu N} v_n, && q^{\nu h_{\delta - \alpha}} v_n = q^{2 \nu N} v_n, \\
&e_\alpha \, v_n = \zeta^{s_1} \kappa_q^{-1} b \, q^{- N} \, v_n, && e_{\delta - \alpha} \, v_n =
\zeta^{s_0} b^\dagger \, v_n.
\end{align*}
These equations suggest to us a homomorphism $\rho: \uqbp \to \Osc_q$ defined by
\begin{align}
\rho(q^{\nu h_\alpha}) &= q^{- 2 \nu N}, & \rho(q^{\nu h_{\delta - \alpha}}) &= q^{2 \nu N}, \label{rhoh} \\
\rho(e_\alpha) &= \kappa_q^{-1} b \, q^{- N}, & \rho(e_{\delta - \alpha}) &= b^\dagger, \label{rhoe}
\end{align}
and the homomorphisms $\rho_\zeta: \uqbp \to \Osc_q$, $\zeta \in \bbC^\times$, as $\rho_\zeta =
\rho \circ \Phi_\zeta$. We can now define the representations
\begin{equation}
\rho^\pm = \chi^\pm \circ \rho, \qquad \rho^\pm_\zeta = \chi^\pm \circ \rho_\zeta \label{rhop}
\end{equation}
of the Borel subalgebra $\uqbp$. For the representations $\rho^\pm_\zeta$ we have explicitly
\begin{align}
&\rho^\pm_\zeta(q^{\nu h_\alpha}) = \chi^\pm(q^{-2 \nu N}), &&
\rho^\pm_\zeta(q^{\nu h_{\delta - \alpha}}) = \chi^\pm(q^{2 \nu N}), \label{rhoph} \\
&\rho^\pm_\zeta(e_\alpha) = \zeta^{s_1} \kappa_q^{-1} \chi^\pm(b q^{- N}), &&
\rho^\pm_\zeta(e_{\delta - \alpha}) = \zeta^{s_0} \chi^\pm(b^\dagger). \label{rhope}
\end{align}
We denote the $\uqbp$-modules corresponding to the representations $\rho^\pm_\zeta$ by $W^\pm_\zeta$.

Let us construct the $L$-operator
\begin{equation*}
L_{\rho, \psi}(\zeta) = L_{\rho, \psi}(\zeta | 1) = (\rho_\zeta \otimes \psi)(\calR)
\end{equation*}
choosing as $\rho$ the homomorphism defined by (\ref{rhoh}) and (\ref{rhoe}), and as $\psi$ the
representation $\varphi^1$. As for the case of monodromy operators, we extend the notion of
$L$-operators allowing for using general homomorphisms instead of representations. To simplify
notation we write instead of $L_{\rho, \psi}(\zeta)$ just $L(\zeta)$.

Having in mind (\ref{vppk}), (\ref{vha}) and (\ref{rhoh}), we observe that
\begin{equation}
(\rho_{\zeta} \otimes \varphi^1)(\calK) = q^{- N} \otimes E_{11} + q^N \otimes E_{22}.
\label{ok1}
\end{equation}
Further, one can easily determine that definition (\ref{epd}) together with (\ref{rhoe}) gives
\begin{equation}
\rho_\zeta(e'_\delta) = \kappa_q^{-1} q^{-1} \zeta^s,
\label{cepd}
\end{equation}
and, using (\ref{eamd}) and (\ref{edmamd}), we immediately obtain
\begin{equation}
\rho_\zeta(e_{\alpha + k \delta}) = 0, \qquad
\rho_\zeta(e_{(\delta - \alpha) + k \delta}) = 0, \qquad k \ge 1.
\label{ceamd1}
\end{equation}
Taking into account (\ref{pi1f}) and (\ref{eek}), we come to
\begin{gather}
(\rho_{\zeta} \otimes \varphi^1)(\calR_{\prec \delta})
= 1 \otimes 1 + b \, q^{-N} \zeta^{s_1} \otimes E_{21}, \label{orld1a} \\
(\rho_{\zeta} \otimes \varphi^1)(\calR_{\succ \delta})
= 1 \otimes 1 + \kappa_q \, b^\dagger \, \zeta^{s - s_1} \otimes E_{12}.
\label{orld1b}
\end{gather}
Definition (\ref{epkd}) and equations (\ref{ceamd1}) give
\[
\rho_\zeta(e'_{k \delta}) = 0, \qquad k > 1,
\]
and one easily finds that
\[
\rho_\zeta(e_{k \delta}) = (-1)^{k - 1} \kappa_q^{-1} q^{-k} \, \frac{\zeta^{k s}}{k}.
\]
Now, using relations (\ref{Rsd}) and (\ref{fmd}), we obtain
\begin{equation}
(\rho_{\zeta} \otimes \varphi^1)(\calR_{\sim \delta})
= \rme^{\lambda_2(q^{-1} \zeta^s)} [1 \otimes E_{11} + (1 - \zeta^s) \otimes E_{22}],
\label{orpd1}
\end{equation}
where the function $\lambda_2(\zeta)$ is defined by (\ref{fa1}).

Multiplying expressions (\ref{ok1}), (\ref{orld1a}), (\ref{orld1b}) and (\ref{orpd1}) in the
prescribed order, we come to the following $L$-operator:
\begin{multline*}
L(\zeta) = \rme^{\lambda_2(q^{-1} \zeta^s)} [q^{- N} \otimes E_{11} \\
+ \kappa_q \,  b^\dagger \, q^N \zeta^{s - s_1} \otimes E_{12}
+ b \, q^{- 2 N} \zeta^{s_1} \otimes E_{21} + (q^N - q^{-2} q^{- N} \zeta^s) \otimes E_{22}].
\end{multline*}
In the matrix form it looks as
\begin{equation*}
\mathbb L(\zeta) = \rme^{\lambda_2(q^{-1} \zeta^s)} \left( \begin{array}{cc}
q^{- N} & \kappa_q \,  b^\dagger \, q^N \zeta^{s - s_1} \\[.5em]
b \, q^{- 2 N} \zeta^{s_1} & q^N - q^{-2} q^{- N} \zeta^s
\end{array} \right).
\end{equation*}

It is evident that the relations
\begin{gather*}
\sigma(h_0) = h_1, \qquad \sigma(h_1) = h_0, \\
\sigma(e_0) = e_1, \qquad \sigma(e_1) = e_0, \qquad \sigma(f_0) = f_1, \qquad \sigma(f_1) = f_0
\end{gather*}
define an automorphism of $\uqlsl$ and, via the restriction, an automorphism of $\uqbp$. Therefore,
the mappings
\begin{equation*}
\overline \rho = \rho \circ \sigma, \qquad \overline \rho_\zeta = \overline \rho \circ \Phi_\zeta
\end{equation*}
are homomorphisms from $\uqbp$ to $\Osc_q$, and the mappings
\begin{equation}
\overline \rho^\pm = \chi^\pm \circ \overline \rho, \qquad
\overline \rho^\pm_\zeta = \chi^\pm \circ \overline \rho_\zeta \label{brhop}
\end{equation}
are representations of $\uqbp$. We denote the $\uqbp$-modules corresponding to the representations
$\overline \rho^\pm_\zeta$ by $\overline W^\pm_\zeta$.

Let us find the expression for the $L$-operator
\begin{equation*}
\overline L(\zeta) = L_{\overline \rho, \varphi^1}(\zeta).
\end{equation*}
Calculations give
\begin{gather*}
\overline \rho_\zeta(e_{\alpha + k \delta})
= (-1)^k q^{-k} q^{-2k N} b^\dagger \zeta^{s_1 + k s}, \\
\overline \rho_\zeta(e_{(\delta - \alpha) + k \delta})
= \kappa_q^{-1} (-1)^k q^{-k} b q^{-(2k + 1)N}\zeta^{s_0 + ks}, \\
\overline \rho_\zeta(e_{k \delta})
= \kappa_q^{-1} (-1)^k q^{-k} [(1 + q^{-2k}) q^{-2k N} - 1] \frac{\zeta^{ks}}{k}.
\end{gather*}
Using these equations, we obtain
\begin{gather*}
(\overline \rho_\zeta \otimes \varphi^1)(\calR_{\prec \delta})
= 1 \otimes 1 + \kappa_q b^\dagger (1 - q^{-2} q^{- 2N} \zeta^s)^{-1} \zeta^{s_1} \otimes E_{21}, \\*
(\overline \rho_\zeta \otimes \varphi^1)(\calR_{\sim \delta})
= \rme^{\lambda_2(q^{-1} \zeta^s)}[(1 - q^{-2} q^{- 2N} \zeta^s)
\otimes E_{11} + (1 - \zeta^s)(1 - q^{- 2N} \zeta^s)^{-1} \otimes E_{22}], \\*
(\overline \rho_\zeta \otimes \varphi^1)(\calR_{\succ \delta})
= 1 \otimes 1 + b q^{- N} (1 - q^{- 2N} \zeta^s)^{-1} \zeta^{s - s_1} \otimes E_{12}, \\
(\overline \rho_{\zeta} \otimes \varphi^1)(\calK) = q^N \otimes E_{11} + q^{- N} \otimes E_{22}
\end{gather*}
Multiplying these expressions in the order prescribed by (\ref{RRRK}), we determine that
\begin{multline*}
\overline L(\zeta) = \rme^{\lambda_2(q^{-1} \zeta^s)} [(q^N - q^{-2} q^{- N} \zeta^s) \otimes E_{11} \\
+ b \, q^{-2N} \zeta^{s - s_1} \otimes E_{12} + \kappa_q \, b^\dagger \, q^N \zeta^{s_1} \otimes
E_{21} + q^{- N} \otimes E_{22}],
\end{multline*}
or in the matrix form
\begin{equation*}
\overline{\mathbb L}(\zeta) = \rme^{\lambda_2(q^{-1} \zeta^s)} \left( \begin{array}{cc}
q^N - q^{-2} q^{- N} \zeta^s & b \, q^{-2N} \zeta^{s - s_1} \\[.5em]
\kappa_q \, b^\dagger \, q^N \zeta^{s_1} & q^{- N}
\end{array} \right).
\end{equation*}

\subsection{\texorpdfstring{$Q$-operators}{Q-operators}}

We again use the twist element
\begin{equation*}
t = q^{\phi h_\alpha},
\end{equation*}
where $\phi$ is a complex number. Using the representations $\chi^+$ and $\chi^-$, we can define two traces
\begin{equation*}
\tr_+ = \tr_{W^+} \circ \chi^+, \qquad \tr_- = \tr_{W^-} \circ \chi^-
\end{equation*}
on the algebra $\Osc_q$. In fact, these traces differ only by an overall sign. We introduce two $Q$-operators,
\begin{gather*}
Q'(\zeta) = Q_{\rho^+, \varphi^1}(\zeta), \qquad \overline Q'(\zeta) = Q_{\overline \rho^-, \varphi^1}(\zeta).
\end{gather*}
We use the prime here because we slightly redefine the $Q$-operators below. Taking into account that
\begin{equation*}
\rho(t) = q^{-2 \phi N}, \qquad \overline \rho(t) = q^{2 \phi N},
\end{equation*}
we obtain
\begin{equation*}
Q'(\zeta) = (\tr_+ \otimes \id)(L(\zeta)(q^{-2 \phi N} \otimes 1)), \qquad \overline Q'(\zeta) =
(\tr_- \otimes \id)(\overline L(\zeta)(q^{2 \phi N} \otimes 1)).
\end{equation*}
In terms of the corresponding matrices we have
\begin{equation*}
\mathbb Q'(\zeta) = \tr_+ (\mathbb L(\zeta) \, q^{- 2 \phi N}), \qquad \overline{\mathbb Q}'(\zeta)
= \tr_- (\overline{\mathbb L}(\zeta) \, q^{2 \phi N}),
\end{equation*}
where $\tr_+$ is applied to the matrix entries, and, in general,
\begin{gather*}
\mathbb Q'(\zeta | \eta_1, \ldots, \eta_n)
= \tr_+((\mathbb L(\zeta \eta_1^{-1}) \boxtimes \ldots
\boxtimes \mathbb L(\zeta \eta_n^{-1})) q^{- 2 \phi N}), \\[.5em]
\overline{\mathbb Q}'(\zeta | \eta_1, \ldots, \eta_n)
= \tr_-((\overline{\mathbb L}(\zeta \eta_1^{-1}) \boxtimes \ldots
\boxtimes \overline{\mathbb L}(\zeta \eta_n^{-1})) q^{2 \phi N}).
\end{gather*}

Using (\ref{Nvk}) and (\ref{bdvk}), we see that for $|q| < 1$ one has
\begin{gather*}
\tr_+ (q^{\nu N}) = \frac{1}{1 - q^\nu}, \\
\tr_+ ((b^\dagger)^{r + 1} q^{\nu N}) = 0, \qquad \tr_+ (b^{r + 1} q^{\nu N}) = 0
\end{gather*}
for any $\nu \in \bbC$ and $r \in \bbZ_{\ge 0}$. For $|q| > 1$ we define the trace $\tr_+$ by
analytic continuation. Using the above relations, for $n = 1$ we obtain
\begin{equation*}
\mathbb Q'(\zeta) = \rme^{\lambda_2(q^{-1} \zeta^s)} \left( \begin{array}{cc}
\displaystyle \frac{1}{1 - q^{- 1 - 2 \phi}} & 0 \\[.5em]
0 & \displaystyle \frac{1}{1 - q^{1 - 2 \phi}} - \frac{1}{1 - q^{- 1 - 2 \phi}} \, q^{-2} \zeta^s
\end{array} \right)
\end{equation*}
and
\begin{equation*}
\overline{\mathbb Q}'(\zeta) = - \rme^{\lambda_2(q^{-1} \zeta^s)} \left( \begin{array}{cc}
\displaystyle \frac{1}{1 - q^{1 + 2 \phi}} - \frac{1}{1 - q^{- 1 + 2 \phi}} q^{-2} \zeta^s & 0 \\[.5em]
0 & \displaystyle \frac{1}{1 - q^{- 1 + 2 \phi}}
\end{array} \right).
\end{equation*}
As well as for the higher transfer matrices $\widetilde{\mathbb T}_\mu(\zeta | \eta_1, \ldots,
\eta_n)$, one sees that the matrices $\mathbb Q'(\zeta | \eta_1, \ldots, \eta_n)$ and  $\overline{\mathbb
Q}'(\zeta | \eta_1, \ldots, \eta_n)$ are finite in the zero-twist limit for an odd $n$, and
singular for an even~$n$.

\section{\texorpdfstring{Functional relations for $\uqlsl$}{Functional relations for UqLsl2}} \label{s:fr}

In this section we derive certain relations satisfied by the universal transfer operators
\begin{equation*}
\widetilde \calT_\mu(\zeta) = \calT_{\widetilde \varphi^\mu}(\zeta), \qquad \calT_m(\zeta) =
\calT_{\varphi^m}(\zeta),
\end{equation*}
and the universal $Q$-operators
\begin{equation*}
\calQ'(\zeta) = \calQ_{\rho^+}(\zeta), \qquad \overline \calQ'(\zeta) = \calQ_{\overline \rho^-}(\zeta),
\end{equation*}
where the representations $\widetilde \varphi^\mu$, $\varphi^m$ of $\uqlsl$ are defined in section
\ref{s:ero}, and the representations $\rho^+$, $\overline \rho^-$ of $\uqbp$ in section
\ref{s:elo}. These relations, known as functional relations, appear to be very useful for
investigation of the corresponding integrable systems.

There are relations which are due only to the fact that the universal transfer operators and
universal $Q$-operators are constructed from the universal $R$-matrices. Another set of relations
depends on the structure of the representations used for their construction. Here, to analyse the
products of the operators, we should analyse the tensor products of the corresponding
representations, see, for example, relations (\ref{Q1Q2}) and (\ref{T1Q2}).

\subsection{Tensor product of representations}

\subsubsection{\texorpdfstring{Tensor product of representations
$\rho^+_{\zeta_1}$ and $\rho^+_{\zeta_2}$}{Tensor product of representation r+z and r+z2}}

To analyse the product of the universal $Q$-operators $\calQ'(\zeta_1)$ and $\calQ'(\zeta_2)$ we
consider the tensor product of the representations $\rho^+_{\zeta_1}$ and $\rho^+_{\zeta_2}$. Here
the representation space is $W^+ \otimes W^+$, which is also the representation space of the
representation $\chi^+ \otimes \chi^+$ of the algebra $\Osc_q \otimes \Osc_q$.

It is not difficult to see that\footnote{Remind that $h_0 = h_{\delta - \alpha}$, $h_1 = h_\alpha$,
$e_0 = e_{\delta - \alpha}$ and $e_1 = e_\alpha$.}
\begin{multline*}
(\rho^+_{\zeta_1} \otimes_\Delta \rho^+_{\zeta_2})(q^{\nu h_0})
= (\rho^+_{\zeta_1} \otimes \rho^+_{\zeta_2})(\Delta(q^{\nu h_0}))
\eqWithRef{dqx} (\rho^+_{\zeta_1} \otimes \rho^+_{\zeta_2})(q^{\nu h_0} \otimes q^{\nu h_0}) \\[.5em]
\eqWithRef{rhoh} \chi^+(q^{2 \nu N}) \otimes \chi^+(q^{2 \nu N})
= (\chi^+ \otimes \chi^+) (q^{2 \nu N} \otimes q^{2 \nu N}).
\end{multline*}
Similarly, we obtain
\begin{equation*}
(\rho^+_{\zeta_1} \otimes_\Delta \rho^+_{\zeta_2})(q^{\nu h_1})
= (\chi^+ \otimes \chi^+) (q^{- 2 \nu N} \otimes q^{- 2 \nu N}).
\end{equation*}
Below we denote
\begin{gather*}
q^{\mu N_A + \nu N_B} = q^{\mu N} \otimes q^{\nu N}, \qquad \mu, \nu \in \bbC, \\[.3em]
b_A = b \otimes 1, \qquad b_B = 1 \otimes b, \qquad b_A^\dagger
= b^\dagger \otimes 1, \qquad b_B^\dagger = 1 \otimes b^\dagger.
\end{gather*}
Now, using module notation, we can write
\begin{equation*}
q^{\nu h_0} \, w = q^{2 \nu (N_A + N_B)} \, w, \qquad q^{\nu h_1} \, w = q^{ - 2 \nu (N_A + N_B)} \, w
\end{equation*}
for any $w \in W^+ \otimes W^+$. Further, we obtain
\begin{align}
e_0 \, w &= (b_A^\dagger \, \zeta_1^{s_0} + b_B^\dagger \, q^{- 2 N_A} \, \zeta_2^{s_0}) \, w, \\
e_1 \, w &= \kappa_q^{-1} (b_A \, q^{- N_A} \, \zeta_1^{s_1} + b_B \, q^{2 N_A - N_B} \,
\zeta_2^{s_1}) \, w. \label{ezwiti}
\end{align}

Remind that $W^+$ is the free vector space generated by the set $\{v_n\}_{n \in \bbZ_{\ge 0}}$. The
vectors $v_n \otimes v_k$, $n, k \in \bbZ_{\ge 0}$, form a basis of $W^+ \otimes W^+$. Consider
another basis $\{w_{n, k} \}_{n, k \in \bbZ_{\ge 0}}$, where
\begin{equation*}
w_{n, k} = (e_0)^n \, (\zeta_2^{s_1} b_A^\dagger)^k \, (v_0 \otimes v_0).
\end{equation*}
Let us show that
\begin{align}
&q^{\nu h_0} \, w_{n, k} = q^{2 \nu (n + k)} \, w_{n, k}, \label{qnhzwnk} \\[.5em]
&q^{\nu h_1} \, w_{n, k} = q^{- 2 \nu (n + k)} \, w_{n, k}, \\[.5em]
&e_0 \, w_{n, k} = w_{n + 1, k}, \\[.5em]
&e_1 \, w_{n, k} = \kappa_q^{-1} q^{-2 n - k} [k]_q \, (\zeta_1 \zeta_2)^{s_1} \, w_{n, k - 1} \notag \\*[.5em]
& \hspace{3em} {} + \kappa_q^{-1} q^{- n} [n]_q \, (\zeta_1^s + \zeta_2^s)
\, w_{n - 1, k} + q^{-1} [n]_q [n - 1]_q \, (\zeta_1 \zeta_2)^{s_0} \, w_{n - 2, k + 1}. \label{eownk}
\end{align}
In fact, the first three equations are evident, and one should prove only the last one. To this
end, we move $e_1$ through $e_0$ and introduce some operators arising during this process. Then we
move these operators through the remaining factors $e_0$, and so on. The process terminates when we
arrive at an operator which can be moved through $e_0$ without introducing new operators. To
finish, we determine the action of $e_1$ and all new operators on the vectors of the form~$w_{0,
k}$.

We start with defining an operator $x$ by the equation
\begin{equation}
x \, w = (e_1 \, e_0 - q^{-2} e_0 \, e_1) \, w \label{dx}
\end{equation}
for any $w \in W^+ \otimes W^+$. Explicitly, we have
\begin{equation}
x \, w = \left[ \kappa_q^{-1} q^{-1} (\zeta_1^s + \zeta_2^s) + (q + q^{-1}) \, \zeta_1^{s_0}
\zeta_2^{s_1} \, b_A^\dagger b_B^{\mathstrut} \, q^{2 N_A - N_B} \right] w. \label{xwiti}
\end{equation}
Now we move $x$ and introduce the operator $y$ as
\begin{equation}
y \, w = (x \, e_0 - e_0 \, x) \, w, \label{dy}
\end{equation}
or, explicitly,
\begin{equation}
y \, w = \left[ q^{-1}(q + q^{-1}) \, \zeta_1^{s_0} \zeta_2^s \, b_A^\dagger + \kappa_q \, q (q +
q^{-1}) \, \zeta_1^{2 s_0} \zeta_2^{s_1}(b_A^\dagger)^2 b_B^{\mathstrut} \, q^{2 N_A - N_B} \right]
w. \label{ywiti}
\end{equation}
One can verify that
\begin{equation}
(y \, e_0 - q^2 e_0 \, y) w = 0. \label{ye0}
\end{equation}
In fact, this equation is a consequence of the Serre relations (\ref{sre}) and its validity does
not depend on the used representation of the quantum group $\uqbp$. Equations (\ref{dx}),
(\ref{dy}) and (\ref{ye0}) give
\begin{equation*}
e_1 (e_0)^n \, w = \left[ q^{-2 n} (e_0)^n e_1 + q^{- n + 1} [n]_q (e_0)^{n - 1} x + (q +
q^{-1})^{-1} [n]_q [n - 1]_q (e_0)^{n - 2} y \right] w,
\end{equation*}
and we obtain
\begin{multline}
e_1 \, w_{n, k} = e_1 (e_0)^n \, w_{0, k} = \Bigl[ q^{-2 n} (e_0)^n e_1 + q^{- n + 1} [n]_q
(e_0)^{n - 1} x \\+ (q + q^{-1})^{-1} [n]_q [n - 1]_q (e_0)^{n - 2} y \Bigr] \, w_{0, k}.
\label{e1wnk}
\end{multline}
Using the explicit relations (\ref{ezwiti}), (\ref{xwiti}) and (\ref{ywiti}), we see that
\begin{align*}
e_1 \, w_{0, k} &= \kappa_q^{-1} q^{-k} [k]_q \, (\zeta_1 \zeta_2)^{s_1} \, w_{0, k - 1}, \\[.5em]
x \, w_{0, k} &= \kappa_q^{-1} q^{-1} (\zeta_1^s + \zeta_2^s) \, w_{0, k}, \\[.5em]
y \, w_{0, k} &= q^{-1} (q + q^{-1}) \, (\zeta_1 \zeta_2)^{s_0} \, w_{0, k + 1}.
\end{align*}
Now, it is easy to see that equation (\ref{eownk}) is true.

\subsubsection{\texorpdfstring{Tensor product of representations $\overline \rho^-_{\zeta_1}$
and $ \overline \rho^-_{\zeta_2}$}{Tensor product of representations r-z1 and r-z2}}

To analyse the product of the universal $Q$-operators $\overline \calQ'(\zeta_1)$ and $\overline
\calQ'(\zeta_2)$, we use the tensor product of the representations $\overline \rho^-_{\zeta_1}$ and
$ \overline \rho^-_{\zeta_2}$. Here we obtain that
\begin{equation*}
q^{\nu h_0} \, w = q^{- 2 \nu (N_A + N_B)} \, w, \qquad q^{\nu h_1} \, w = q^{2 \nu (N_A + N_B)} \, w,
\end{equation*}
and that
\begin{align}
e_0 \, w &= \kappa_q^{-1} (b_A \, q^{- N_A} \, \zeta_1^{s_0} + b_B \, q^{2 N_A - N_B} \,
\zeta_2^{s_0}) \, w, \\ e_1 \, w &= (b_A^\dagger \, \zeta_1^{s_1} + b_B^\dagger \, q^{- 2 N_A} \,
\zeta_2^{s_1}) \, w \label{e1wiitii}
\end{align}
for any $w \in \overline W^- \otimes \overline W^-$. Introduce now a basis
\begin{equation*}
w_{n, k} = (e_0)^n \, (\zeta_1^{s_1} b_B)^k \, (u_0 \otimes u_0).
\end{equation*}
and show that
\begin{align}
&q^{\nu h_0} \, w_{n, k} = q^{2 \nu (n + k + 2)} \, w_{n, k}, \label{qnhzwiitii} \\[.5em]
&q^{\nu h_1} \, w_{n, k} = q^{- 2 \nu (n + k + 2)} \, w_{n, k}, \\[.5em]
&e_0 \, w_{n, k} = w_{n + 1, k}, \\[.5em]
&e_1 \, w_{n, k} = - q^{2 n + 2} \, [k]_q \, (\zeta_1 \zeta_2)^{s_1} \, w_{n, k - 1} \notag \\[.5em]
& \hspace{3em} {} - \kappa_q^{-1} q^n \, [n]_q \, (\zeta_1^s + \zeta_2^s) \, w_{n - 1, k} +
\kappa_q^{-1} q^k \, [n]_q [n - 1]_q \, (\zeta_1 \zeta_2)^{s_0} \, w_{n - 2, k + 1}.
\label{e1wnkiitii}
\end{align}
The first three equations are evident, and to prove the fourth one we introduce the operators $x$ and $y$ by
\begin{equation}
x \, w = (e_1 \, e_0 - q^2 e_0 \, e_1) \, w \label{dx1}
\end{equation}
and (\ref{dy}). Explicitly, we have
\begin{equation*}
x \, w = - \Bigl[ \kappa_q^{-1} q \, (\zeta_1^s + \zeta_2^s) + q^2  (q + q^{-1}) \, \zeta_1^{s_1}
\zeta_2^{s_0} \, b^\dagger_A b_B^{\mathstrut} \, q^{2 N_A - N_B} \Bigr] \, w
\end{equation*}
and
\begin{equation*}
y \, w = \Bigr[ \kappa_q^{-1} q \, (q + q^{-1}) \, \zeta_1^s \zeta_2^{s_0} \, b_B \, q^{2 N_A -
N_B} + q^4 (q + q^{-1}) \, \zeta_1^{s_1} \zeta_2^{2 s_0} \, b^\dagger_A (b_B^{\mathstrut})^2 \,
q^{4 N_A - 2N_B} \Bigl] \, w.
\end{equation*}
It follows from these relations and from (\ref{e1wiitii}) that
\begin{align*}
e_1 \, w_{0, k} &= - q^2 [k]_q \, (\zeta_1 \zeta_2)^{s_1} \, w_{0, k - 1}, \\[.5em]
x \, w_{0, k} &= - \kappa_q^{-1} q (\zeta_1^s + \zeta_2^s) \, w_{0, k}, \\[.5em]
y \, w_{0, k} &= \kappa_q^{-1} q^k (q + q^{-1}) \, (\zeta_1 \zeta_2)^{s_0} \, w_{0, k + 1}.
\end{align*}
Now, instead of (\ref{e1wnk}), we have
\begin{multline}
e_1 \, w_{n, k} = e_1 (e_0)^n \, w_{0, k} = \Bigl[ q^{2 n} (e_0)^n e_1 + q^{n - 1} [n]_q (e_0)^{n -
1} x \\+ (q + q^{-1})^{-1} [n]_q [n - 1]_q (e_0)^{n - 2} y \Bigr] \, w_{0, k}. \label{e1wnk1}
\end{multline}
Using this equation, we conclude that equation (\ref{e1wnkiitii}) is true.

\subsubsection{\texorpdfstring{Tensor product of representations $\rho^+_{\zeta_1}$ and
$\overline \rho^-_{\zeta_2}$}{Tensor product of representations r+z1 and r-z2}}

Now we consider the tensor product of the representations $\rho^+_{\zeta_1}$ and $\overline
\rho^-_{\zeta_2}$ which is necessary to analyse the product of the universal $Q$-operators
$\calQ'(\zeta_1)$ and $\overline \calQ'(\zeta_2)$. For this tensor product we obtain that
\begin{equation*}
q^{\nu h_0} \, w = q^{2 \nu (N_A - N_B)} \, w, \qquad q^{\nu h_1} \, w = q^{- 2 \nu (N_A - N_B)} \, w,
\end{equation*}
and that
\begin{align}
e_0 \, w &=  (\zeta_1^{s_0} \, b^\dagger_A + \kappa_q^{-1} \zeta_2^{s_0} \,
b_B \, q^{- 2 N_A - N_B}) \, w, \\
e_1 \, w &= (\kappa_q^{-1}  \zeta_1^{s_1} \, b_A \, q^{- N_A} + \zeta_2^{s_1} \, b_B^\dagger \,
q^{2 N_A}) \, w. \label{e1witii}
\end{align}
A convenient basis of $W^+ \otimes \overline W^-$ is formed by the vectors
\begin{equation*}
w_{n, k} = ((\zeta_1 \zeta_2)^{- s_0/2} e_0)^n (\zeta_2^{-s_1} b_B)^k \, (v_0 \otimes u_0).
\end{equation*}
Here one obtains
\begin{align}
&q^{\nu h_0} \, w_{n, k} = q^{2 \nu (n + k + 1)} \, w_{n, k}, \label{h0wnkitii} \\[.5em]
&q^{\nu h_1} \, w_{n, k} = q^{- 2 \nu (n + k + 1)} \, w_{n, k}, \label{h1wnkitii} \\[.5em]
&e_0 \, w_{n, k} = (\zeta_1 \zeta_2)^{s_0/2} \, w_{n + 1, k}, \label{e0wnkitii} \\[.5em]
&e_1 \, w_{n, k} = - q^{2 n} \, [k]_q \, w_{n, k - 1} + \kappa_q^{-1} [n]_q \, (\zeta_1 \zeta_2)^{-
s_0/2} q^{-n} (\zeta_1^{s_0 + s_1} - q^{2 n} \zeta_2^{s_0 + s_1}) \, w_{n - 1, k}.
\label{e1wnkitii}
\end{align}
Let us prove the last equation. The operators $x$ and $y$ defined by relations (\ref{dx1}) and
(\ref{dy}), act on a vector $w$ of $W^+ \otimes \overline W^-$ as
\begin{multline*}
x \, w = \Bigl[ \kappa_q^{-1} (q + q^{-1}) \, \zeta_1^s \, q^{-2 N_A} \\- \kappa_q^{-1} q \,
(\zeta_1^s + \zeta_2^s) - \kappa_q^{-1} (q + q^{-1}) \, q^2 \, \zeta_1^{s_1}\, \zeta_2^{s_0} \, b_A
\, b_B \, q^{- 3 N_A - N_B} \Bigr] w
\end{multline*}
and
\begin{multline*}
y \, w = - \Bigl[ q^{-1} (q + q^{-1}) \, \zeta_1^{s + s_0} \, b_A^\dagger \, q^{- 2 N_A} \\*[.5em]
+ \kappa_q^{-1} (q + q^{-1})^2 \, \zeta_1^s \, \zeta_2^{s_0} \, b_B \, q^{- 4 N_A - N_B}
- \kappa_q^{-1} q \, (q + q^{-1}) \, \zeta_1^s \, \zeta_2^{s_0} \, b_B \, q^{- 2 N_A - N_B} \\*[.5em]
- \kappa_q^{-1} q^4 (q + q^{-1}) \, \zeta_1^{s_1} \, \zeta_2^{2 s_0} \, b_A (b_B)^2 \, q^{- 5 N_A - 2 N_B} \Bigr] w.
\end{multline*}
It follows from these equations and from (\ref{e1witii}) that
\begin{align*}
e_1 \, w_{0, k} &= - [k]_q \, w_{0, k - 1}, \\[.5em]
x \, w_{0, k} &= \kappa_q^{-1} q^{-1} (\zeta_1^s - q^2 \zeta_2^s) \, w_{0, k}, \\[.5em]
y \, w_{0, k} &= - q^{-1} (q + q^{-1}) \, \zeta_1^s \, e_0 \, w_{0, k}.
\end{align*}
Using relation (\ref{e1wnk1}), we see that equation (\ref{e1wnkitii}) is true.

Introduce the parameters $\zeta$ and $\mu$ such that
\begin{equation}
\zeta = (\zeta_1 \zeta_2)^{1/2}, \qquad q^{\mu + 1} = (\zeta_1 / \zeta_2)^{(s_0 + s_1)/2}.\label{zeta}
\end{equation}
The inverse transformation to the parameters $\zeta_1$ and $\zeta_2$ is
\begin{equation*}
\zeta_1 = q^{(\mu + 1)/s} \zeta, \qquad \zeta_2 = q^{-(\mu + 1)/s} \zeta.
\end{equation*}
In terms of the new parameters equations (\ref{e0wnkitii}) and (\ref{e1wnkitii}) take the form
\begin{align*}
e_0 \, w_{n, k} &= \zeta^{s_0} \, w_{n + 1, k}, \\[.5em]
e_1 \, w_{n, k} &= - q^{2 n} [k]_q \, w_{n, k - 1} + \zeta^{s_1} [n]_q [\mu - n + 1]_q \, w_{n - 1, k}.
\end{align*}
Thus, we have an increasing filtration
\begin{equation*}
\{0\} = (W^+_{\zeta_1} \otimes \overline W{}^-_{\zeta_2})^{\mathstrut}_{-1} \subset (W^+_{\zeta_1}
\otimes \overline W{}^-_{\zeta_2})^{\mathstrut}_0 \subset (W^+_{\zeta_1} \otimes \overline
W{}^-_{\zeta_2})^{\mathstrut}_1 \subset \ldots
\end{equation*}
formed by the submodules
\begin{equation*}
(W^+_{\zeta_1} \otimes \overline W{}^-_{\zeta_2})^{\mathstrut}_k = \bigoplus_{\ell = 0}^k
\bigoplus_{n = 0}^\infty \bbC \, w_{n, \ell}
\end{equation*}
with the quotient modules
\begin{equation}
(W^+_{\zeta_1} \otimes \overline W{}^-_{\zeta_2})^{\mathstrut}_k / (W^+_{\zeta_1} \otimes \overline
W{}^-_{\zeta_2})^{\mathstrut}_{k - 1} \cong \widetilde V^\mu_\zeta[\xi_k]. \label{pmqm}
\end{equation}
Here $\xi_k \in \widetilde \gothh^*$ are determined by the relations
\begin{equation}
\xi_k(h_0) = \mu + 2 k + 2, \qquad \xi_k(h_1) = - \mu - 2 k - 2, \label{xik}
\end{equation}
compare (\ref{h0wnkitii}) and (\ref{h1wnkitii}) with (\ref{shqh}).

\subsubsection{\texorpdfstring{Tensor product of representations $\overline \rho^-_{\zeta_2}$
and $\rho^+_{\zeta_1}$}{Tensor product of representations r-z2 and r+z1}}

Finally, we consider the tensor pro\-duct of the representations $\overline \rho^-_{\zeta_2}$ and
$\rho^+_{\zeta_1}$. Here we see that
\begin{equation*}
q^{\nu h_0} \, w = q^{- 2 \nu (N_A - N_B)} \, w, \qquad q^{\nu h_1} \, w = q^{2 \nu (N_A - N_B)} \, w,
\end{equation*}
and that
\begin{align*}
e_0 \, w &= (\kappa^{-1}_q \zeta_2^{s_0} \, b_A \,  q^{- N_A}
+ \zeta_1^{s_0} \, b^\dagger_B \, q^{2 N_A}) \, w, \\[.5em]
e_1 \, w &= (\zeta_2^{s_1} \, b_A^\dagger + \kappa^{-1}_q \zeta_1^{s_1}\, b_B \, q^{- 2N_A - N_B}) \, w
\end{align*}
for any $w \in \overline W^- \otimes W^+$. To construct a convenient basis, we introduce an operator
$f$ acting on a vector $w \in \overline W^- \otimes W^+$ as
\begin{equation*}
f \, w = (\zeta_1^{s_1} \, b_A \, q^{- N_A - 2 N_B} + \kappa_q \, \zeta_2^{s_1} \, b_B^\dagger) \, w.
\end{equation*}
One can verify that
\begin{equation*}
(e_0 f - f e_0) \, w = 0, \qquad (e_1 f - f e_1) \, w = 0.
\end{equation*}
The basis in question is formed by the vectors
\begin{equation*}
w_{n, k} = ((\zeta_1 \zeta_2)^{-s_0 / 2} e_0)^n \, f^k \, (u_0 \otimes v_0),
\end{equation*}
and one can show that
\begin{align}
&q^{\nu h_0} \, w_{n, k} = q^{2 \nu (n + k + 1)} \, w_{n, k}, \\[.5em]
&q^{\nu h_1} \, w_{n, k} = q^{- 2 \nu (n + k + 1)} \, w_{n, k}, \\[.5em]
&e_0 \, w_{n, k} = (\zeta_1 \zeta_2)^{s_0 / 2} \, w_{n + 1, k}, \label{e0wnkiiti} \\[.5em]
&e_1 \, w_{n, k} = \kappa^{-1}_q (\zeta_1 \zeta_2)^{-s_0 / 2} q^{-n} (\zeta_1^s - q^{2 n} \zeta_2^s)
[n]_q \, w_{n - 1, k} \notag \\[.5em]
& \hspace{18em} + \kappa_q^{-1} q^{-1} [n]_q [n-1]_q \, w_{n - 2, k + 1}. \label{e1wnkiiti}
\end{align}
As before, the first three relations are evident, and to prove the last one we define the operators
$x$ and $y$ by (\ref{dx1}) and (\ref{dy}). These operators act on a vector $w \in \overline W^-
\otimes W^+$ as
\begin{equation*}
x \, w = (\kappa_q^{-1} (q + q^{-1}) \, \zeta_1^s \, q^{-2 N_B} - \kappa_q^{-1} q \,
(\zeta_1^s + \zeta_2^s) - \kappa_q \, (q + q^{-1}) \, q^2 \, \zeta_1^{s_0} \, \zeta_2^{s_1}
\, b_A^\dagger \, b_B^\dagger \, q^{2 N_A}) \, w
\end{equation*}
and
\begin{multline*}
y \, w = ((q + q^{-1}) \, q \, \zeta_1^{s_0} \, \zeta_2^s \, b_B^\dagger \, q^{2 N_A}\\[.5em]
- (q + q^{-1}) \, q^{-1} \, \zeta_1^{s + s_0} \, b_B^\dagger \, q^{2 N_A - 2 N_B} + \kappa_q^2 (q +
q^{-1}) \, q^3 \, \zeta_1^{2 s_0} \, \zeta_2^{s_1} \, b_A^\dagger (b_B^\dagger)^2 \, q^{4 N_A}) \,
w.
\end{multline*}
One can verify that
\begin{equation*}
[x, f] = 0, \qquad [y, f] = 0,
\end{equation*}
and, having in mind equation (\ref{e1wnk1}), we see that we only need to determine the action of
$e_1$, $x$ and $y$ on the vector $w_{0, 0}$. The explicit form of the action of these operators on
an arbitrary vector $\overline W^- \otimes W^+$ implies that
\begin{align*}
e_1 \, w_{0, 0} &= 0, \\[.5em]
x \, w_{0, 0} &= \kappa_q^{-1} q^{-1} (\zeta_1^s - q^2 \zeta_2^s) \, w_{0, 0}, \\[.5em]
y \, w_{0, 0} &= - (q + q^{-1}) q^{-1} \zeta_1^s e_0 \, w_{0, 0} + \kappa_q^{-1} (q + q^{-1}) q^{-1}
(\zeta_1 \zeta_2)^{s_0} f \, w_{0, 0}.
\end{align*}
Now, using (\ref{e1wnk1}), one can be convinced in the validity of (\ref{e1wnkiiti}).

Introducing the parameters $\zeta$ and $\mu$ with the help of (\ref{zeta}), we write equations
(\ref{e0wnkiiti}) and (\ref{e1wnkiiti}) as
\begin{align*}
e_0 \, w_{n, k} &= \zeta^{s_0} \, w_{n + 1, k}, \\[.5em]
e_1 \, w_{n, k} &= \zeta^{s_1} [n]_q [\mu - n + 1]_q \, w_{n - 1, k}
+ \kappa_q^{-1} q^{-1} [n]_q [n - 1]_q \, w_{n - 2, k + 1}.
\end{align*}
Thus, we have a decreasing filtration
\begin{equation*}
 \overline W^-_{\zeta_2} \otimes W{}^+_{\zeta_1} = (\overline W^-_{\zeta_2}
 \otimes W{}^+_{\zeta_1})^{\mathstrut}_{-1} \supset (\overline W^-_{\zeta_2}
 \otimes W{}^+_{\zeta_1})^{\mathstrut}_0 \supset (\overline W^-_{\zeta_2}
 \otimes W{}^+_{\zeta_1})^{\mathstrut}_1 \supset \ldots
\end{equation*}
with the submodules
\begin{equation*}
(\overline W^-_{\zeta_2} \otimes W{}^+_{\zeta_1})^{\mathstrut}_k
= \bigoplus_{\ell = k + 1}^\infty \bigoplus_{n = 0}^\infty \bbC \, w_{n, \ell}
\end{equation*}
and the quotient modules
\begin{equation}
(\overline W^-_{\zeta_2} \otimes W{}^+_{\zeta_1})^{\mathstrut}_{k - 1} /
(\overline W^-_{\zeta_2} \otimes W{}^+_{\zeta_1})^{\mathstrut}_k
\cong \widetilde V^\mu_\zeta[\xi_k], \label{mpqm}
\end{equation}
where $\xi_k \in \widetilde \gothh^*$ are determined by relations (\ref{xik}).

\subsection{Commutativity relations}

First, it is worth to note that since for any $\nu \in \bbC$ the element $q^{\nu h_1}$ is an invertible
group-like element of $\uqlsl$ commuting with the twist element $q^{t h_1}$, we have
\begin{equation*}
[q^{\nu h_1}, \calT_\mu(\zeta)] = 0, \qquad [q^{\nu h_1}, \calQ'(\zeta)] = 0, \qquad [q^{\nu h_1},
\overline \calQ'(\zeta)] = 0,
\end{equation*}
see sections \ref{s:Uto} and \ref{sss:uqo}.

As we noted before, there are functional relations which are due only to the fact that the
universal transfer operators and universal $Q$-operators are constructed from the universal
$R$-matrices. These are the commutativity relations for the universal transfer operators
\begin{equation*}
[\widetilde \calT_{\mu_1}(\zeta_1), \widetilde \calT_{\mu_2}(\zeta_2)] = 0, \qquad
[\calT_{m_1}(\zeta_1), \calT_{m_2}(\zeta_2)] = 0, \qquad [\widetilde \calT_\mu(\zeta_1),
\calT_m(\zeta_2)] = 0,
\end{equation*}
see relation (\ref{T1T2}), and the commutativity of the universal transfer operators and
the universal $Q$-operators
\begin{gather*}
[\widetilde \calT_\mu(\zeta_1), \calQ'(\zeta_2)] = 0, \qquad
[\widetilde \calT_\mu(\zeta_1), \overline \calQ'(\zeta_2)] = 0, \\[.5em]
[\calT_m(\zeta_1), \calQ'(\zeta_2)] = 0, \qquad [\calT_m(\zeta_1), \overline \calQ'(\zeta_2)] = 0,
\end{gather*}
see relation (\ref{QTTQ}).

Another set of commutativity relations follows from the properties of the representations used to
define the universal transfer operators and universal $Q$-operators. Note that relations
(\ref{qnhzwnk})--(\ref{eownk}) are symmetric with respect to $\zeta_1$ and $\zeta_2$. It is not
difficult to understand that this fact implies the equation
\begin{equation*}
[\calQ'(\zeta_1), \calQ'(\zeta_2)] = 0.
\end{equation*}
Similarly, relations (\ref{qnhzwiitii})--(\ref{e1wiitii}) are symmetric with respect to $\zeta_1$
and $\zeta_2$, therefore,
\begin{equation*}
[\overline \calQ'(\zeta_1), \overline \calQ'(\zeta_2)] = 0.
\end{equation*}
Further, comparing (\ref{pmqm}) and (\ref{mpqm}) we conclude that
\begin{equation*}
[\calQ'(\zeta_1), \overline \calQ'(\zeta_2)] = 0.
\end{equation*}

\subsection{\texorpdfstring{Universal $TQ$-relations}{Universal TQ-relations}}

It follows from relations (\ref{pmqm}) and (\ref{tsh}) that
\begin{multline*}
\calQ'(q^{(\mu + 1)/s} \zeta) \overline \calQ'(q^{-(\mu + 1)/s} \zeta) \\
= \widetilde \calT_\mu(\zeta) \, q^{-(\mu / 2 + 1)(h_1 + 2 \phi)}
\sum_{k = 0}^\infty q^{- (h_1 + 2 \phi) k} = \widetilde \calT_\mu(\zeta) \,
\frac{q^{-(\mu / 2 + 1)(h_1 + 2 \phi)}}{1 - q^{- h_1 - 2 \phi}},
\end{multline*}
where the parameters $\zeta$ and $\mu$ are defined by relations (\ref{zeta}). We can write
\begin{equation}
\widetilde \calT_\mu(\zeta) = \calC \, q^{(\mu + 1) (h_1 / 2 + \phi)}
\calQ'(q^{(\mu + 1)/s} \zeta) \overline \calQ'(q^{-(\mu + 1)/s} \zeta), \label{tqpqp}
\end{equation}
where
\begin{equation*}
\calC = q^{h_1/2 + \phi} - q^{- h_1/2 - \phi}.
\end{equation*}
It is convenient to redefine the universal $Q$-operators as
\begin{equation*}
\calQ(\zeta) = \zeta^{s h_1/4} \calQ'(\zeta), \qquad \overline \calQ(\zeta)
= \zeta^{- s h_1/4} \, \overline \calQ'(\zeta).
\end{equation*}
In accordance with our convention (\ref{zetazetaw}), we assume that
\begin{equation*}
\zeta^{\nu h_1} = q^{\nu u h_1}.
\end{equation*}
The new universal $Q$-operators commute with the universal transfer operators:
\begin{gather}
[\widetilde \calT_\mu(\zeta_1), \calQ(\zeta_2)] = 0, \qquad
[\widetilde \calT_\mu(\zeta_1), \overline \calQ(\zeta_2)] = 0, \label{comtt} \\[.5em]
[\calT_m(\zeta_1), \calQ(\zeta_2)] = 0, \qquad [\calT_m(\zeta_1), \overline \calQ(\zeta_2)] = 0,
\end{gather}
and among themselves:
\begin{gather}
[\calQ(\zeta_1), \calQ(\zeta_2)] = 0, \qquad [\overline \calQ(\zeta_1), \overline \calQ(\zeta_2)] = 0, \\
[\calQ(\zeta_1), \overline \calQ(\zeta_2)] = 0. \label{comqq}
\end{gather}
Equation (\ref{tqpqp}) takes the form
\begin{equation}
\widetilde \calT_\mu(\zeta) = q^{(\mu + 1) \phi} \calC \, \calQ(q^{(\mu + 1)/s} \zeta) \overline
\calQ(q^{-(\mu + 1)/s} \zeta), \label{ttmu}
\end{equation}
and we write
\begin{equation}
\widetilde \calT_\mu(q^{\nu/s} \zeta) = q^{(\mu + 1) \phi} \calC \, \calQ(q^{(\mu + \nu + 1)/s}
\zeta) \overline \calQ(q^{- (\mu - \nu + 1)/s} \zeta). \label{tctqq}
\end{equation}
Now, introducing new parameters
\begin{equation}
\alpha = \mu + 1 + \nu, \qquad \beta = - (\mu + 1) + \nu, \label{pab}
\end{equation}
so that
\begin{equation*}
\mu = (\alpha - \beta) / 2 - 1, \qquad \nu = (\alpha + \beta) / 2,
\end{equation*}
we come to the equation
\begin{equation}
\widetilde \calT_{(\alpha - \beta)/2 - 1}(q^{(\alpha + \beta)/2s} \zeta)
= q^{(\alpha - \beta) \phi / 2} \calC \, \calQ(q^{\alpha/s}  \zeta)
\overline \calQ(q^{\beta/s} \zeta). \label{ttmu1}
\end{equation}
Using (\ref{ttmu}), we easily obtain that
\begin{equation}
q^{\gamma \phi / 2} \widetilde \calT_{(\alpha - \beta)/2 - 1}(q^{(\alpha + \beta)/2s} \zeta)
\calQ(q^{\gamma/s} \zeta) = q^{\alpha \phi / 2} \widetilde \calT_{(\gamma - \beta)/2 -
1}(q^{(\gamma + \beta)/2s} \zeta) \calQ(q^{\alpha/s} \zeta) \label{ttq}
\end{equation}
and that
\begin{multline}
q^{- \gamma \phi / 2} \widetilde \calT_{(\alpha - \beta)/2 - 1}(q^{(\alpha + \beta)/2s} \zeta)
\overline \calQ(q^{\gamma/s} \zeta) \\= q^{- \beta \phi / 2} \widetilde \calT_{(\alpha - \gamma)/2
- 1}(q^{(\alpha + \gamma)/2s} \zeta) \overline \calQ(q^{\beta/s} \zeta). \label{ttbq}
\end{multline}

Introduce the universal transfer operators $\calT_\mu(\zeta)$ defined with the help of the trace
$\tr_\mu$ given by equation (\ref{trmu}). It is clear that
\begin{equation}
\calT_\mu(\zeta) = \widetilde \calT_\mu(\zeta) - \widetilde \calT_{- \mu - 2}(\zeta). \label{ctmu}
\end{equation}
The universal transfer operators $\calT_\mu(\zeta)$ possess the evident property
\begin{equation*}
\calT_{- \mu - 2}(\zeta) = - \calT_\mu(\zeta).
\end{equation*}
This gives, in particular, that $\calT_{-1} = 0$. It worth to note that, as follows from the
explicit expression for the universal $R$-matrix, $\calT_0 = 1$.

Equations (\ref{ctmu}) and (\ref{ttmu}) give
\begin{multline}
\calT_\mu(\zeta) = \calC \Bigl[ q^{(\mu + 1)\phi} \calQ(q^{(\mu + 1)/s}  \zeta) \overline
\calQ(q^{- (\mu + 1)/s} \zeta) \\*- q^{-(\mu + 1) \phi} \calQ(q^{-(\mu + 1)/s}  \zeta) \overline
\calQ(q^{(\mu + 1)/s} \zeta) \Bigr]. \label{ctqq}
\end{multline}
In particular, for $\mu = 0$ we have the Wronskian-type relation
\begin{equation*}
\calC \Bigl[ q^\phi \calQ(q^{1/s}  \zeta) \overline \calQ(q^{- 1/s} \zeta) - q^{-\phi}
\calQ(q^{-1/s} \zeta) \overline \calQ(q^{1/s} \zeta) \Bigr] = 1.
\end{equation*}

It is easy to obtain from (\ref{ctqq}) the equation
\begin{multline*}
\calT_{(\alpha - \beta)/2 - 1}(q^{(\alpha + \beta)/2s} \zeta) = \calC \Bigl[ q^{(\alpha -
\beta)\phi / 2} \calQ(q^{\alpha/s}  \zeta) \, \overline \calQ(q^{\beta/s} \zeta) \\- q^{(\beta -
\alpha) \phi / 2} \calQ(q^{\beta/s} \zeta) \, \overline \calQ(q^{\alpha/s} \zeta) \Bigr],
\end{multline*}
which implies that
\begin{multline}
q^{\gamma \phi / 2} \calT_{(\alpha - \beta)/2 - 1}(q^{(\alpha + \beta)/2s} \zeta) \calQ(q^{\gamma /
s} \zeta) + q^{\alpha \phi / 2} \calT_{(\beta - \gamma)/2 - 1}(q^{(\beta + \gamma)/2s} \zeta)
\calQ(q^{\alpha / s} \zeta) \\*[.5em] + q^{\beta \phi / 2} \calT_{(\gamma - \alpha)/2 -
1}(q^{(\gamma + \alpha)/2s} \zeta) \calQ(q^{\beta / s} \zeta) = 0 \label{utqr}
\end{multline}
and
\begin{multline}
q^{- \gamma \phi / 2} \calT_{(\alpha - \beta)/2 - 1}(q^{(\alpha + \beta)/2s} \zeta) \overline
\calQ(q^{\gamma / s} \zeta) + q^{-\alpha \phi / 2} \calT_{(\beta - \gamma)/2 - 1}(q^{(\beta +
\gamma)/2s} \zeta) \overline \calQ(q^{\alpha / s} \zeta) \\*[.5em] + q^{-\beta \phi / 2}
\calT_{(\gamma - \alpha)/2 - 1}(q^{(\gamma + \alpha)/2s} \zeta) \overline \calQ(q^{\beta / s}
\zeta) = 0. \label{utbqr}
\end{multline}
We call equations (\ref{utqr}) and (\ref{utbqr}) the {\em universal $TQ$-relations\/}. Putting
\begin{equation*}
\alpha = \gamma - 2, \qquad \beta = \gamma + 2,
\end{equation*}
we obtain the relations of more usual form,
\begin{gather}
\calT(\zeta) \calQ(\zeta) = q^\phi \calQ(q^{2 / s} \zeta) + q^{-\phi} \calQ(q^{- 2 / s} \zeta),
\label{ctcq} \\[.5em]
\calT(\zeta) \overline \calQ(\zeta) = q^{-\phi} \overline \calQ(q^{2 / s} \zeta) + q^{\phi}
\overline \calQ(q^{- 2 / s} \zeta), \label{ctcbq}
\end{gather}
where
\begin{equation*}
\calT(\zeta) = \calT_1(\zeta) = - \calT_{-3}(\zeta).
\end{equation*}

\subsection{\texorpdfstring{Universal $TT$-relations}{Universal TT-relations}}

Using relation (\ref{ttmu1}), we obtain from (\ref{ttq}), or from (\ref{ttbq}), the equation
\begin{multline*}
\widetilde \calT_{(\alpha - \beta)/2 - 1}(q^{(\alpha + \beta)/2s} \zeta) \widetilde \calT_{(\gamma
- \delta)/2 - 1}(q^{(\gamma + \delta)/2s} \zeta) \\= \widetilde \calT_{(\gamma - \beta)/2 -
1}(q^{(\gamma + \beta)/2s} \zeta) \widetilde \calT_{(\alpha - \delta)/2 - 1}(q^{(\alpha +
\delta)/2s} \zeta).
\end{multline*}
For the universal transfer operators $\calT_\mu(\zeta)$ defined by (\ref{ctmu}) we obtain
\begin{multline*}
\calT_{(\alpha - \beta)/2 - 1}(q^{(\alpha + \beta)/2s} \zeta) \calT_{(\gamma - \delta)/2 -
1}(q^{(\gamma + \delta)/2s} \zeta) \\*[.5em] = \calT_{(\alpha - \gamma)/2 - 1}(q^{(\alpha +
\gamma)/2s} \zeta) \calT_{(\beta - \delta)/2 - 1}(q^{(\beta + \delta)/2s} \zeta) \\*[.5em]-
\calT_{(\beta - \gamma)/2 - 1}(q^{(\beta + \gamma)/2s} \zeta) \calT_{(\alpha - \delta)/2 -
1}(q^{(\alpha + \delta)/2s} \zeta).
\end{multline*}
We call these relations the {\em universal $TT$-relations\/}. There are two interesting special
cases of these relations. In the first case we put
\begin{equation*}
\alpha = \gamma + 2, \qquad \beta = \delta + 2
\end{equation*}
and obtain
\begin{equation}
\calT_\mu(q^{1/s} \zeta) \calT_\mu(q^{-1/s} \zeta)
= 1 + \calT_{\mu - 1}(\zeta) \calT_{\mu + 1}(\zeta), \label{tt1tt}
\end{equation}
where $\mu = (\gamma - \delta) / 2 - 1$. In the second case we put
\begin{equation*}
\alpha = \gamma + 2, \qquad \beta = \gamma - 2
\end{equation*}
and obtain
\begin{equation}
\calT(\zeta) \calT_\mu(q^{-(\mu + 1)/s} \zeta) = \calT_{\mu + 1}(q^{- \mu/s} \zeta) + \calT_{\mu -
1}(q^{-(\mu + 2)/s} \zeta), \label{tttt}
\end{equation}
where again $\mu = (\gamma - \delta)/2 - 1$.

\subsection{Six-vertex model}

The six-vertex model arises when we use for the second factor of the tensor product $\uqlsl \otimes
\uqlsl$ the representation $\varphi^1_{\eta_1} \otimes_{\Delta^\op} \ldots \otimes_{\Delta^\op}
\varphi^1_{\eta_n}$. It is convenient to introduce in this case the transfer matrices
\begin{equation*}
\bbT^{\rmp}_\mu(\zeta | \eta_1, \ldots, \eta_n) = q^{n/2} \prod_{i = 1}^n \Bigl[ (\zeta
\eta^{-1}_i)^{-s/2} \, \rme^{- \lambda_2(q^\mu (\zeta \eta^{-1}_i)^s) - \lambda_2(q^{-\mu - 2}
(\zeta \eta^{-1}_i)^s)} \Bigr] \, \bbT_\mu(\zeta | \eta_1, \ldots, \eta_n),
\end{equation*}
being a Laurent polynomial in $\zeta^{s/2}$. Similarly, we define the $Q$-operators
\begin{equation*}
\bbQ^{\rmp}(\zeta | \eta_1, \ldots, \eta_n) = \zeta^{- n s / 4} \, \prod_{i = 1}^n \rme^{-
\lambda_2(q^{-1}(\zeta \eta^{-1}_i)^s)} \, \bbQ(\zeta | \eta_1, \ldots, \eta_n)
\end{equation*}
and
\begin{equation*}
\overline \bbQ^{\rmp}(\zeta | \eta_1, \ldots, \eta_n) = \zeta^{- n s / 4} \prod_{i
= 1}^n \rme^{- \lambda_2(q^{-1}(\zeta \eta^{-1}_i)^s)} \, \overline \bbQ(\zeta | \eta_1,
\ldots, \eta_n)
\end{equation*}
also being Laurent polynomials in $\zeta^{s/2}$. It is clear that the introduced transfer matrices
and $Q$-operators satisfy the commutativity relations which follow from
(\ref{comtt})--(\ref{comqq}). Now, starting from (\ref{ctcq}) and using identity (\ref{l2l2}), we
obtain
\begin{multline}
\bbT^{\rmp}(\zeta | \eta_1, \ldots, \eta_n) \bbQ^{\rmp}(\zeta | \eta_1, \ldots, \eta_n)
\\*= q^\phi \prod_{i = 1}^n a((\zeta \eta^{-1}_i)^{-s/ 2}) \,
\bbQ^{\rmp}(q^{2 / s} \zeta | \eta_1, \ldots, \eta_n) \\*+ q^{-\phi}
\prod_{i = 1}^n b((\zeta \eta^{-1}_i)^{- s / 2}) \, \bbQ^{\rmp}(q^{- 2 / s} \zeta | \eta_1, \ldots, \eta_n),
\label{tpq}
\end{multline}
while relation (\ref{ctcbq}) gives
\begin{multline*}
\bbT^{\rmp}(\zeta | \eta_1, \ldots, \eta_n) \overline \bbQ^{\rmp}(\zeta | \eta_1, \ldots, \eta_n)
\\*= q^{-\phi} \prod_{i = 1}^n a((\zeta \eta^{-1}_i)^{-s / 2}) \,
\overline \bbQ^{\rmp}(q^{2 / s} \zeta | \eta_1, \ldots, \eta_n)
\\*+ q^\phi \prod_{i = 1}^n b((\zeta \eta^{-1}_i)^{- s / 2}) \,
\overline \bbQ^{\rmp}(q^{- 2 / s} \zeta | \eta_1, \ldots, \eta_n).
\end{multline*}
Here the functions $a(\zeta)$ and $b(\zeta)$ are defined by (\ref{bw}). For $s = -2$ equation
(\ref{tpq}) coincides with the Baxter's $TQ$-equation (\ref{TQ}).

Similarly, we obtain from (\ref{tt1tt}) the relation
\begin{multline*}
\mathbb T^{\rmp}_\mu(q^{1/s} \zeta| \eta_1, \ldots, \eta_n) \mathbb T^{\rmp}_\mu(q^{-1/s} \zeta|
\eta_1, \ldots, \eta_n) \\= \prod_{i = 1}^n a(q^{\mu / 2} (\zeta \eta^{-1}_i)^{-s / 2}) b(q^{-\mu /
2} (\zeta \eta^{-1}_i)^{-s / 2}) \\+ \mathbb T^{\rmp}_{\mu - 1}(\zeta | \eta_1, \ldots, \eta_n)
\mathbb T^{\rmp}_{\mu + 1}(\zeta | \eta_1, \ldots, \eta_n),
\end{multline*}
and from (\ref{tttt}) the relation
\begin{multline*}
\bbT^{\rmp}(\zeta | \eta_1, \ldots, \eta_n) \bbT^{\rmp}_\mu(q^{-(\mu + 1)/s} \zeta | \eta_1,
\ldots, \eta_n) \\*
= \prod_{i = 1}^n a((\zeta \eta^{-1}_i)^{-s / 2}) \, \bbT^{\rmp}_{\mu + 1}(q^{-
\mu / s} \zeta | \eta_1, \ldots, \eta_n) \\*
+ \prod_{i = 1}^n b((\zeta \eta^{-1}_i)^{- s / 2}) \,
\bbT^{\rmp}_{\mu - 1} (q^{-(\mu + 2) / s} \zeta | \eta_1, \ldots, \eta_n).
\end{multline*}
Thus, we obtain all known functional relations for the six-vertex model as a consequence of the
universal $TQ$- and $TT$-relations.

As we noted before, in the case of an even $n$ the limit $\phi \to 0$ is singular for the matrices
$\bbQ(\zeta | \eta_1, \ldots, \eta_n)$ and $\overline \bbQ(\zeta | \eta_1, \ldots, \eta_n)$.
However, some linear combinations of these operators with matrix coefficients are finite. One can
find such combinations using the observation made by Pronko \cite{Pro00} on the relation of
transfer matrices and $Q$-operators, see the paper \cite{BazLukMenSta10} for the limiting case $q =
1$.

\section{Conclusions}

We made an attempt to collect and organize general definitions and facts on the application of
quantum groups to the construction of functional relations in the theory of integrable systems. As
an example, we reconsidered the case of the quantum group $\uqlsl$ related to the six-vertex model.
We proved the full set of the functional relations in the form independent of the representation of
the quantum group in the quantum space and specialized them to the case of the six-vertex model.
There are three sets of functional relations. The first set consists of commutativity relations
satisfied by the universal transfer operators and universal $Q$-operators. The second set is formed
by the universal $TQ$-relations which are the origin of the Baxter's $TQ$-equations. The third set
is formed by the universal $TT$-relations generating various fusion relations. The specialization
of the universal $TQ$-relations and universal $TT$-relations to the case of the six-vertex model in
the limiting case $q = 1$ was obtained by other methods in the papers \cite{DerMan06,
BazLukMenSta10}. In fact, the universal $TQ$-relations and universal $TT$-relations have similar
structures. It seems that they can be combined into one set of relations, see the paper
\cite{BazFraLukMenSta11} for the case of integrable systems related to Yangians $Y(\gothgl_n)$.

\vskip 5mm

{\em Acknowledgments.\/} This work was supported in part by the DFG grant KL \hbox{645/10-1} and
by the Volkswagen Foundation. Kh.S.N. and A.V.R. were supported in part by the RFBR grants
\#~10-01-00300 and \#~13-01-00217. Kh.S.N. was supported in part by the grant N.Sh.-5590.2012.2.
A.V.R. would like to thank the Max Planck Institute for Mathematics in Bonn for the hospitality
extended to him during his stay in February-May 2012.

\appendix

\section{Endomorphism algebra} \label{a:ea}

Let $V$ be a finite-dimensional vector space of dimension $n$, and $\{e_a\}$ a basis of $V$. For
any pair of indices $a$ and $b$ define an endomorphism $E_{a b} \in \End(V)$ by the equation
\begin{equation*}
E_{a b} e_c = e_a \delta_{b c}.
\end{equation*}
The endomorphisms $E_{ab}$ satisfy the relation
\begin{equation}
E_{ab} E_{cd} = \delta_{bc} E_{ad}. \label{EE}
\end{equation}
One can verify that $\{E_{ab}\}$ is a basis of $\End(V)$, so that any endomorphism $M \in \End(V)$
has a unique representation of the form
\begin{equation*}
M = \sum_{a, b} E_{ab} M_{ab}
\end{equation*}
for some $M_{a b} \in \bbC$. It is not difficult to see that
\begin{equation*}
M e_a = \sum_b e_b M_{ba}.
\end{equation*}
One can consider $M_{ab}$ as the entries of an $n \times n$ matrix called the matrix of $M$ with
respect to the basis $\{e_a\}$. From the other hand, any $n \times n$ matrix $(M_{ab})$, via the
above relation, defines an element of $\End(V)$.

In the case when $V = \bbC^n$ the algebra $\End(V)$ is identified with the algebra $\Mat_n(\bbC)$.
Here we assume that $\{e_a\}$ is the standard basis of $\bbC^n$. Hence, $E_{ab}$ in this case are
the standard matrix units.

Let $V$ and $U$ be finite-dimensional vector spaces, $\{e_a\}$ and $\{e_i\}$ their bases. One can
show that $\{e_a \otimes e_i\}$ is a basis of the vector space $V \otimes U$. Let $\{E_{ai | bj}\}$
be a basis of $\End(V \otimes U)$ corresponding to the basis $\{e_a \otimes e_i\}$, so that any
endomorphism $M \in \End(V \otimes U)$ can be uniquely represented as
\begin{equation*}
M = \sum_{a, i, b, j}E_{ai | bj} M_{ai | bj}.
\end{equation*}
It is easy to see that
\begin{equation}
E_{ai | bj} = E_{ab} \otimes E_{ij}. \label{EeqEE}
\end{equation}
This equation is a reflection of the natural isomorphism
\begin{equation*}
\End(V \otimes U) \cong \End(V) \otimes \End(U).
\end{equation*}

Now let $V$ be a finite-dimensional vector space of dimension $n$, $\{e_a\}$ a basis of $V$,
$\{E_{ab}\}$ the corresponding basis of $\End(V)$, and $A$ an algebra. It is clear that any element
$M$ of $\End(V) \otimes A$ has a unique representation of the form
\begin{equation*}
M = \sum_{a, b} E_{ab} \otimes M_{ab},
\end{equation*}
where $M_{ab}$ are elements of $A$. As before, one can consider $M_{ab}$ as the entries of $n
\times n$ matrix called the matrix of the endomorphism $M$ with respect to the basis $\{e_a\}$. Now
it is an element of $\Mat_n(A)$. Thus, we have a correspondence between the elements of $\End(V)
\otimes A$ and $\Mat_n(A)$ which is an isomorphism of algebras. If $A$ is the algebra $\End(U)$ for
some vector space $U$ we obtain the isomorphism of the algebras $\End(V \otimes U)$ and
$\Mat_n(\End(U))$. Here one can uniquely represent a general element $w \in V \otimes U$ as
\begin{equation*}
w = \sum_a e_a \otimes w_a,
\end{equation*}
where $w_a$ are elements of $U$. One can consider $w_a$ as the entries of $n \times 1$ matrix.
Hence we can identify $V \otimes U$ with the module $\Mat_{n, 1}(U)$. Here the action of an element
$M \in \End(V \otimes U)$ on an element $w \in V \otimes U$ corresponds to matrix multiplication.

The similar consideration can be performed for the case $A \otimes \End(V)$.

Introduce two useful operations for matrices with entries in algebras. First, let $M = (M_{ab}) \in
\Mat_n(A)$ and $N = (N_{ab}) \in \Mat_n(B)$, were $A$ and $B$ are some algebras. We denote
\begin{equation}
M \boxdot N = \bigl( \sum_c M_{ac} \otimes N_{cb} \bigr). \label{boxdot}
\end{equation}
The matrix $M \boxdot N$ is an element of $\Mat_n(A \otimes B)$.

Further, let $A$ be an algebra, $K = (K_{ij}) \in \Mat_m(A)$, and $L = (L_{rs}) \in \Mat_\ell(A)$. Denote
\begin{equation}
K \boxtimes L = ((K \boxtimes L)_{ir | js}) = (K_{ij} L_{rs}). \label{boxtimes}
\end{equation}
The operation $\boxtimes$ is a natural generalization of the Kronecker product of matrices to the
case of matrices with entries in a noncommutative algebra. It is clear that $K \boxtimes L \in
\Mat_{m \ell}(A)$.

\section{Symmetric group and tensor products} \label{a:sgtp}

Let $A_1$, \ldots, $A_n$ be algebras, and
\begin{equation*}
A = A_1 \otimes \ldots \otimes A_n.
\end{equation*}
Given an element $s$ of the symmetric group $\mathrm S_n$, we define
\begin{equation*}
A^s = A_{s^{-1}(1)} \otimes \ldots \otimes A_{s^{-1}(n)}.
\end{equation*}
For any $t, s \in \mathrm S_n$ we define $\Pi^t$ as an isomorphism from $A^s$ to $A^{ts}$ by the equation
\begin{equation*}
\Pi^t (a_1 \otimes \ldots \otimes a_n) = a_{t^{-1}(1)} \otimes \ldots \otimes a_{t^{-1}(n)}.
\end{equation*}
It is not difficult to show that
\begin{equation*}
\Pi^{t_1} \circ \Pi^{t_2} = \Pi^{t_1 t_2}.
\end{equation*}
for all $t_1, t_2 \in \mathrm S_n$.

Let $j_1, j_2, \ldots, j_m$ be distinct integers in the range from $1$ to $n$, and $M$ be an
element of the tensor product $A_{j_1} \otimes \ldots \otimes A_{j_m}$. Represent $M$ as a sum
\begin{equation*}
M = \sum_r a_1^r \otimes \ldots \otimes a_m^r,
\end{equation*}
where $a_\ell^r \in A_{j_\ell}$. Let $i_1, i_2, \ldots, i_m$ be another set of distinct integers in
the range from $1$ to $n$. We denote by $M^{i_1 \ldots i_m}$ the element of $A = A_1 \otimes \ldots
\otimes A_n$ which is the sum over $r$ of monomials having for each $\ell = 1, \ldots, m$ the
element $a_\ell^r$ as the factor with the number $i_\ell$, and $1$ as all remaining factors. Here
for any $s \in \mathrm S_n$ we have
\begin{equation}
\Pi^s(M^{i_1 i_2 \ldots i_m}) = M^{s(i_1) s(i_2) \ldots s(i_m)}. \label{pim}
\end{equation}
We assume certainly that $A_{j_\ell} = A_{i_\ell}$.

Now let $V_1$, \ldots, $V_n$ be vector spaces, and
\begin{equation*}
V = V_1 \otimes \ldots \otimes V_n.
\end{equation*}
Given an element $s$ of the symmetric group $\mathrm S_n$, we define
\begin{equation*}
V^s = V_{s^{-1}(1)} \otimes \ldots \otimes V_{s^{-1}(n)}.
\end{equation*}
For any $t, s \in \mathrm S_n$, we define an isomorphism $P^t$ from $V^s$ to $V^{st}$ by
\begin{equation*}
P^t(v_1 \otimes \ldots \otimes v_n) = v_{t^{-1}(1)} \otimes
\ldots \otimes v_{t^{-1}(n)}.
\end{equation*}
In fact, the definitions of $\Pi^t$ and $P^t$ coincide. We use the notation $\Pi^t$ when the tensor
products of algebras are considered and $P^t$ for the tensor products of vector spaces.

When $A_i = \End(V_i)$, we have the relation
\begin{equation}
\Pi^t (M^{i_1 \ldots i_k}) = P^t M^{i_1 \ldots i_k} (P^t)^{-1} \label{aPiM}
\end{equation}
which implies the equation
\begin{equation}
P^t M^{i_1 \ldots i_k} = M^{t(i_1) \ldots t(i_k)} P^t. \label{aPM}
\end{equation}

If $t$ is a transposition $(i j)$ we write $\Pi^{ij}$ and $P^{ij}$ instead of $\Pi^{(i j)}$ and
$P^{(i j)}$ respectively. If $n = 2$ we denote $\Pi = \Pi^{12}$ and $P = P^{12}$.

Let $\varphi_1 \colon A_1 \to B_1$ and $\varphi_2 \colon A_2 \to B_2$ be homomorphisms of algebras.
One can show that
\begin{equation}
\Pi \circ (\varphi_1 \otimes \varphi_2) = (\varphi_2 \otimes \varphi_1) \circ \Pi, \label{Pivarphi}
\end{equation}
where $\Pi \in \Hom(B_1 \otimes B_2, \, B_2 \otimes B_1)$ at the left hand side of the equation and
$\Pi \in \Hom(A_1 \otimes A_2, \, A_2 \otimes A_1)$ at the right hand side.

\section{Quasitriangular Hopf algebras} \label{a:qha}

Let $A$ be a Hopf algebra with comultiplication $\Delta$. One can show that $A$ is also a Hopf
algebra with comultiplication
\begin{equation}
\Delta^\op = \Pi \circ \Delta. \label{deltaop}
\end{equation}
The Hopf algebra $A$ is said to be {\em almost cocommutative\/} if there exists an invertible
element $\calR \in A \otimes A$ such that
\begin{equation}
\Delta^{\op}(a) = \calR \, \Delta(a) \, \calR^{-1} \label{auniR}
\end{equation}
for all $a \in A$. An almost cocommutative Hopf algebra $A$ is called {\em quasitriangular\/} if
\begin{gather}
(\Delta \otimes \id) (\calR) = \calR^{13} \calR^{23}, \label{aDeltaid} \\
(\id \otimes \Delta) (\calR) = \calR^{13} \calR^{12}. \label{aidDelta}
\end{gather}
In this case the element $\calR$ is called the {\em universal $R$-matrix\/}.

Write $\calR$ in the form $\calR = \sum_i a_i \otimes b_i$. Multiplying both sides of equation
(\ref{aDeltaid}) from the left by $\calR_{12}$, we obtain
\begin{multline*}
\calR^{12} \calR^{13} \calR^{23} = \calR^{12} (\Delta \otimes \id)(\calR) = \sum_i \calR \,
\Delta(a_i) \otimes b_i \\ \eqWithRef{auniR} \sum_i \Delta^\op(a_i) \calR \otimes b_i = \bigl(
\sum_i \Delta^\op(a_i) \otimes b_i \bigr) (\calR \otimes 1) = (\Delta^\op \otimes \id)(\calR)
\calR^{12}.
\end{multline*}
Applying now the mapping $\Pi^{12}$ to both sides of the same equation, we come to the relation
\begin{equation*}
(\Delta^\op \otimes \id)(\calR) = \calR^{23} \calR^{13}.
\end{equation*}
Hence, we see that the universal $R$-matrix satisfies the Yang-Baxter equation
\begin{equation}
\calR^{12} \, \calR^{13} \, \calR^{23} = \calR^{23} \, \calR^{13} \, \calR^{12}.
\label{aRRR}
\end{equation}

\section{\texorpdfstring{$\bbZ$-graded Hopf algebras}{Z-graded Hopf algebras}} \label{a:zgha}

If a Hopf algebra $A$ is represented as a direct sum of linear subspaces
\begin{equation*}
A = \bigoplus_{n \in \bbZ} A_n,
\end{equation*}
where
\begin{equation*}
A_n A_m \subset A_{n + m}
\end{equation*}
and
\begin{equation*}
\Delta(A_n) \subset \bigoplus_{m \in \bbZ} A_{n - m} \otimes A_m,
\end{equation*}
one says that $A$ is $\bbZ$-graded. Note that $A_0$ is a subalgebra of $A$.

Any element $a$ of a $\bbZ$-graded Hopf algebra $A$ can be uniquely represented as a sum
\begin{equation*}
a = \sum_{n \in \bbZ} a_n
\end{equation*}
with $a_n \in A_n$. Given $\nu \in \bbC^\times$, define a mapping $\Phi_\nu: A \to A$ by the equation
\begin{equation}
\Phi_\nu(a) = \sum_{n \in \bbZ} \nu^n a_n. \label{aPhinu}
\end{equation}
It is clear that $a \in A_n$ if and only if $\Phi_\nu(a) = \nu^n a$. It is also not difficult to verify that
\begin{equation}
\Phi_{\nu_1 \nu_2} = \Phi_{\nu_1} \circ \Phi_{\nu_2}. \label{aPhinunu}
\end{equation}

Any $\bbZ$-gradation of a Hopf algebra $A$ induces a $\bbZ$-gradation of the Hopf algebra $A \otimes A$, for which
\begin{equation*}
(A \otimes A)_n = \bigoplus_{m \in \bbZ} A_{n - m} \otimes A_m = \bigoplus_{m \in \bbZ} A_m \otimes A_{n - m}.
\end{equation*}
Here the role of the automorphism $\Phi_\nu$ is played by the automorphism $\Phi_\nu \otimes \Phi_\nu$.

\section{Traces on algebras} \label{a:ta}

Recall that the usual trace of a matrix $M = (M_{a b}) \in \Mat_n(\bbC)$ is defined as
\[
\tr (M) = \sum_a M_{a a}.
\]
The basic property of the trace is its cyclicity
\[
\tr(M N) = \tr(N M)
\]
for any two matrices $M, N \in \Mat_n(\bbC)$. Note also that for any two matrices $M \in
\Mat_n(\bbC)$ and $N \in \Mat_m(\bbC)$ we have
\begin{equation}
\tr(M \otimes N) = \tr(M) \, \tr(N). \label{atrPtimesQ}
\end{equation}

Let $V$ be a finite-dimensional vector space, $\{e_a\}$ a basis of $V$, and $\{E_{a b}\}$ the
corresponding basis of $\End(V)$. The trace of an element $M = \sum_{a, b} E_{a b} M_{a b}$ of
$\End(V)$ can be defined as
\begin{equation}
\tr(M) = \sum_a M_{a a}. \label{trM}
\end{equation}
It can be easily shown that $\tr(M)$ does not depend on the choice of a basis. Hence, the trace of
$M$ coincides with the standard trace of its matrix with respect to any basis of $V$. We often
denote this trace as $\tr_V$ Here, the basic property of the trace holds, namely,
\[
\tr_V(M N) = \tr_V(N M)
\]
for all $M, N \in \End(V)$. If $U$ is another finite-dimensional vector space, then
\begin{equation*}
\tr_{V \otimes U}(M \otimes N) = \tr_V(M) \tr_U(N)
\end{equation*}
for all $M \in \End(V)$ and $N \in \End(U)$.

More generally, a trace on an algebra $A$ is a linear mapping $\tr$ from $A$ to $\bbC$, which
satisfies the equation
\begin{equation}
\tr(a b) = \tr(b a) \label{trab}
\end{equation}
for all $a, b \in A$.

Multiplying a trace by a complex number we again obtain a trace. Up to this freedom the trace on
$\End(V)$, where $V$ is a finite-dimensional vector space of dimension $n$ given by (\ref{trM}), is
unique. Indeed, assume that $\tr$ is a trace on $\End(V)$. Relation (\ref{EE}) gives
\begin{equation*}
E_{a c} E_{c b} = E_{ab},
\end{equation*}
therefore,
\begin{equation*}
\tr_V(E_{a c} E_{c b}) \eqWithRef{trab} \tr_V(E_{c b} E_{a c}) \eqWithRef{EE} \delta_{a b}
\tr_V(E_{c c}) = \tr_V(E_{a b}).
\end{equation*}
The last equation and the evident identity
\begin{equation*}
\id_V = \sum_c E_{c c}
\end{equation*}
give
\begin{equation*}
n \, \tr_V(E_{a b}) = \delta_{a b} \, \tr_V(\id_V),
\end{equation*}
and we obtain the trace defined by (\ref{trM}) if we assume that $\tr (\id_V) = n$.

Let $\tr_A$ and $\tr_B$ be traces on algebras $A$ and $B$ respectively. Then $\tr_{A \otimes B} =
\tr_A \otimes \tr_B$ is a trace on $A \otimes B$ and we have
\begin{equation}
\tr_{A \otimes B}(a \otimes b) = \tr_A(a) \tr_B(b) \label{traotb}
\end{equation}
for any $a \in A$ and $b \in B$. One can also consider the partial traces $\tr_A \otimes \id$ and
$\id \otimes \tr_B$. The basic property of trace does not hold here, however,
\begin{equation}
(\tr_A \otimes \id)((a \otimes 1) c) = (\tr_A \otimes \id)(c (a \otimes 1)) \label{traid}
\end{equation}
for any $a \in A$ and $c \in A \otimes B$, and
\[
(\id \otimes \tr_B)((1 \otimes b) c) = (\id \otimes \tr_B)(c (1 \otimes b))
\]
for any $b \in B$ and $c \in A \otimes B$.

If $\varphi$ is a homomorphism from an algebra $A$ to an algebra $B$ and $\tr_B$ is a trace on $B$,
then
\begin{equation*}
\tr_A = \tr_B \circ \varphi
\end{equation*}
is a trace on $A$. In particular, if $\varphi$ is a representation of $A$ in a finite-dimensional
vector space $V$, then the mapping $\tr_V \circ \varphi$ is a trace on $A$. Note that there are
traces on algebras which cannot be obtained in this way.

In the case where $V$ is an infinite-dimensional vector space, the trace is not defined for all
elements of $\End(V)$. It particular, it is not defined for the identity mapping.

\bibliographystyle{amsrusunsrt}

\bibliography{IntegrableSystems}

\end{document}